\documentclass{aa}  

\usepackage{graphicx}
\usepackage{multirow}
\usepackage{morefloats}
\usepackage[caption = false]{subfig}
\usepackage{adjustbox}
\usepackage{xcolor}

\usepackage{txfonts}
\def\BibTeX{{\rm B\kern-.05em{\sc i\kern-.025em b}\kern-.08em
    T\kern-.1667em\lower.7ex\hbox{E}\kern-.125emX}}
\bibpunct{(}{)}{;}{a}{}{,}  
\begin{document} 

\title{Carbon stars in the X-Shooter Spectral Library: \\II. Comparison with models \thanks{
Based on observations collected at the European Southern Observatory, Paranal, Chile, Prog. ID 084.B-0869(A/B), 085.B-0751(A/B), 189.B-0925(A/B/C/D).}}

\author{A. Gonneau\inst{\ref{inst1},\ref{inst2}}
\and A. Lan\c{c}on\inst{\ref{inst1}}
\and S.~C. Trager\inst{\ref{inst2}}
\and B. Aringer\inst{\ref{inst3},\ref{inst4}}
\and W. Nowotny\inst{\ref{inst3}} 
\and R.~F. Peletier\inst{\ref{inst2}}
\and \\ P. Prugniel\inst{\ref{inst5}}
\and Y.-P. Chen\inst{\ref{inst6}}
\and M. Lyubenova\inst{\ref{inst2}}
}

\institute{Observatoire Astronomique de Strasbourg, Universit\'e de
  Strasbourg, CNRS, UMR 7550, 11 rue de l'Universit\'e, \\F-67000 Strasbourg, France\label{inst1} \\
  \email{anais.gonneau@astro.unistra.fr}
\and
Kapteyn Astronomical Institute, University of Groningen, Postbus 800, 9700 AV, Groningen, The Netherlands\label{inst2}
\and 
University of Vienna, Department of Astrophysics, T\"urkenschanzstra{\ss}e 17, 1180 Wien, Austria\label{inst3}
\and 
Dipartimento di Fisica e Astronomia Galileo Galilei,
Universit\`a di Padova, Vicolo dell'Osservatorio 3, I-35122 Padova, Italy\label{inst4}
\and 
CRAL-Observatoire de Lyon, Universit\'e de Lyon, Lyon I,  CNRS, UMR5574, France\label{inst5}
\and
New York University Abu Dhabi, Abu Dhabi, P.O. Box 129188, Abu Dhabi, United Arab Emirates\label{inst6}
}

\date{Received: 19 September 2016 ; Accepted: 20 Januray 2017}

 
  \abstract{
In a previous paper, we assembled a collection of medium-resolution spectra of 35 carbon stars,
covering optical and near-infrared wavelengths from 400 to 2400\,nm. The sample includes stars from the Milky Way and the Magellanic Clouds, with a variety of $(J-K_s)$ colors and pulsation properties. In the present paper, we compare these observations to a new set of high-resolution synthetic spectra, based on hydrostatic model atmospheres. \\
We find that the broad-band colors and the molecular-band strengths measured by spectrophotometric indices match those of the models when $(J-K_s)$ is bluer than about 1.6, 
while the redder stars require either additional reddening or dust emission or both. 
Using a grid of models to fit the full observed spectra, we estimate the most likely atmospheric parameters $T_\mathrm{eff}$, $\log(g)$, $[\mathrm{Fe/H}]$ and C/O. 
These parameters derived independently in the 
optical and near-infrared are generally consistent when $(J-K_s)<1.6$. 
The temperatures found based on either wavelength range are typically within $\pm$100K of each other, and $\log(g)$ and $[\mathrm{Fe/H}]$ are consistent with the values expected for this sample.\\
The reddest stars ($(J-K_s)$ $>$ 1.6) are divided into two families, 
characterized by the presence or absence of an absorption feature at 1.53\,$\mu$m, generally associated with HCN and C$_2$H$_2$. 
Stars from the first family begin to be more affected by circumstellar extinction. The parameters found using optical or near-infrared wavelengths are still compatible with each other, but the error bars become larger. 
In stars showing the 1.53\,$\mu$m feature, which are all large-amplitude variables, 
the effects of pulsation are strong and the spectra are poorly matched with hydrostatic models. 
For these, atmospheric parameters could not be derived reliably, and dynamical models are needed for proper interpretation.}


\maketitle
%

\section{Introduction}

Modeling the spectra of luminous red stars, such as red supergiants or 
luminous asymptotic giant branch stars, remains an immense challenge. 
These stars have hugely extended atmospheres that host molecules
and sometimes dust. Their heavy element abundance ratios are non-solar 
as the result of dredge-up episodes in their previous evolutionary history
\citep[e.g.][]{Iben83}. 
This is particularly true for carbon stars (C stars), whose
atmospheres have carbon-to-oxygen abundance ratios higher than 1.
Many -- if not all -- luminous red stars 
are photometric variables as a result of pulsation of the stellar interior \citep[e.g.][]{Wood15},
which triggers shock waves that pro\-pagate through the atmospheres.
In addition, interferometric observations have 
demonstrated significant departures from spherical symmetry in some 
of these objects, which have been interpreted as the signatures of 
large-scale convective cells. In the case of red supergiants,
these inhomogeneities have been identified as a plausible cause
of differences between the stellar effective temperatures estimated 
from optical mo\-le\-cu\-lar bands on one hand, and from the spectral energy
distribution on the other hand \citep[e.g.][]{Davies13}. Qualitatively similar inhomogeneities
have been disco\-vered in asymptotic giant branch stars \citep{vanBelle13}.

Luminous red giants can be observed at very large distances, 
and they contribute significantly to the light of ga\-la\-xies
especially at red and near-infrared (NIR) wavelengths \citep[e.g.][]{Melbourne12}. 
Their relative numbers are predicted to be sensitive functions of the age and initial 
metallicity of the host systems. 
Underestimating the contribution of the thermally pul\-sing asymptotic giant branch (TP-AGB) stars to galaxy spectral energy distributions (SEDs) can bias determinations of their properties, such as stellar masses \citep{Ilbert10}.
Their importance in studies of the stellar populations 
of local and extragalactic galaxies justifies the continuous efforts 
devoted to empirically characterizating and to modeling them \citep[e.g.][]{Lyubenova10,Lyubenova12,Zibetti13}.

In a previous article \citep[][hereafter Paper\,I]{Gonneau16}, 
we described spectroscopic observations of carbon stars 
obtained with the ESO/VLT/X-Shooter instrument as part of the 
XSL project (X-Shooter Spectral Library, PI. S.~C. Trager).
This collection is the first to provide simultaneous optical and
NIR observations at a resolving power of
$R = \lambda/\delta \lambda \sim 8\,000$.
Hence it offers a unique possibility to test 
synthetic spectra of C stars. For stars that can be reproduced
in a satisfactory way, estimates of the fundamental stellar
parameters may be derived. These estimates are essential if the empirical
spectra are to be used as templates in future studies 
of the stellar populations of galaxies.

A recent series of four articles has provided the largest 
current collection of model atmospheres and low-resolution 
theore\-tical spectra for C stars (\citealt{Aringer09}; \citealt{Nowotny11,Nowotny13}; \citealt{Eriksson14}; they are respectively referred to as Papers T1 to T4 hereafter). 
While in T1, the authors investigated synthetic spectra and photometry based on a grid of hydrostatic model atmospheres, T2 to T4 
are based on dynamical model atmospheres.
These papers demonstrate that only the SEDs of rela\-tively blue 
asymptotic giant branch C stars ($J-K\leq 1.5$) can be reproduced by 
hydrostatic mo\-dels. The SEDs of redder objects 
result from the combination of photospheric emission 
and reprocessing of this radiation by the dusty circumstellar material, which is
produced naturally as a result of pulsation.
Dynamic model atmospheres taking into account the effects of pulsation-enhanced 
dust-driven winds are needed to reproduce the obser\-vable pro\-per\-ties of such 
evolved objects (e.g. their location in a number of color--magnitude and 
color--color diagrams, their wind velocities and mass-loss rates, cf T2 and following).

In this article, we focus on the hydrostatic models presented in T1. 
\citet{Paladini11} used early synthetic spectra based on these models
to analyze low-resolution spectra ($400<R<1\,800$)
of C stars with no circumstellar dust in the near-infrared 
($0.9<\lambda<4.2\,\mu$m), together with interferometric 
observations. These models allowed the authors to select a preferred treatment
of the C$_2$ opacity. They concluded that the C$_2$H$_2$ feature at
3.1\,$\mu$m was the spectral signature most sensitive to changes in effective temperature.
A large grid of new high-resolution theore\-tical spectra
based on the hydrostatic models of T1 has been computed for the
present article.  High-resolution theore\-tical spectra for the dynamical models
of T2, T3, T4 are not yet available. 

The following questions guide our comparisons with the X\-Shooter observations.
Can the full spectra (from the optical to the near-infrared) be matched reasonably well, at least for
stars with little or no evidence of circumstellar dust?
To what extent are parameters estimated from optical wavelengths 
compatible with those obtained from NIR wavelengths?
In this article, the analysis was performed at an intermediate spectral resolution: $R \sim 2\,000$.
At the full resolution of XSL, line profiles become difficult to model, in part because of instrumental effects and in part because the velocity field in
the atmospheres of long-period va\-ria\-bles (LPVs) has noticeable consequences
\citep{Nowotny10}. This interesting aspect of the study of C-star spectra
is postponed to future articles.

The paper is organized as follows. 
Section 2 describes the observed carbon-rich spectra, and Section 3 presents the theo\-re\-ti\-cal grid of models. Section 4 shows the first part of the study: a comparison between models that helps evaluate to what precision stellar parameters can be recovered from ideal C-star spectra. 
In Section 5 we use the results from Section 4 and compare our observations with the grid of models. 
In Section 6 we summarize the
results and conclude.


\section{Observations}

For our sample of 35 carbon stars, we obtained medium-resolution spectra by using 
the European Southern Observatory (ESO/VLT) spectrograph X-Shooter \citep{Vernet11}. 
This instrument allows simultaneous acquisition
of spectra from 0.3 to 2.5\,$\mu$m, using two dichroics to split the beam into three arms: ultraviolet-blue (UVB), visible (VIS), and near-infrared (NIR). 

The C-star spectra were acquired as part of the X-Shooter Spectral Library \citep[hereafter XSL,][]{Chen14}, through an ESO Large Programme \citep{Messenger14}. This empirical library contains about 700 stars, observed at a moderate resolving power ($7\,700 \leq R \leq 11\,000$ depending
on the arm) and covering a wide range of stellar atmospheric parameters. 

Our sample of carbon stars includes stars from the Milky Way (MW) and
the Large and Small Magellanic Clouds (LMC, SMC). Details about the star selection and the data reduction can be found in Paper I. 

Our sample presents quite a diversity in global SED and absorption-line characteristics.
It exhibits a bimodal behavior of carbon stars with relatively 
red near-infrared co\-lors. Some of our carbon stars with $(J-K_s) > 1.6$ display an absorption band 
at 1.53\,$\mu$m, for which HCN and C$_2$H$_2$ are usually consi\-dered responsible \citep{Loidl04}. 
In our sample, the appearance of the 1.53\,$\mu$m feature is also associated with
a smoother aspect of the near-infrared spectrum and an energy distribution with two components, one peaking at red optical wavelengths, the other at longer wavelengths (cf. Figure 18 of Paper I). Paper I noted that all stars displaying the 1.53\,$\mu$m feature in the sample 
are large-amplitude variables, but that large-amplitude variability does not systematically imply the presence of that feature in the spectrum.

In Paper I, our sample of spectra was divided into four groups, numbered 1 to 4, based on the $(J-K_s)$ color of the corres\-ponding target. 
In the following, the discussion again uses four groups, but this time with a different focus 
because we concentrate on the spectral features in the redder objects. We reclassify our groups 1 to 4 
in Paper I into groups A to D as follows. 

The first group (A) contains the bluest C stars from our sample, with $(J-K_s) < 1.2$. It remains the same as Group 1. The second group (B) contains the classical C stars with $1.2 < (J-K_s) < 1.6$, as did Group 2. The last two groups contain all the carbon stars with $(J-K_s) > 1.6$.
We separate these stars based on the presence or absence of the 1.53\,$\mu$m absorption feature.
The stars from the former Groups 3 and 4 without this absorption feature are placed in Group C. 
All other red stars with the absorption feature are placed in Group D. 

Table~\ref{table_sample} lists the observed carbon stars used in this paper. 
All spectra are plotted in the appendix of Paper I.
Star V CrA, a star of type R Coronae Borealis (R CrB), are not be discussed in this work
because it was undergoing an obscuration event at the time of observation\footnote{See light curve, Appendix A in Paper I.}. Its spectrum is dominated by circumstellar emission to the point of not showing any photospheric features. 
T Cae, a former star from Group C, was removed from our sample as its spectrum appears to be partially saturated in the near-infrared wavelength range.



\begin{table}
\caption{\label{table_sample}Main properties of the sample of observed carbon stars}
\centering
\begin{tabular}{lccrr}
\hline\hline
Name 					& Host 	& $(J-K_s)$ 	& Gr. 	& Gr. \\
 						&			& [mag]				& 	I	&  II \\
\hline
\hline

HE 1428-1950 			& MW 		&  0.71 			& 1 	& A		\\
HD 202851 				& MW 		&  0.83			& 1 	& A		\\
Cl* NGC 121 T V8 		& SMC 		&  1.06			& 1		& A \\
SHV 0517337-725738 		& LMC 		&  1.13			& 1 	& A \\
SHV 0518161-683543 		& LMC 		&  1.16			& 1 	& A \\

\hline

2MASS J00571648-7310527 & SMC 		& 1.31			& 2		& B \\
2MASS J01003150-7307237 & SMC 		& 1.33 			& 2 	& B \\
2MASS J00563906-7304529 & SMC 		& 1.37 			& 2 	& B \\
2MASS J00530765-7307477 & SMC 		& 1.43			& 2 	& B \\
2MASS J00493262-7317523 & SMC 		& 1.44 			& 2 	& B \\
2MASS J00490032-7322238 & SMC 		& 1.50 			& 2 	& B \\
2MASS J00571214-7307045 & SMC 		& 1.54			& 2 	& B \\

\hline


2MASS J00570070-7307505 & SMC 		& 1.66 			& 3 		& C \\ 
{[}W65] c2 				& MW 		& 1.71			& 3 		& C \\
2MASS J00564478-7314347 & SMC 		& 1.77			& 3 		& C \\
2MASS J00542265-7301057 & SMC 		& 1.92			& 3 		& C\\ 
Cl* NGC 419 LE 27 		& SMC 		& 1.98			& 3 		& C \\
IRAS 09484-6242 			& MW 		& 2.02			& 3 		& C\\
Cl* NGC 419 LE 35 		& SMC 		& 2.09			& 3  	& C \\
2MASS J00553091-7310186 & SMC 		& 2.11			& 3 		& C \\
SHV 0520427-693637 		& LMC 		& 2.11			& 3  	& C \\
SHV 0504353-712622 		& LMC 		& 2.17			& 3 	& C \\ 
{[}ABC89] Pup 42 		& MW 		& 2.30			& 4 	& C \\
{[}ABC89] Cir 18 		& MW 		& 2.45 			& 4 	& C \\
{[}ABC89] Cir 18 		& MW 		& 2.52 			& 4 	& C \\ 
  
\hline

SHV 0500412-684054 		& LMC 		& 1.84 			& 3 	& D \\
SHV 0502469-692418 		& LMC 		& 1.97			& 3 	& D \\
SHV 0520505-705019 		& LMC 		& 2.37			& 4 	& D \\ 
SHV 0518222-750327 		& LMC 		& 2.52			& 4  	& D \\
SHV 0527072-701238 		& LMC 		& 2.55			& 4 	& D \\
SHV 0525478-690944	 	& LMC 		& 3.02 			& 4 	& D \\
SHV 0536139-701604 		& LMC 		& 3.12			& 4  	& D \\
SHV 0528537-695119 		& LMC 		& 3.23			& 4  	& D \\

\hline
\end{tabular}
\end{table}



\section{Models}

We use a grid of synthetic spectra computed specifically for this study and based on the C-rich COMARCS model atmospheres of T1, in an updated version as presented in \citet{Aringer16}. These represent hydrostatic dust-free carbon-rich giants under the assumption of spherical symmetry.

\subsection{Original spectral library}

The original series of models (T1) covers a wide range of effective temperatures, surface gravities, carbon-to-oxygen ratios, and a few different masses. In addition, subgrids with various metallicities were computed to reproduce stars in the Milky Way and the Magellanic Clouds.

In T1 and subsequent papers, the hydrostatic COMARCS atmospheres were then used to compute a grid of synthetic spectra covering the range between 0.444 and 25.0\,$\mu$m with a resolution of R = 10\,000. Owing to the statistical nature of the opacity sampling in these calculations, only the average over a large number of wavelength points (usually 20 to 100) gives a realistic representation of  observed stellar spectra, which reduces the useful resolution from 10 000 to a few hundred.

\subsection{New grid of synthetic spectra}

For the resolution of the X-Shooter spectra ($\simeq$ 8\,000), the o\-ri\-gi\-nal resolution 
of the synthetic spectra was insufficient. A new grid of theoretical spectra was computed using the existing atmospheric models, this time with a resolving power of 200\,000. We then smoothed the spectra to match the resolution of the observations.

The atmosphere models used for this grid assume a stellar mass of 1.0 $M_\odot$
and surface gravities of $-0.4$, 0 or 2 (values of $\log(g)$, with $g$ in cm s$^{-2}$).
The effective temperatures range from 2500 to 4000 K, with a step of 100 K.  
The main model grid has $[\mathrm{Fe/H}]=0$, and a subset of models with $[\mathrm{Fe/H}]=-0.5$ was also computed. The spectral synthesis assumes a microturbulent velocity of 2.5\,km/s, which is consistent with the opacities used to construct the models. 
Solar-scaled abundances are adopted except for carbon, which is enhanced at a given  $[\mathrm{Fe/H}]$ to sample a range of carbon-to-oxygen ratios: C/O $=$ 1.01, 1.05, 1.10, 1.40 and 2.0.

\begin{figure}
\begin{center}
	\includegraphics[trim=10 60 90 280, clip, width=\hsize]{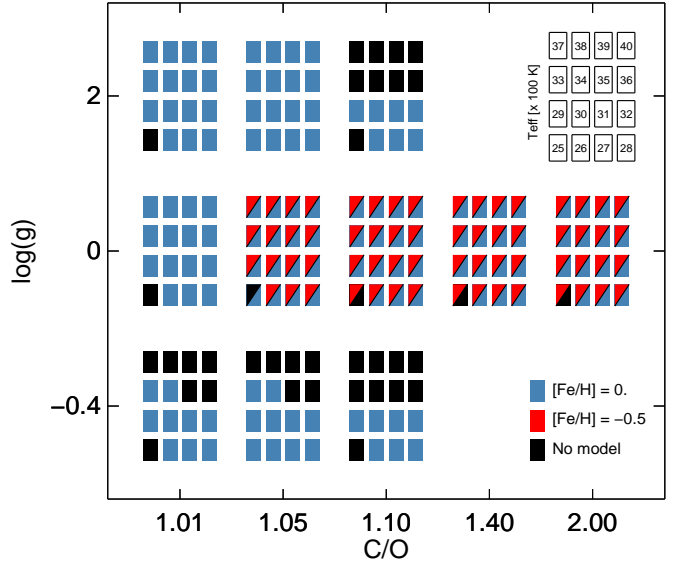}
	       \caption{Schematic overview showing the parameters of our grid of synthetic spectra. Each square corresponds to a given temperature. }
    \label{grid_models}
\end{center}
\end{figure}

Figure~\ref{grid_models} shows a schematic representation of all the synthetic spectra available, as a function of the parameters listed above.
It is important to keep in mind that although this grid is a good starting point, it does not vary all the relevant parameters. The turbulent velocity parameters and the nitrogen abundance,
for instance, are expected also to affect the relative strengths of molecular bands, 
but are set to fixed values here.

\subsection{Illustrative examples of synthetic spectra}

\begin{figure}
\begin{center}
	\includegraphics[trim=60 30 80 80, clip, width=\hsize]{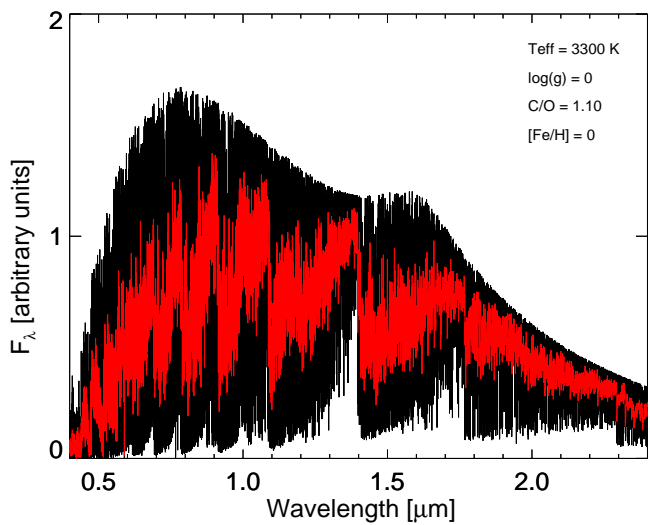}
	       \caption{Example of a synthetic spectrum at high resolution (R = 200\,000, black spectrum) and smoothed to XSL resolution (R $\simeq$ 8\,000, red spectrum).}
    \label{plot_mod_res}
\end{center}
\end{figure}

Figure~\ref{plot_mod_res} shows an example of a synthetic spectrum at high re\-so\-lu\-tion (R = 200\,000) as a black spectrum and its smoothed version (R $\simeq$ 8\,000) as a red spectrum. 
By downgrading the spectral resolution, it is worth noting that we lose direct access to the
continuum.

\begin{figure}
\begin{center}
	\includegraphics[trim=60 40 80 80, clip, width=\hsize]{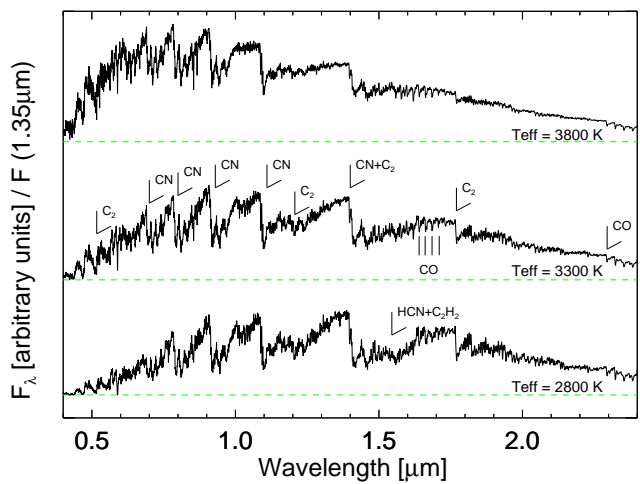}
	       \caption{Examples of synthetic spectra of C-rich giants with different temperatures ($\log(g)$ = 0, C/O = 1.10 and $[\mathrm{Fe/H}] = 0$). The spectra are smoothed to R = 2\,000 and normalized around 1.35\,$\mu$m for display purposes.}
    \label{plot_evol_teff}
\end{center}
\end{figure}

Figure~\ref{plot_evol_teff} shows three hydrostatic models that share the same properties, except for the effective temperature. The temperature decreases from the top to the bottom. 
The peak of the SED shifts from the blue to the red -- as the stars become cooler. 
The strongest bandheads and the ragged aspect of the spectra are mostly due to numerous lines from CN and C$_2$, which tend to increase in intensity with decreasing temperature. The signatures with a weaker dependence on $T_\mathrm{eff}$ over the plotted range, such as those of CO around 1.65 and 2.3\,$\mu$m, are progressively masked by the forest of other features when $T_\mathrm{eff}$ decreases.

The effects of the fundamental parameters on the spectra are not obvious to the eye in plots of the whole spectrum. They are more efficiently summarized by index measurements, as shown below.

\subsection{Molecular indices based on model spectra}
\label{index_model}

In Paper I (see Table 2), we defined spectroscopic indices for carbon stars, 
to quantify the strengths of selected spectral features. We also calculated the values of these indices for all the spectra in the model grid.

The index $DIP153$ was based on the ratio between the flux
measured near the center of the 1.53\,$\mu$m feature and the flux measured
on its short wavelength side. A plot of this index
versus $(J-K_s)$ separates the stars that display
this molecular band well (Paper I), but at a given molecular band
strength the value of the index depends strongly on color. This
dependence can be quantified by fitting the indices of
(observed) stars with no 1.53\,$\mu$m feature as a function of color.
We define a new (almost) color-independent index by sub\-trac\-ting
this trend from the original index.
\begin{equation}
DIP153b=DIP153-0.132[1.06-(J-K_s)]
\label{eq_new_dip}
\end{equation}

\begin{figure*}
\begin{center}
\subfloat{\includegraphics[trim=5 35 5 220, clip, width=0.25\hsize]{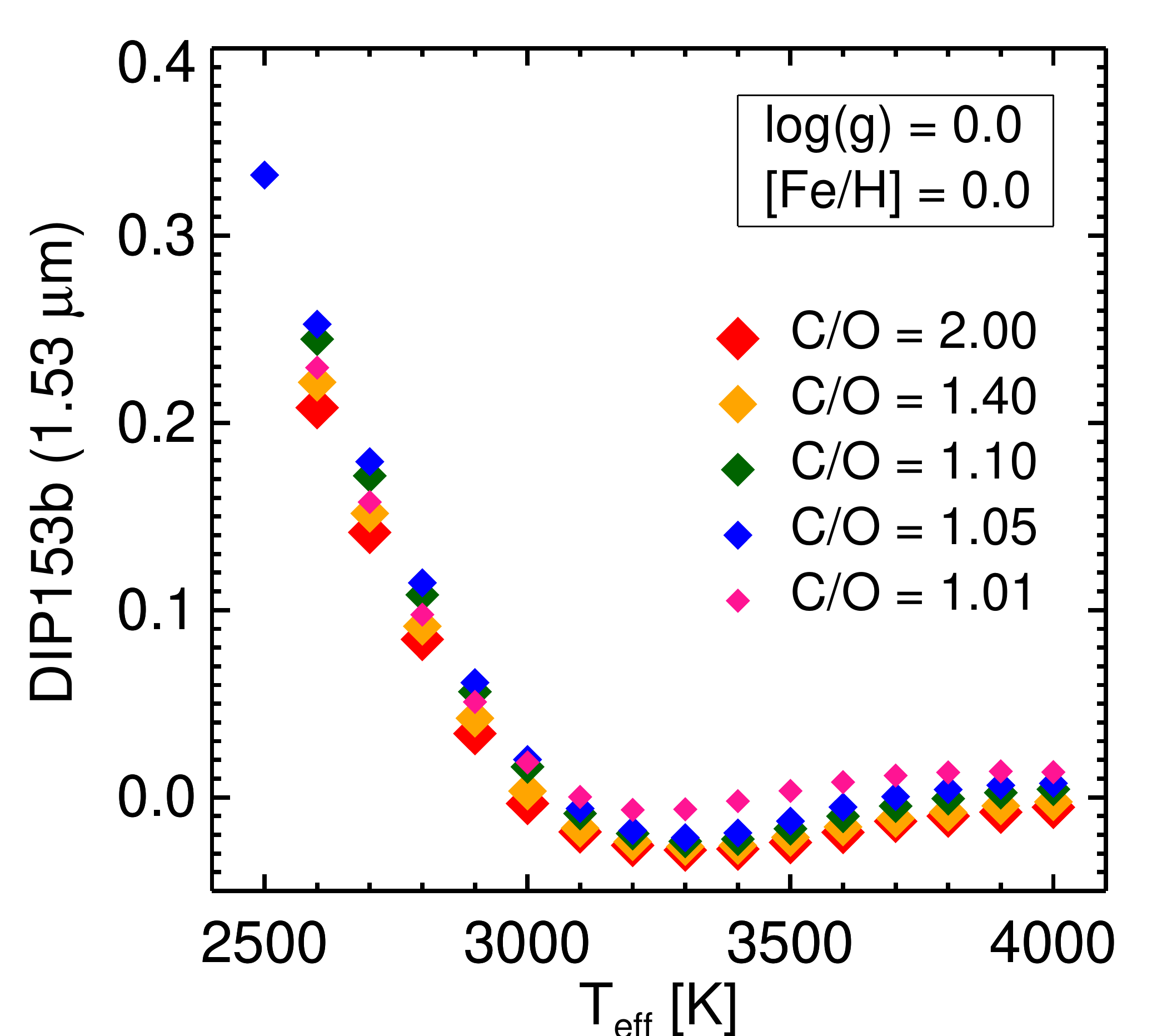}}
\subfloat{\includegraphics[trim=5 35 5 220, clip, width=0.25\hsize]{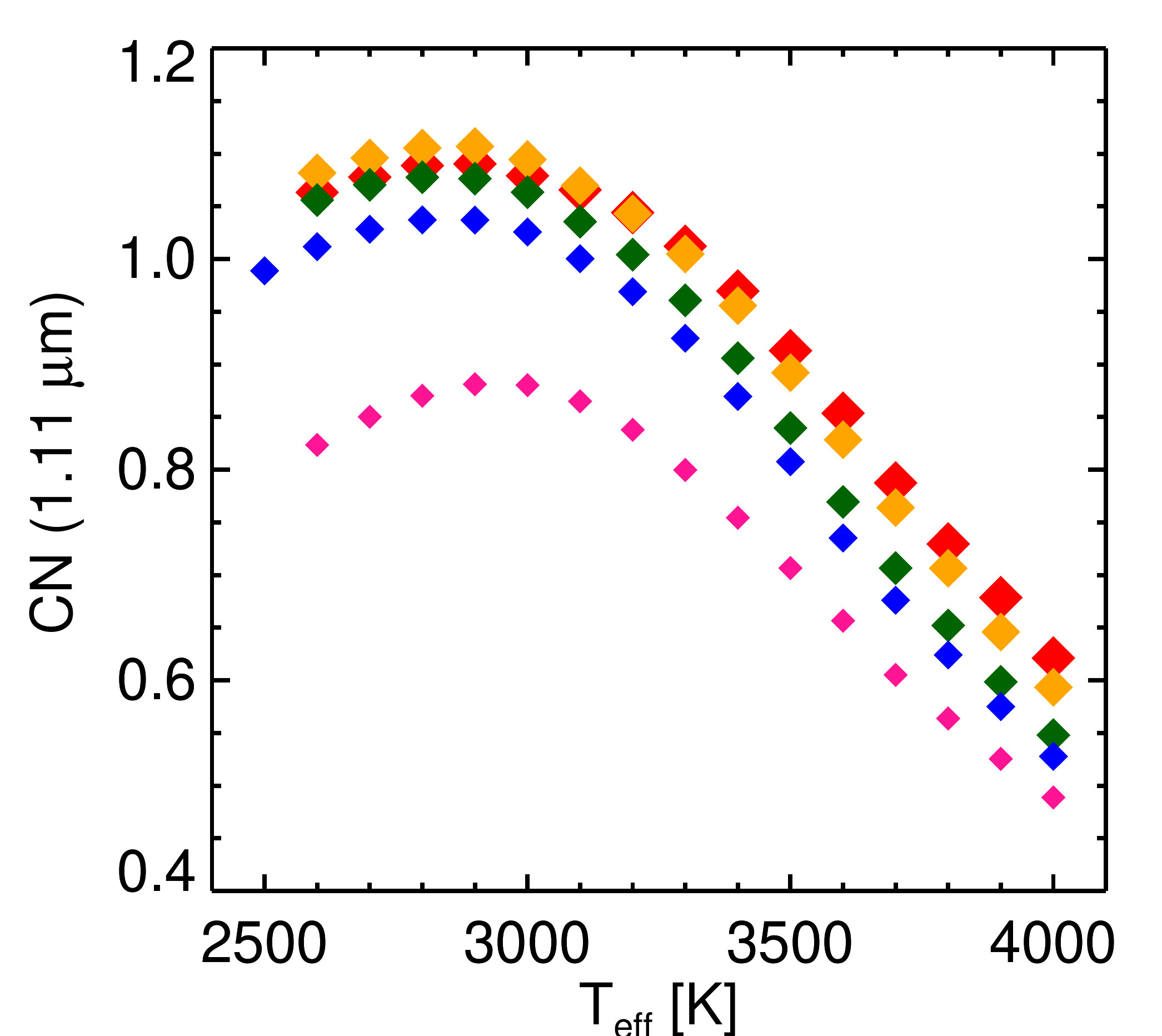}} 
\subfloat{\includegraphics[trim=5 35 5 220, clip, width=0.25\hsize]{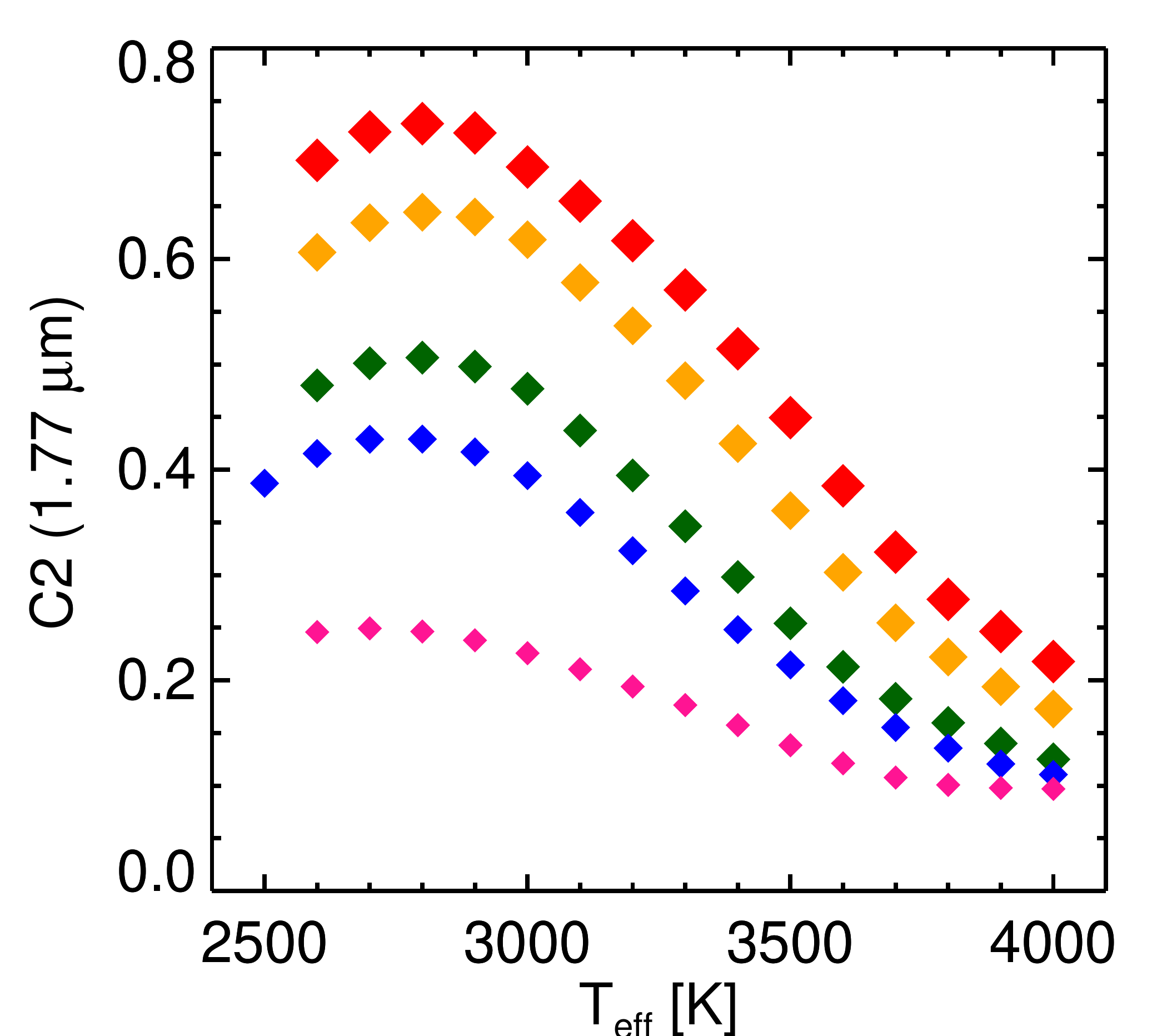}}
\subfloat{\includegraphics[trim=5 35 5 220, clip, width=0.25\hsize]{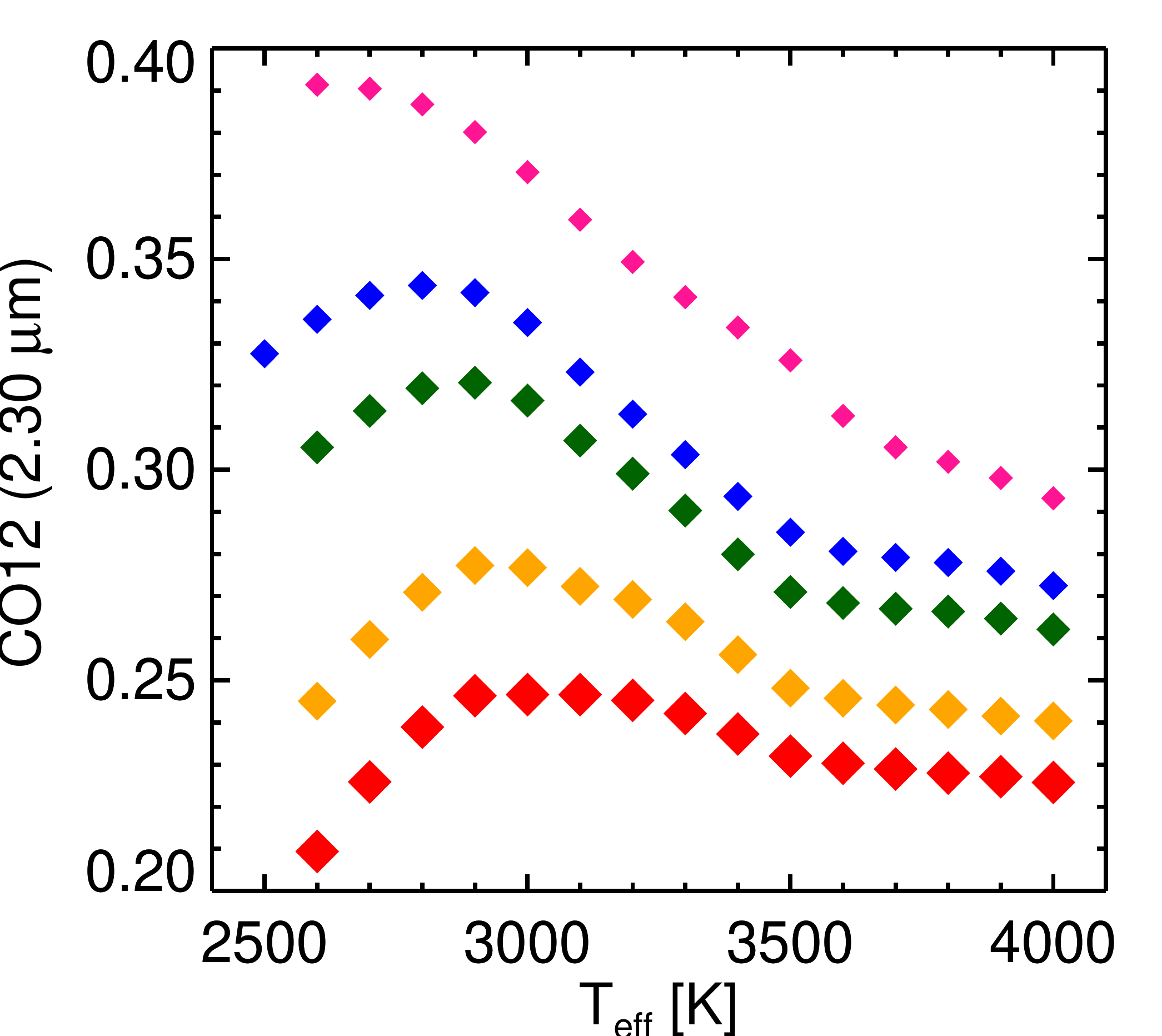}} \\
\subfloat{\includegraphics[trim=5 35 5 220, clip, width=0.25\hsize]{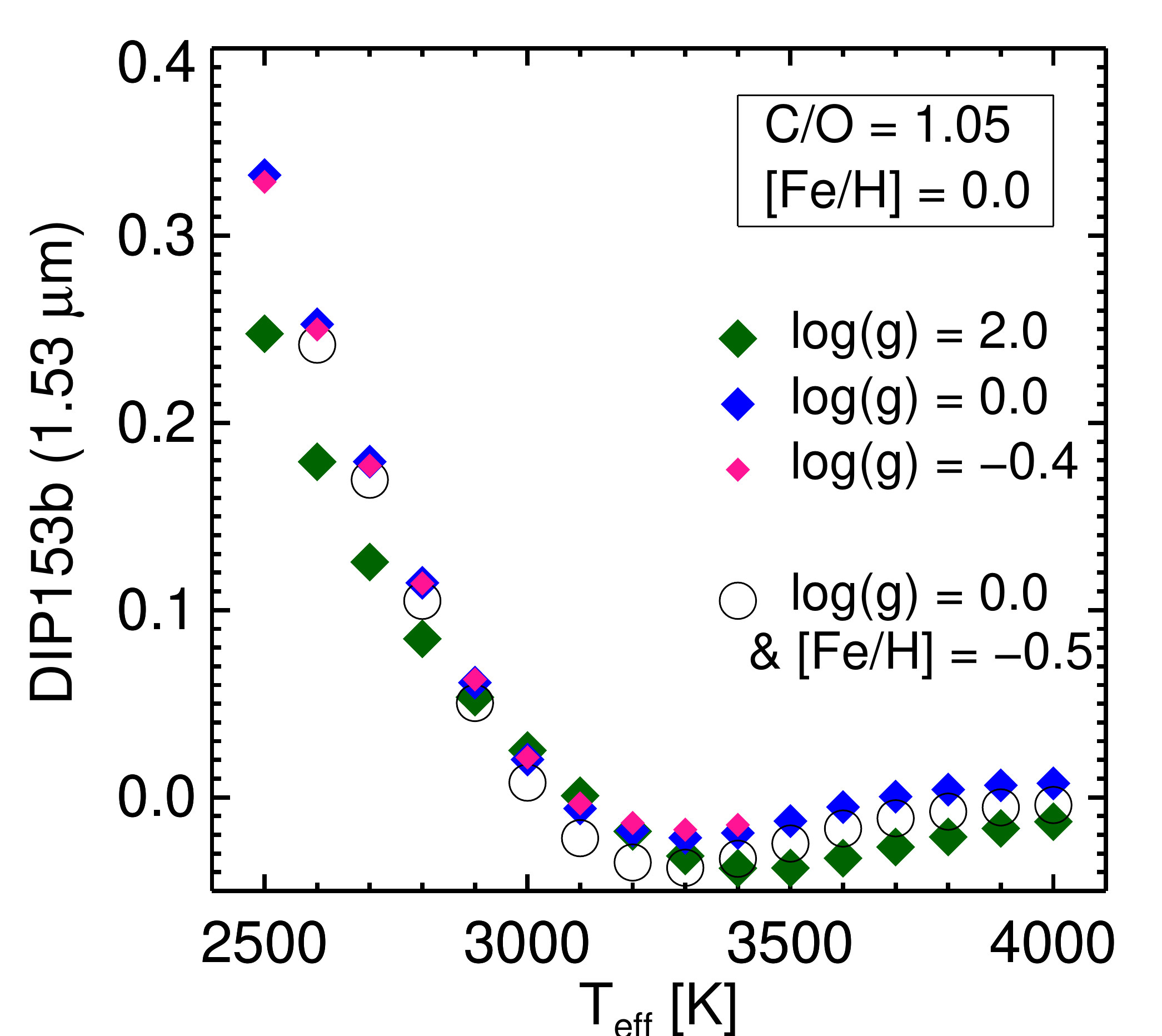}}
\subfloat{\includegraphics[trim=5 35 5 220, clip, width=0.25\hsize]{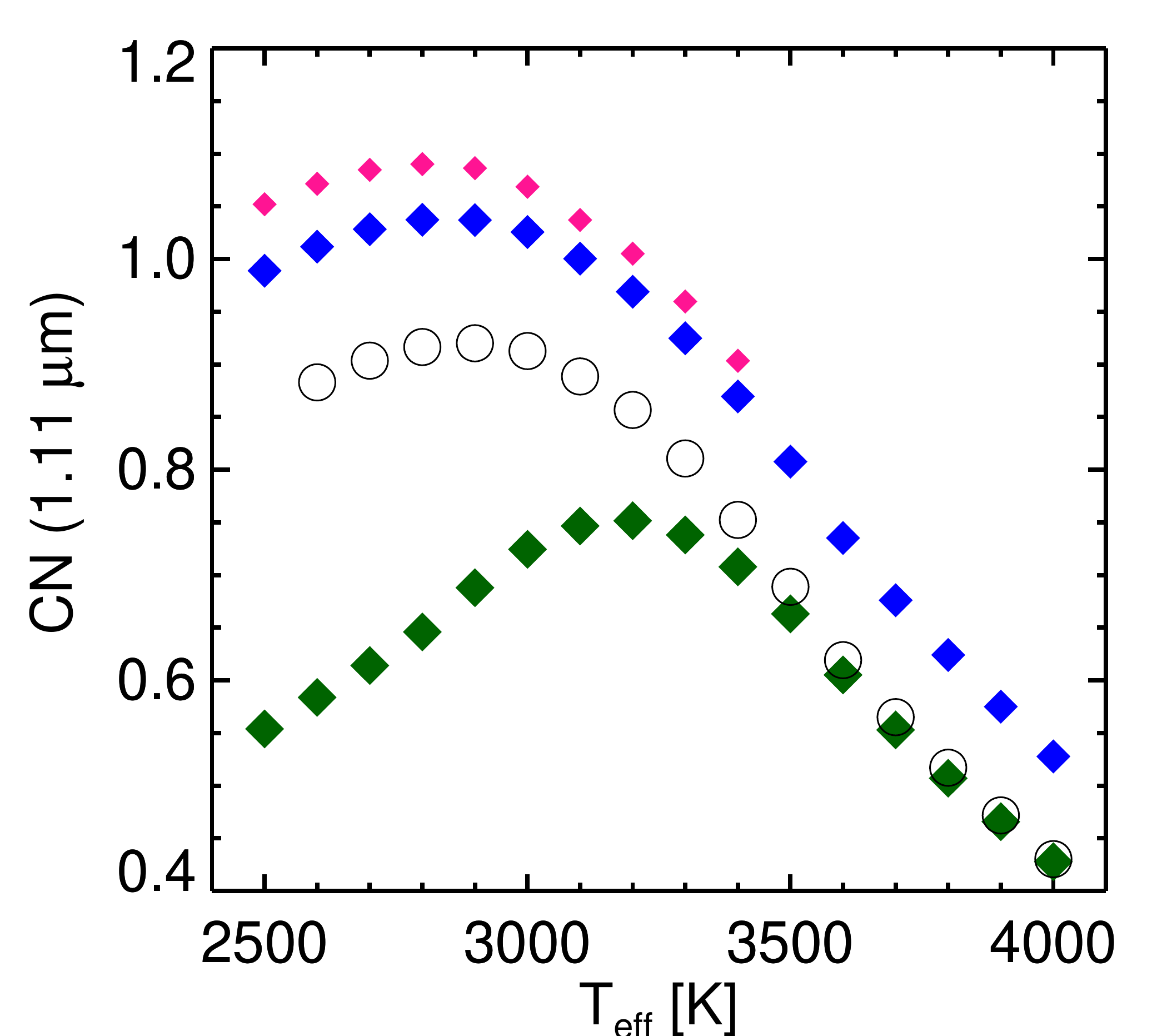}} 
\subfloat{\includegraphics[trim=5 35 5 220, clip, width=0.25\hsize]{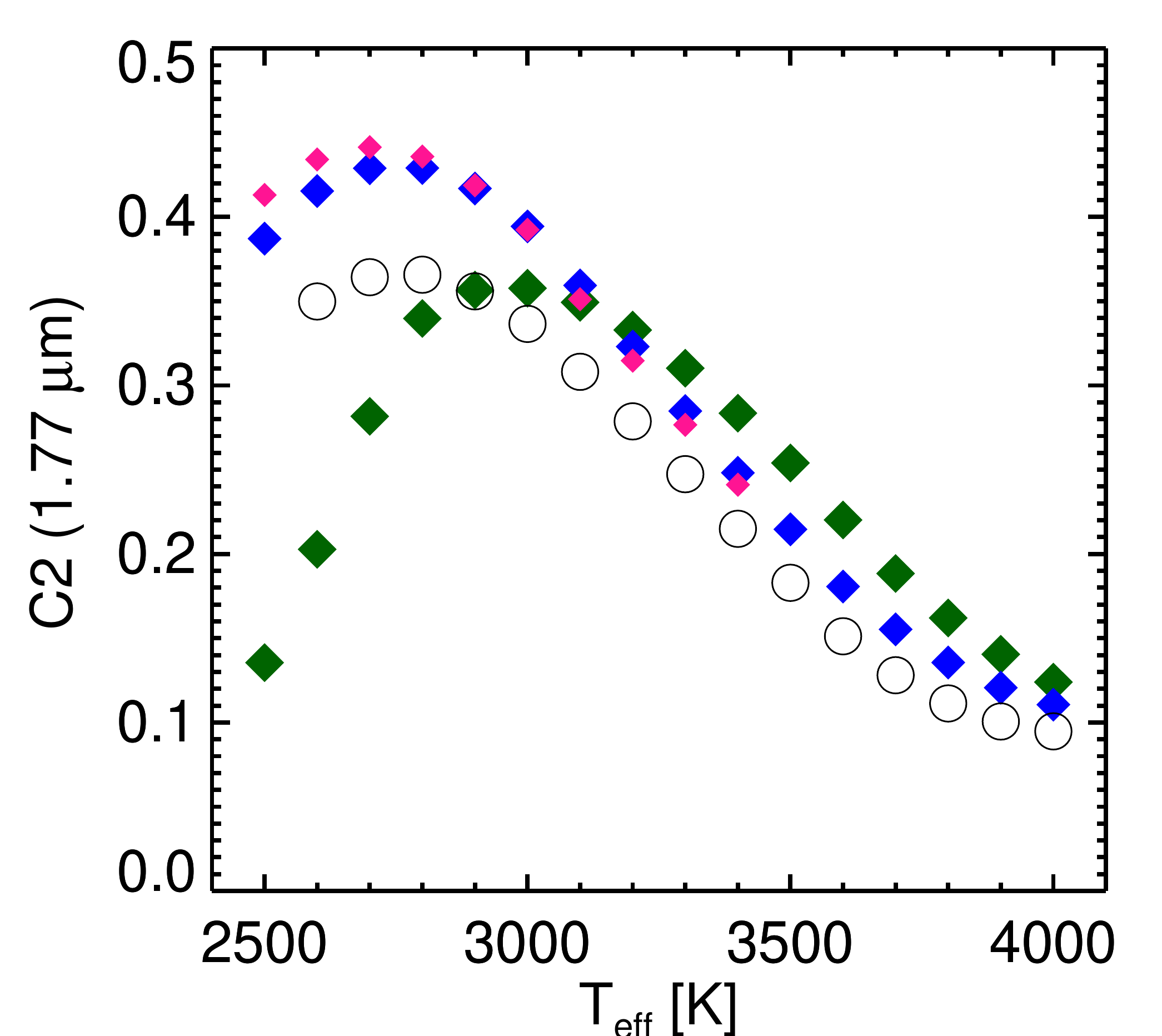}}
\subfloat{\includegraphics[trim=5 35 5 220, clip, width=0.25\hsize]{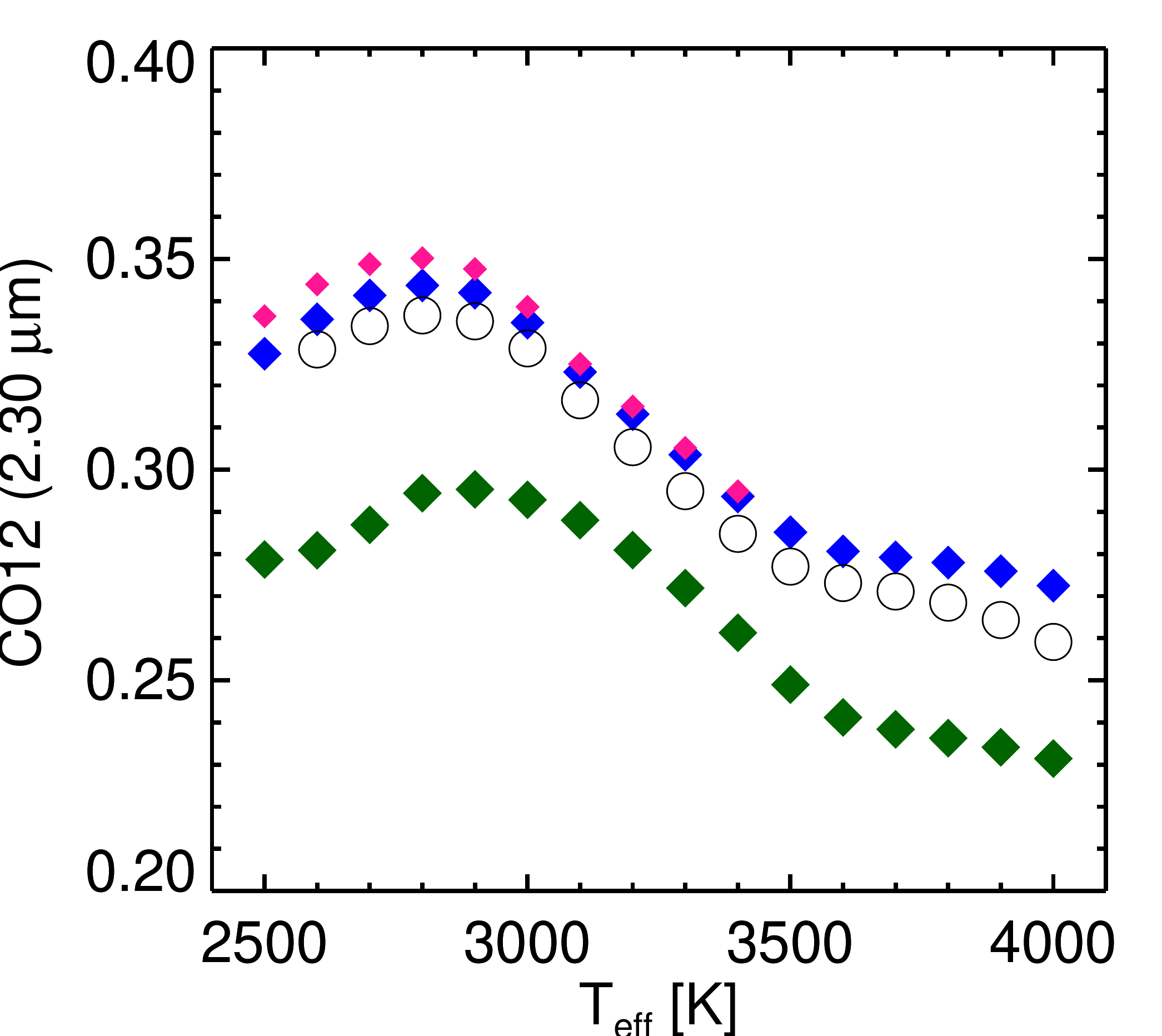}} \\
\caption{Spectro-photometric indices vs $T_\mathrm{eff}$ for a representative set of models. 
The color code of the symbols correspond to different values of C/O in the upper panels and of $\log(g)$ in the lower panels. Figure~\ref{plot_color_index} shows more indices.}
\label{plot_prop_models}
\end{center}
\end{figure*}

Figure~\ref{plot_prop_models} shows the values of four representative near-infrared indices as 
a function of the effective temperature of the models. In addition to $DIP153b$, we plot $CN$, which measures the strength of the CN molecule at 1.11\,$\mu$m, $C2$, which measures the bandhead of the C$_2$ molecule at 1.77\,$\mu$m,  and $CO12$, which measures the first overtone ro-vibrational band of CO at 2.3\,$\mu$m. The symbol sizes represent model C/O ratios in the upper pa\-nels and model surface gravities in the lower panels. 

As a general trend, molecular bandhead strengths increase with decreasing temperatures. This is the case for CN, C$_2$ and CO. At the lowest temperatures, contamination of the index passbands by lines from other molecules (or other bands of the same molecule) weakens the index values. 

At a given $T_\mathrm{eff}$, the bands of CN and C$_2$ increase with C/O, while CO decreases. 
Concerning the surface gravity, the strengths of the bands increase with decreasing $\log(g)$.
For $T_\mathrm{eff}$ $<$ 3000\,K, the high-gravity models ($\log(g)$=2) differ strongly from models with lower gravities.

The cooler models ($T_\mathrm{eff}$ $<$ 3000\,K) display the 1.53\,$\mu$m absorption band, as seen in the left panels of Figure~\ref{plot_prop_models} ($DIP153b$ as a function of temperature). The feature shows weak dependence on the C/O ratio or gravity.

\begin{figure*}
\begin{center}
\subfloat{\includegraphics[trim=5 35 5 240, clip, width=0.33\hsize]{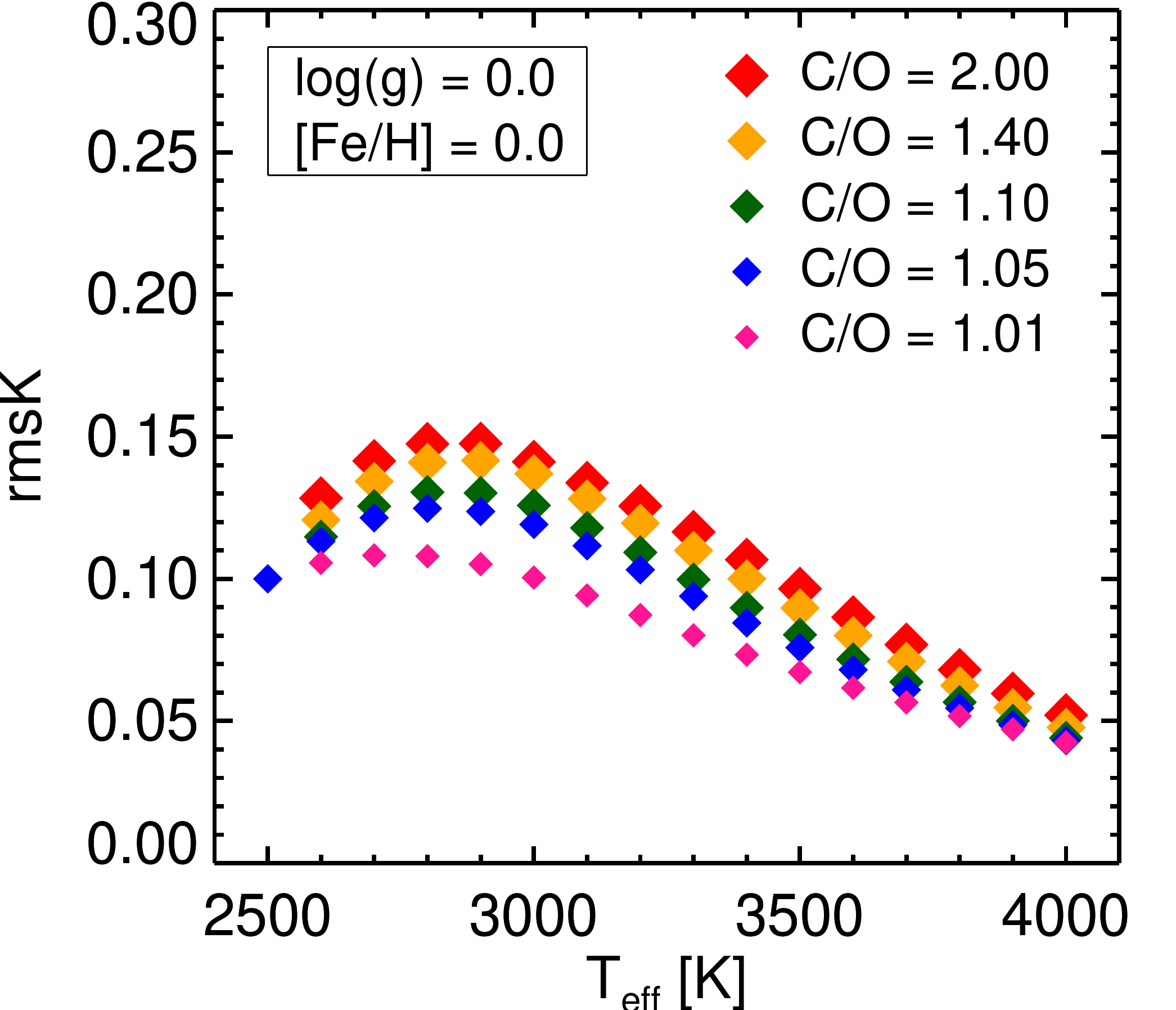}}
\subfloat{\includegraphics[trim=5 35 5 240, clip, width=0.33\hsize]{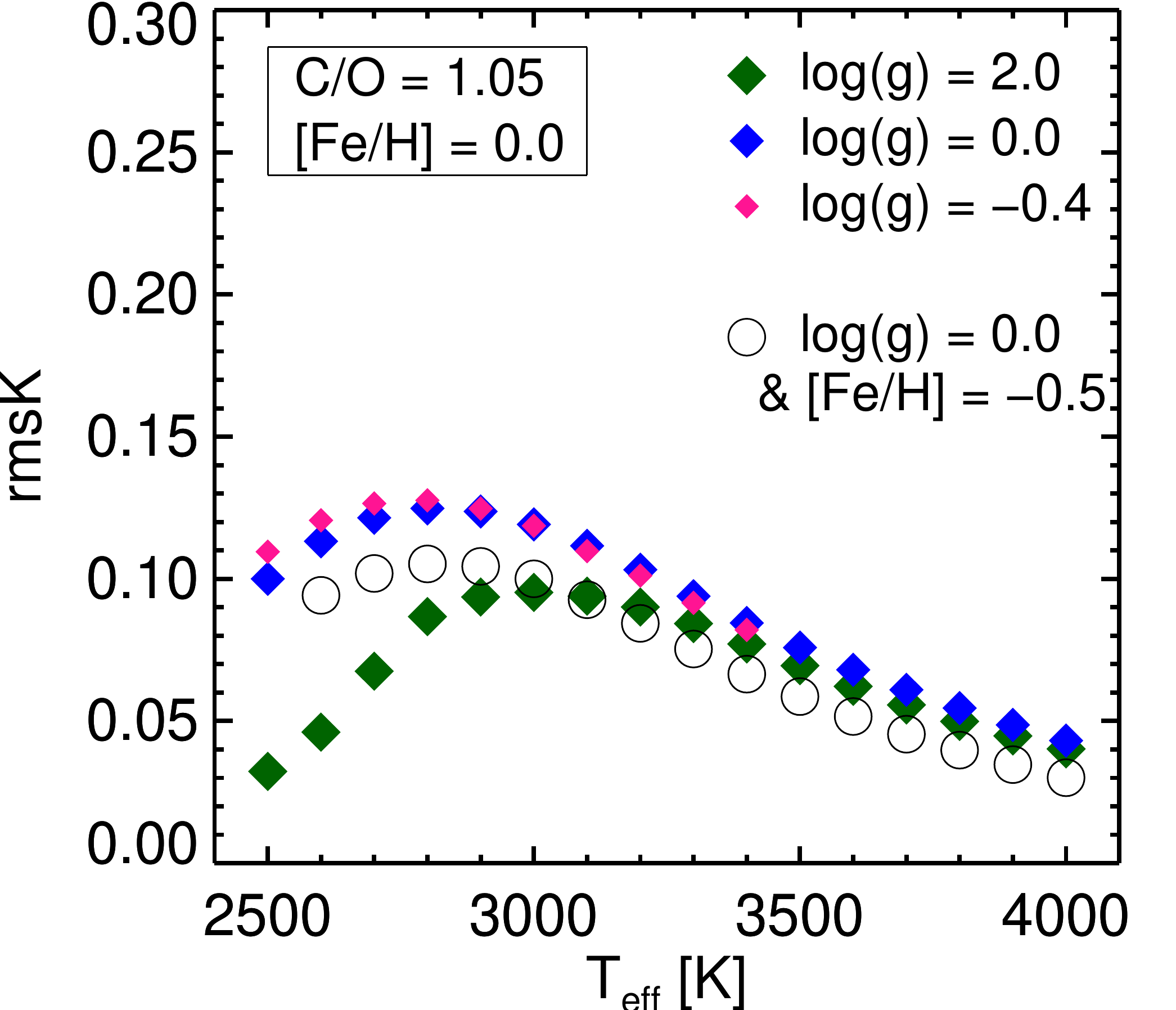}} 
\subfloat{\includegraphics[trim=5 35 5 240, clip, width=0.33\hsize]{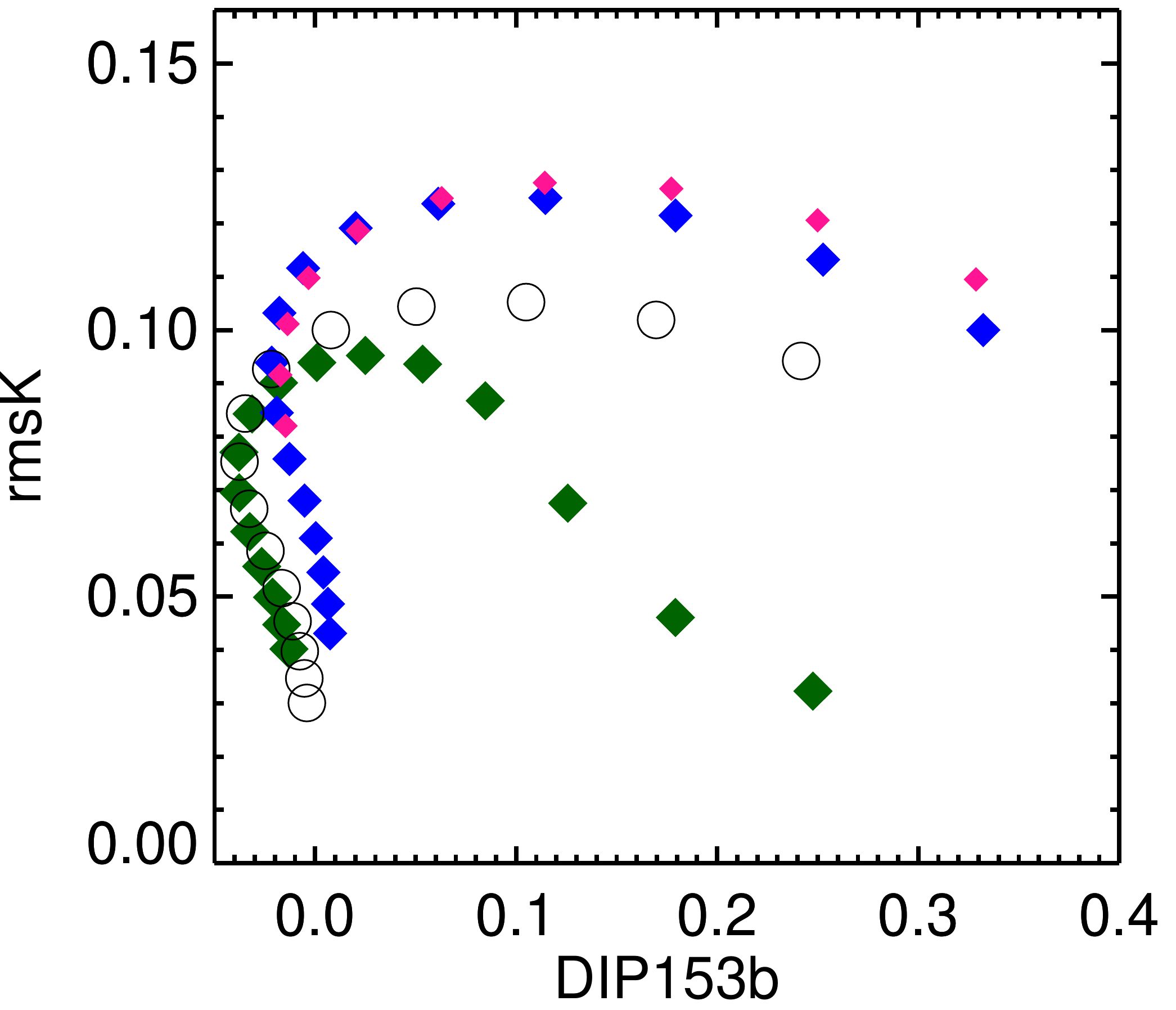}} 
\end{center}
\caption{$rmsK$ index vs $T_\mathrm{eff}$ and $DIP153b$ for a representative set of models. The symbols are for C/O in the left panel and for $\log(g)$ in the two right panels.}
\label{plot_prop_models2}
\end{figure*}

Paper I defined $rmsH$ and $rmsK$ to measure the apparent strength of the forest of lines in the $H$ and $K$ windows. They are computed as the ratio of the local standard deviation around the mean flux in small regions in the $H$ and $K$ windows, in units of that local mean. Figure~\ref{plot_prop_models2} shows the evolution of $rmsK$ as a function of $T_\mathrm{eff}$ and $DIP153b$.
At low tem\-pe\-ra\-tu\-res, $rmsK$ is sensitive to gravity. Therefore, the theore\-tical spectra
appear smoother at $\log(g)$ = 2 than at low gravities, gravities which are more
typical of luminous giants. In the hydrostatic models, the pre\-sen\-ce
of the absorption feature at 1.53\,$\mu$m feature combined with
a ``smooth'' appearance of the NIR spectrum (i.e., a low value of $rmsK$) is found at $\log(g)$ = 2. $rmsH$ behaves in the same way as $rmsK$.


\section{Comparison of pseudo-observations and models}
\label{mod_mod}


Before comparing our observations to the grid of models, it is important to 
evaluate the amount of information present in the models themselves. If the models were perfect representations of reality, with what uncertainties could we estimate the va\-lues of the fundamental parameters? 
To answer this question, we selected a subset of models that we analyzed with the full model grid as if they were observations. We refer to the subset of six models as pseudo-observations. Table~\ref{table_models} summarizes the para\-meters of the pseudo-observations.

Before analysis, the pseudo-observations were smoothed to a resolving power comparable to X-Shooter spectra, that is, R\,$\sim$\,8\,000, and they were resampled. Artificial Gaussian noise was added, to reach a signal-to-noise ratio of 75 per pixel.
Because these models are hydrostatic and thus dust-free, we reddened them artificially with $A_{\rm V}$ = 1, using the extinction law of \citet{Cardelli89}. The shape of the extinction law matters little in this exercise, considering the way we later
analyze the pseudo-observations (Eq.~\ref{eq_chi2}). The added extinction mainly serves to test color-based details of our analysis code and is not essential to results on parameter estimation.

\begin{table}
\caption{\label{table_models}Properties of the subset of models selected as pseudo-observations}
\centering
\begin{tabular}{lc}
\hline
\hline
Input parameters & Values \\
\hline 

$T_\mathrm{eff}$ [K]  		& 2600 / 3300 / 3800  \\
$\log(g)$ [cm/s$^2$] 	& 0.0  	\\
C/O					& 1.05 / 1.40 \\
$[\mathrm{Fe/H}]$ 			& 0.0 \\
$A_{\rm V}$				& 1 \\

\hline
\end{tabular}
\end{table}



\subsection{Method}
\label{methodology}

To compare our pseudo-observations with the models, we performed a $\chi^2$ minimization in a four-dimensional space with the following parameters: the effective temperature ($T_\mathrm{eff}$), the surface gravity ($\log(g)$), the ratio of carbon over oxygen (C/O), and the metallicity ($[\mathrm{Fe/H}]$).

The useful range of our X-Shooter observations extends from 0.4 to 2.4\,$\mu$m. We chose to perform two comparisons for each observation: one over the visible wavelength range (0.4--1.0\,$\mu$m), which we refer to as VIS, and one over near-infrared wavelength ranges (1.0--2.4\,$\mu$m), which we refer to as NIR.
The aim is to determine which wavelength range more strongly constrains the various stellar parameters. When applied to observations, these separate studies will allow us to explore whether the parameters derived from optical and near-infrared wavelengths using hydrostatic models are consistent. 

To identify the model that best fits a pseudo-observation (and later an X-Shooter observation), we used the following step-by-step procedure in each of the VIS and NIR wavelength ranges.

First, in order to avoid unphysical dereddening of the mo\-dels, we excluded any models  intrinsically redder than the ana\-lyzed spectrum, based on $(R-I)$ for the VIS range or $(J-K_s)$ for the NIR. 
At this step, a tolerance of 10\% on these colors is included to allow for spectrophotometric errors in the observations.

We then used the ULySS package\footnote{\url{http://ulyss.univ-lyon1.fr/}} \citep{Koleva09} to determine the velocity and the velocity dispersion differences between the pseudo-observations and each model. Each model is convolved with the corresponding Gaussian kernel in velocity
space and resampled to match the analyzed spectrum.

Finally, we computed the reduced $\chi^2$, expressed as
\begin{equation}
\chi^2_{red} = \frac{1}{N} \sum_{i=1}^{M} W(i) \times \frac{ [F_{obs}(i) - P(i) \times F_{mod}(i) ]^2}{\sigma^2_{F_{obs}}(i)}.
\label{eq_chi2}
\end{equation}

Here, $F_{obs}$ and $F_{mod}$ are the fluxes of the analyzed and theo\-retical spectra, $\sigma$ is the noise associated with the observation, and $W$ is the weight assigned to each pixel.
We set the weights to 0 in the regions of strong telluric absorption and 1 elsewhere.
We masked the following regions:  0.634--0.639\,$\mu$m, 0.994--1.02\,$\mu$m, 1.11--1.15\,$\mu$m, 1.34--1.475\,$\mu$m, 1.8--1.98\,$\mu$m, and 2.26--2.28\,$\mu$m.
$M$ is the total number of pixels and $N$ is defined as the sum of the weights.
$P$ is the multiplicative spline polynomial that minimizes $\chi^2$ . It absorbs reddening effects as well as any resi\-dual flux calibration errors. The adopted implementation has eight spline
nodes in the VIS and seven in the NIR. It is important to note that this polynomial can only mimic multiplicative effects of circumstellar dust such as absorption, and not additive effects such as any thermal emission by dust.

\subsection{$\chi^2$ maps and results}


\begin{figure}
\begin{center}
	\includegraphics[trim=5 35 200 160, clip, width=\hsize]{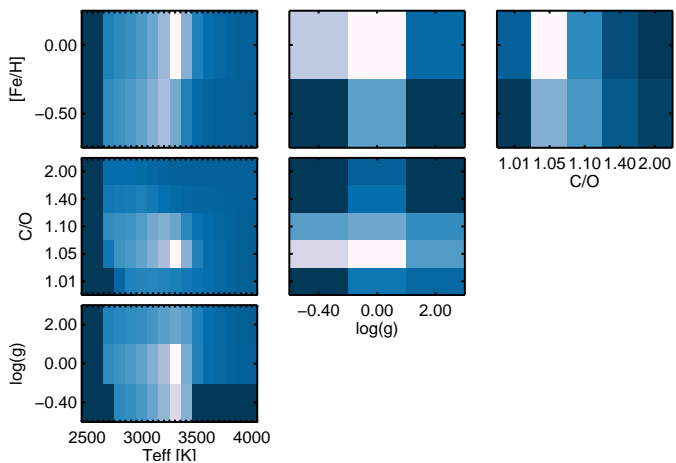}
	       \caption{Example of $\chi^2$ maps for one of our pseudo-observations ($T_\mathrm{eff}$ = 3300 K, $\log(g)$ = 0, C/O = 1.05, $[\mathrm{Fe/H}] = 0$) for the VIS wavelength range. 
The best values (i.e., the values that minimize the $\chi^2$ calculation) are represented in white.}
    \label{plot_chi2map_vis}
\end{center}
\end{figure}

\begin{figure}
\begin{center}
	\includegraphics[trim=5 35 200 160, clip, width=\hsize]{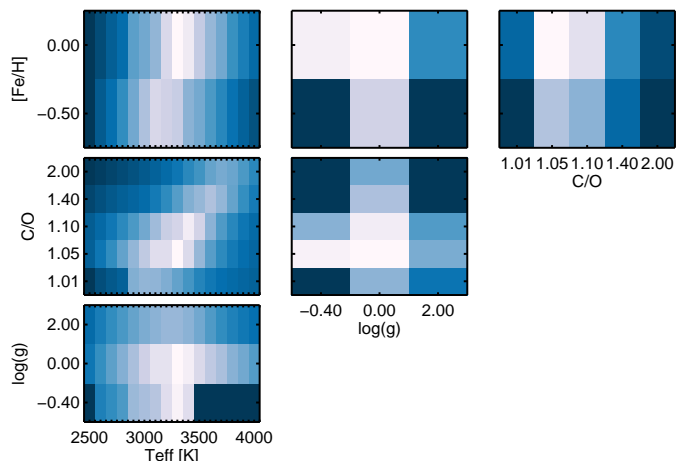}
	       \caption{Same as Fig.~\ref{plot_chi2map_vis} but for the NIR wavelength range.}
    \label{plot_chi2map_nir}
\end{center}
\end{figure}

For each pseudo-observation, we computed a $\chi^2$ map showing the distribution of the best models in the parameter space of the models. Figures~\ref{plot_chi2map_vis} and~\ref{plot_chi2map_nir} provide examples of such maps: we fit the VIS and NIR wavelength ranges of a pseudo-observation in these figures, respectively. The best models, which correspond to the lowest values of $\chi^2$, are plotted in white.  

The $\chi^2$ maps demonstrate that the classical degeneracy 
between metallicity and effective temperature exists for carbon stars: fitting with models at underestimated metallicities leads to underestimated temperatures. 
In the near-infrared, another de\-ge\-ne\-ra\-cy is found to link the effective 
temperature and the C/O ratio (Fig.\,\ref{plot_chi2map_nir}). However, this second degeneracy is 
not seen in the optical range (Fig.\,\ref{plot_chi2map_vis}).  The maps for other
pseudo-observations confirm these trends.  The grid of models available to date is 
too small to identify any other noteworthy systematics.

In the controlled context of pseudo-observations, 
the para\-meters of the analyzed spectrum can all be recovered better than the grid sampling. 
However, although the number of degrees of freedom are similar while fitting the NIR and VIS ranges, 
the $\chi^2$ valleys are much shallower for the NIR than they are for the VIS range. 
This indicates that the optical spectra of carbon stars more tightly constrain the 
stellar parameters than their near-infrared counterpart.

We can illustrate this by exa\-mining the distribution of mo\-dels for which 
the $\chi^2$ distance to a given pseudo-observation is smaller than a threshold.
For instance, the condition $\chi^2<3$ is typically fulfilled for only one model
in the VIS range, while about five to ten models satisfy this criterion in the NIR.  


\section{Comparisons of observations and models}
\label{comp_obs}

In the following, we compare the observed targets with the mo\-de\-ling results in different ways.  
Section~\ref{section_color} presents some color indices in standard broadband filters. 
Section~\ref{section_index} focuses on spectro-photometric indices.
Section~\ref{section_fit} details how we fit our observed spectra and discusses the stellar para\-meter estimations.

\subsection{Broadband colors}
\label{section_color}

A comparison of the colors of the XSL targets and those of the models gives information on the stellar energy distribution as well as on the effects of dust. When the effects of dust are limi\-ted, color indices involving an optical and a near-infrared passband are good first-order indicators of the effective temperature, with a low sensitivity to other fundamental parameters (T1). This pro\-perty is rapidly lost when circumstellar material becomes important.

For this comparison, we computed synthetic photometry for the model grid and the X-Shooter spectra, using the \citet{Bessel90} filters $R$ and $I$ and the 2MASS near-infrared filters  $J$, $H$ and $K_s$ \citep{Cohen03}. Figure~\ref{plot_comp_color} and Figure~\ref{plot_color_color} of the appendix display the resulting near-infrared color indices.
Mira-type stars are identified by green filled symbols.\footnote{We use the classification from Paper I, Table B.1, Column 6.}
The spread between Miras and non-Miras is similar to the spread found by \citet{Whitelock06}. 
On the red side of the two-color diagrams, Mira-type stars tend to spread to much redder colors because of their higher mass-loss rates and the resulting circumstellar shells. On the blue side, it is more difficult to separate the two groups. 
The extinction vector in Figure~\ref{plot_comp_color} guides the eye to the locus of reddened models and suffices to suggest that some observed spectra re not represented properly with hydrostatic mo\-dels and a simple extinction law.

\begin{figure}
	\begin{center}
		\includegraphics[trim=10 50 10 235, clip, width=\hsize]{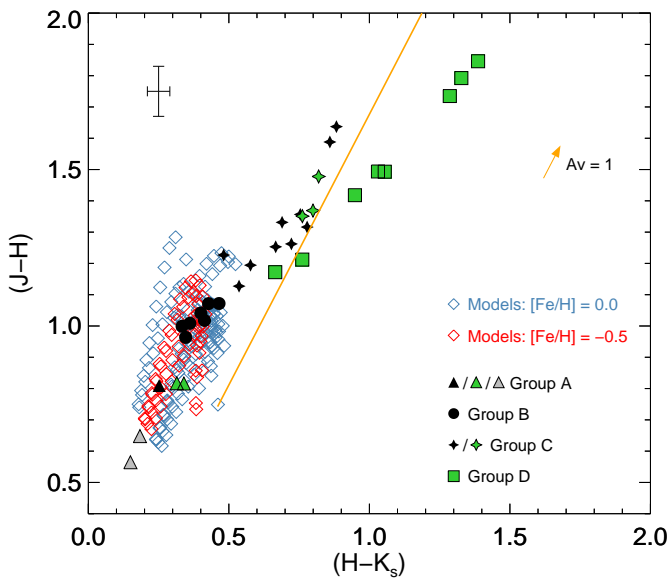}
	     \caption{Colors (no correction for interstellar reddening) for our sample of carbon stars (filled symbols), using the 2MASS filters. 
The green symbols indicate the Mira-type stars, while the gray symbols are for the stars outside of the color range of the models.
The synthetic colors (open diamonds) of our grid of hydrostatic models are overplotted for comparison. 
The orange line shows the effect of reddening using a simple extinction law \citep{Cardelli89}.
The bars plotted show the $\pm1\sigma$ root-mean-square deviation
 of our photometry with respect to the literature (large-amplitude variables 						  excluded). This is an upper limit of the uncertainties 
 in the flux calibration and any possible residual varia\-bility. }
    \label{plot_comp_color}
    	\end{center}
\end{figure}

Group A gathers the bluest stars of our sample, represented as filled triangles in Figure~\ref{plot_comp_color}. Two of the stars, HD 202851 and HE 1428-1950, represented as gray triangles, lie outside the color range of the models.
Estimates of the temperatures of these two stars are available in the literature. 
\citet{Bergeat02} found $T_\mathrm{eff}$ = 4780 K for HD 202851, and \citet{Placco11} found $T_\mathrm{eff}$ = 4562 K for HE 1428-1950. 
These temperatures are higher than expected for classical carbon stars on the asymptotic giant branch, and in Paper I we suggested that these objects may be extrinsic carbon stars.
As the highest temperature of our model grid is $T_\mathrm{eff}$ = 4000 K, 
these objects are expected to be out of the grid.
They are not considered in the comparison with the models hereafter.

Stars from Group B should be well reproduced by the grid of models as the colors of both sets overlap well. 
Stars from Group C start to be more affected by circumstellar dust than stars from the previous groups and we are closer to the limits of what can be done with hydrostatic models. By red\-de\-ning the models, we should be able to reproduce most or all of these observations.

Group D gathers all the stars with the absorption band at 1.53\,$\mu$m. Although a small number of hydrostatic models contain this feature, the current grid combined with simple extinction does not reach values of $(H-K_s)$ red enough to explain all the observations.

\subsection{Molecular indices versus color}
\label{section_index}

Spectrophotometric indices provide a good overview of the mo\-dels with respect to the XSL data, at a somewhat higher resolution than broadband colors. For brevity, we only discuss the loci as a function of $(J-K_s)$, as already done in Paper I.

\begin{figure*}
	\begin{center}
		\includegraphics[trim=20 40 30 90,width=\textwidth]{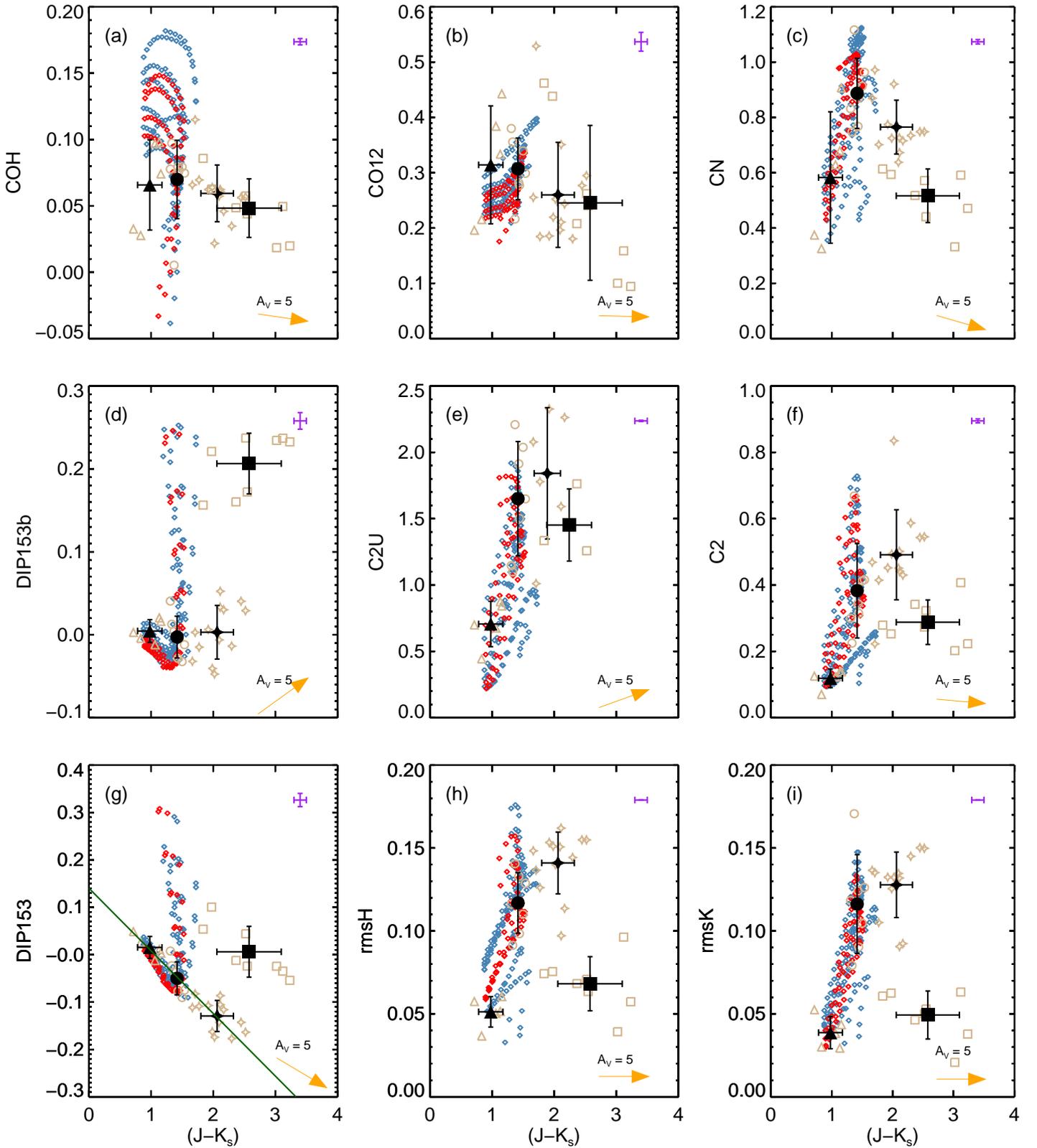}
                \caption{Spectro-photometric indices derived for our sample of
                  carbon stars (brown symbols) and the grid of models (color symbols) as a function of $(J-K_s)$. The triangles are for stars from Group A, the circles for Group B, the stars for Group ,C and the squares for Group D. 
The filled symbols represent the averaged values of our indices in the groups, and the bars
measure the dispersion within the bin.                   
The open diamonds represent the models of solar (in blue) and subsolar metalliticity (red).                
The models were smoothed to R $\simeq$ 8\,000 for the purpose of this figure. 
Typical uncertainties on individual measurements are shown in purple.
The orange vector represents the extinction vector (computed for $A_{\rm V} = 5$).
In panel (g), the green line was used to define the new index $DIP153b$ (see Equation~\ref{eq_new_dip}).}
    \label{plot_color_index}
    	\end{center}
\end{figure*}

Figure~\ref{plot_color_index} displays the values of the molecular indices for both the observed targets and the model atmospheres (cf Section~\ref{index_model}).
The filled symbols stand for our observations, while the open diamonds represent the models.

The error bars shown in Figure~\ref{plot_color_index} account for the noise per resolution element and uncertainties in the shape of the spectrograph's response curve in the scale of molecular features. The latter component is usually dominant because the signal-to-noise ratio of the spectra is on the order of 100. Uncertainties in the response curve on the relevant scales are due mostly to imperfect modeling of the telluric absorption that affects the spectro-photometric standard star observations.

Both the data and the model indices present a large dispersion within color bins.
In general, the model loci agree well with the locus of the observations of our sample with $(J-K_s)<1.6$. This is true in particular for the CN bands. 

The CO bands in the $H$ window tend to be too strong in the models compared to the data. These features are sensitive to surface gravity at a given effective temperature. The relative strengths of the first- and second-overtone CO bands (measured by $CO12$ and $COH$) are also sensitive to the microturbulent velocity \citep{Origlia97,Lancon10}. 
Furthermore, the CO bands are very sensitive to dynamical effects cau\-sing emission components in the lines \citep{Nowotny10}. This may cause weaker bands in some of the observed variable stars.
It is unclear as yet what the predominant cause of the systematic difference could be.

Some of the models display the 1.53\,$\mu$m feature (panels d and g). However, low-gravity models among them do not display damped line forests in the $H$ and $K$ window, as seen in the corresponding observations (panels h and i). In Paper I, the damping of the high-frequency structure in spectra with the 1.53\,$\mu$m feature was interpreted as veiling by circumstellar dust. The index plots for the dust-free static models are consistent with this picture.

\subsection{Full spectral fitting}
\label{section_fit}

\subsubsection{Method}

For a direct comparison between the observed and the model spectra, we use a method similar to the method we described in Sect.~\ref{methodology}. The diffe\-rences are as follows.

The noise spectra of our observations are those that come out of the X-Shooter reduction pipeline. We propagate the errors through the reduction process, including the correction of the telluric features and the flux calibration. 

For this study, we degraded the resolution of the XSL spectra to $R \sim 2\,000$. Before smoothing, the ve\-lo\-ci\-ty resolution of the XSL data is $\sim$ 30 km/s (R $\sim$ 10000). In Mira-type variables, velocity discontinuities with amplitudes larger than 10 km/s are expected as shocks propagate through the atmospheres \citep{Nowotny05,Nowotny10}. They produce significant wavelength shifts that may differ for various molecular bands. Since hydrostatic models cannot cover such effects, we postpone the study of these high-resolution effects to a future article.

For each observed star, the best-fitting model mini\-mizes the $\chi^2$ calculation.
Models for which $\chi^2/\chi^2_{min} < 1.1$ are 
consi\-dered similarly acceptable, and the distribution of their parameters
provides our estimate of the uncertainties on the star's physical properties.
It is worth noting that the best model does not always fit the data well. Unfortunately,
because the absolute levels of the error spectra produced by the 
X-Shooter pipeline are sometimes unreliable, and because the correlations
between noise in neighboring pi\-xels are not completely characterized, 
the numerical value of the minimum $\chi^2$ cannot be
reliably used as a direct measure of the quality of the fitting process. 
The exa\-mination of the fitting residuals is safer for this assessment.

\subsubsection{Results of model fitting}
\label{results_obs}

Appendix~\ref{obs_plots_all} shows the models that best fit 
our X-Shooter spectra, ordered by increasing $(J-K_s)$.
As mentioned above, the visi\-ble wavelength range runs in principle from 
0.4\,$\mu$m to 1.0\,$\mu$m. However, the reddest carbon stars could not be measured below
0.5 or 0.6\,$\mu$m.  The corresponding part of their spectra 
 was rejected from the fitted models and figures.

The bottom panels of the figures show the residuals and an unsharp-mask filtered version 
of the model spectrum (the difference between the synthetic spectrum and a heavily smoothed version thereof). 
In many cases, the residuals are very small compared to the high- and medium-resolution features seen in the filtered spectrum, showing that the models successfully capture the shapes and relative strengths of the dominant molecular bands (mostly of CN). This is the first time the ability of C-star models to fit observations is demonstrated over such an extended wavelength range.

Tables~\ref{table_grp_a} to~\ref{table_grp_d} list the results of the fitting procedure.
The letter V indicates the values found over the VIS wavelength range, the letter N stands for near-infrared. 
For each parameter ($T_\mathrm{eff}$, $\log(g)$, C/O, $[\mathrm{Fe/H}]$), the first co\-lumn gives the value of the best model (mini\-mum $\chi^2$), the second column the weighted ave\-rage value, and the third and fourth columns the extreme va\-lues compatible with our $\chi^2/\chi^2_{min}$ -threshold. The weighted average is given by
\begin{equation}
param\_weighted = \frac{\sum_{i} param(i) \times \exp[-(\chi^2_i/\chi^2_{min})/2]}{\sum_{i} \exp[-(\chi^2_i/\chi^2_{min})/2]},
\label{eq_wei}
\end{equation} 
where $i$ samples all the available models.

With only a few exceptions, the favored surface gravity for all our observations is log(g) = 0 (rather than 2), and the favored metallicity is $[\mathrm{Fe/H}] = -0.5$ (rather than 0). These
va\-lues are satisfactory for TP-AGB stars and for a sample consis\-ting mostly of LMC, SMC, and Milky Way halo stars. 

\begin{figure}
	\begin{center}
		\includegraphics[trim=25 35 50 20, clip,width=\hsize]{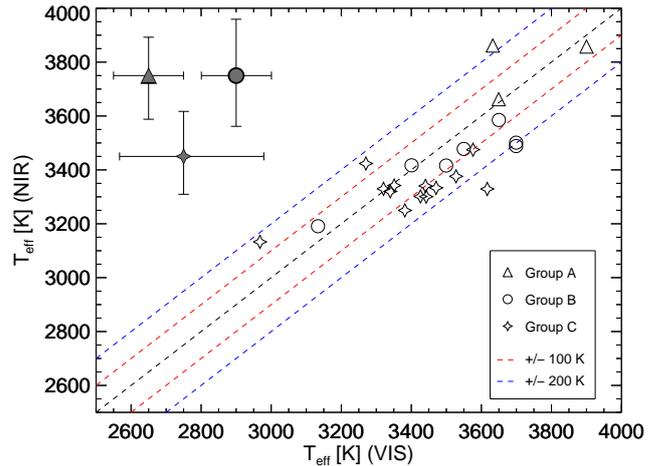}
	     \caption{Comparison of the weighted averaged values found for the effective temperature for both wavelength ranges (VIS and NIR) for Groups A, B, and C. The black dashed line indicates the one-to-one relation, while the red and blue dashed lines correspond to $\pm 100\,K$ and $\pm 200\,K$, respectively. The typical (median) error bars for each group are indicated in the top left corner.}
    \label{plot_best_param_teff}
    	\end{center}
\end{figure}

Figure~\ref{plot_best_param_teff} compares the weighted average values of the effective temperatures estimated by fitting the VIS and NIR wavelength ranges. The black dashed line indicates the one-to-one relation. Different symbols indicate the values for the spectra of groups A, B, and C (Group D is omitted for reasons explained). 
Consi\-dering the typical error bars of $\pm 200\,K$, the NIR and VIS tempe\-ratures
are consistent with each other.
Nevertheless, the temperatures found tend to be warmer in the VIS than in the NIR.
The median of the weighted temperatures is 3376.10 $\pm 160\,K$ for the NIR and 3470.89 $\pm 100\,K$ for the VIS for the three groups. \\

We now consider each group in turn.

\paragraph{Group A}

The parameters listed in Table~\ref{table_grp_a} agree quite well between the VIS and the NIR wavelength ranges. Figures~\ref{fit_beg_a} to~\ref{fit_end_a} show the best-fitting models. Effective temperatures in Group A are above 3600\,K. 

Two of the three stars in Group A display hydrogen lines in emission. They are large-amplitude variables (Paper I), and these lines are interpreted as signatures of shocks that propagate through the atmosphere. For one of these stars, the fitted model is not quite as good as for the others. However, it is remarkable how well hydrostatic models reproduce the medium-resolution spectral features of these warm pulsators in general.

The CO bands in the $H$ window are too strong in the mo\-dels compared to the observations. This is particularly visible in Fig.~\ref{fit_end_a} by comparing the residuals (in green) with the bottom panel curve, which corresponds to the unsharp-mask filtered version of the model (in blue).

\paragraph{Group B}

Stars from Group B are expected to be reproduced well by the grid of models, according to Fig.~\ref{plot_comp_color}. 
Indeed, the mo\-dels fit the data well, and the curves of $\chi^2$ versus model $T_\mathrm{eff}$ are well behaved with narrow minima. The best values are summarized in Table~\ref{table_grp_b}. In all but one case (Fig.~\ref{ex_grp_b_temp}), the range of $T_\mathrm{eff}$ derived from the VIS spectrum is narrower than the range accepted based on the NIR data, in agreement with expectations from Section~\ref{mod_mod}.  The temperatures within Group B range between 3200 and 3800\,K. 
In general, a slightly higher temperature in the VIS than in the NIR
is compensated by a higher C/O ratio. In most cases, the 
optical and near-infrared temperatures are within the uncertainties of each other.

From Figures~\ref{fit_beg_b} to~\ref{fit_end_b}, some small discrepancies appear progressively. 
Some are instrumental, like the lack of data around 0.63\,$\mu$m (due to a bad column).
A real systematic diffe\-rence in the shape and depth of the C$_2$ band at 1.77\,$\mu$m becomes apparent for $(J-K_s) \geq 1.4$. 
This is discussed further for Group C below.

\paragraph{Group C}

\begin{figure}
\begin{center}
\includegraphics[trim=40 25 65 80, clip,width=\hsize]{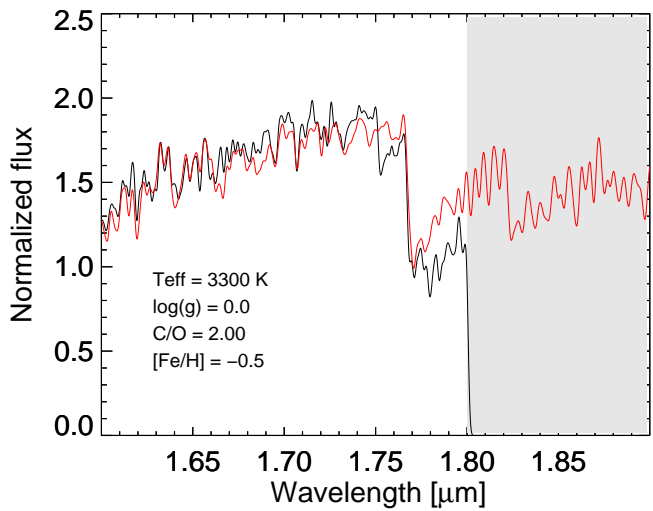}
\caption{Close-up of the C$_2$ features at 1.77\,$\mu$m for [ABC89]\,Cir\,18 (cf. Fig.~\ref{fit_end_c} for full fitting). The black curve corresponds to the stellar spectrum and the red curve to the best
model times the best-fitting polynomial. The gray band masks a region with strong telluric absorption.}
\label{ex_grp_c_177}
\end{center}
\end{figure}

Table~\ref{table_grp_c} summarizes the properties of the best mo\-dels for the stars of Group C. Stars from Group C are stars with $(J-K_s) \geq 1.6$ and are therefore in a color regime 
that hydrostatic models cannot reproduce without any effects of dust \citep{Aringer09}. 
The multiplicative polynomial in our procedure accounts for extinction,
 and we focus on the spectral features.
Fi\-gures~\ref{fit_beg_c} to~\ref{fit_end_c} show the results from fitting.

The range of temperatures found in Group C is similar to that in Group B, despite the redder colors in Group C. 
Again, the NIR and VIS temperatures are compatible (within the error bars) in most cases,
but the error bars tend to become larger than in Group B, and there are a few 
formally incompatible cases, with significantly higher 
VIS than NIR temperatures (e.g., Fig.~\ref{ex_grp_c_temp1}, ~\ref{ex_grp_c_temp2} and ~\ref{ex_grp_c_temp3}).

The fitted models favor a C/O ratio of 2. The effect of C/O on color, at a given $T_\mathrm{eff}$ (above 3000 K), is very small and hence C/O does not explain the redder colors of Group C. The difference in color between Groups B and C is mainly driven by circumstellar extinction. 

The models struggle to reproduce the depth and shape of the C$_2$ features at 1.77\,$\mu$m, as shown in Figure~\ref{ex_grp_c_177}.
The observations show a feature at 1.75\,$\mu$m \citep[H$_2$O + $^{12}$CO, see Figure~3 from][]{Lyubenova12} that correlates with the depth of the 1.77\,$\mu$m bandhead. This is also seen in C-star spectra of \citet{Lancon_Wood} or IRTF \citep{Rayner09}, but is not present in the models.

\paragraph{Group D}

\begin{figure}
\begin{center}
\includegraphics[trim=40 40 65 75, clip,width=\hsize]{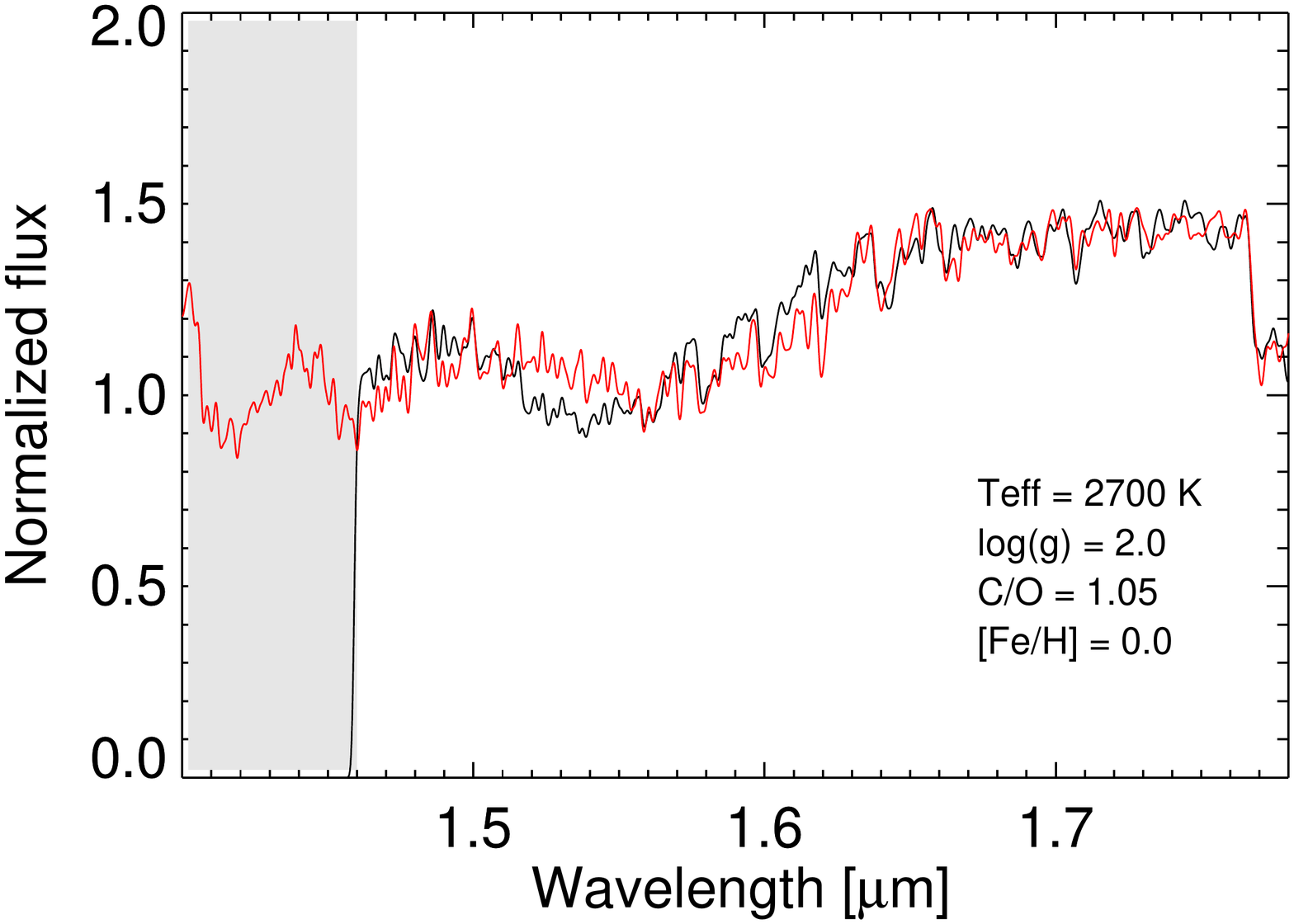}
\caption{Close-up of the HCN\,+\,C$_2$H$_2$ feature at 1.53\,$\mu$m for SHV\,0502469-692418 (cf. Fig.~\ref{ex_grp_d_hcn} for full fitting). The black curve corresponds to the stellar spectrum and the red curve to the best
model times the best-fitting polynomial.  The gray band masks a region with strong telluric absorption.}
\label{ex_grp_d_153}
\end{center}
\end{figure}

Table~\ref{table_grp_d} summarizes the results for the stars from Group D, that is, the stars that display the 1.53\,$\mu$m absorption feature. These stars are affected by pulsation. They are all large-amplitude variables (cf. Paper I), and the use of hydrostatic mo\-dels is a clear limitation. 
Figures~\ref{fit_beg_d} to~\ref{fit_end_d} show the best-fitting models. 
While the $\chi^2$ distributions as a function of effective temperature remain relatively well behaved for most of the VIS spectra, the distributions for the NIR are very flat. The NIR best models provide no or only weak constraints on the parameters. 

The NIR X-Shooter spectra in Group D have a relatively smooth appearance, compared to Groups C and B, as was highlighted in Paper I. As a consequence, the best-fitting models tend to have either relatively high temperatures ($>$ 3500\,K) and $\log(g)=0$, or lower temperatures and $\log(g)=2$ (last panel of Fig.~\ref{plot_prop_models2}).

However, these parameters cannot be relied upon. A hot temperature is difficult to reconcile with the presence of the 1.53\,$\mu$m feature, and indeed the best-fitting model with high temperature fails to reproduce this feature (e.g., Fig.~\ref{ex_grp_d_pb1}). The best models with $\log(g)=2$ combine the pre\-sence of this absorption feature with a smooth appeareance (see plot $rmsK$ index vs $DIP153b$, from Figure~\ref{plot_prop_models2}, Section~\ref{index_model}). A gravity that high is difficult to reconcile with the expected location of large-amplitude C-rich variables on the asymptotic giant branch, however. 
Our favoured interpretation is that these stars have normal AGB gra\-vi\-ties and are not as warm as the NIR models at $\log(g)=0$ may indicate.

Instead, their spectra are profoundly modified by dust that produces both extinction and emission \citep{Nowotny11}. This combination is known as veiling. The (additive) emission component attenuates the equi\-valent widths of molecular absorption features produced in the photosphere, and both extinction and emission affect the energy distribution. Parameter estimations based on fits to the absorption features that are produced near the photosphere then become unreliable. The parameters given in Table~\ref{table_grp_d} based on the NIR spectra should therefore only be used with great caution.

 In principle, the parameters derived in the VIS should be more reliable than those based on the NIR, as dust emission is weaker at shorter wavelengths. 
However, the useful part of the VIS spectrum of highly reddened objects almost exclusively contains signatures of CN, and we should keep in mind that unexplored parameters such as the N abundance could affect them.
 
Finally, we emphasize two systematic spectral issues. The C$_2$ feature at 1.77\,$\mu$m is never matched correctly. More inte\-restingly the 1.53\,$\mu$m feature present in the coolest hydrostatic models does not have the shape of its observed counterpart, as shown in Figure~\ref{ex_grp_d_153}.
This suggests that the line lists for the main carriers (C$_2$ on one hand, HCN and C$_2$H$_2$ on the other) may need revision.


\section{Conclusions}

We have compared optical and near-infrared spectra of carbon stars with hydrostatic models at a spectral resolving power of $\sim$ 2\,000. The models provide impressive matches to most of the features in the optical and NIR range, but it remains difficult to fit all features simultaneously with a single model when only extinction is allowed to alter the energy distribution.

In defining a goodness-of-fit criterion, choices can be made that focus either on the high-frequency structure within the bands or on the medium-resolution energy distribution and the shape of the main molecular bands. In this paper, we have emphasized the latter. XSL spectra contain a wealth of high-frequency information that will be exploited more completely in the future. The main difficulty encountered at the native reso\-lution of the XSL data, R $\sim$ 10\,000 in the VIS and 8000 in the NIR, is the effect of pulsation. Shocks that travel through the atmospheres of LPVs create velocity discontinui\-ties with amplitudes above 10\,km/s, which will affect the location and shape of lines at the resolution of XSL. Moreover, line lists for carbon-bearing molecules such as C$_2$ or HCN are incomplete and
known to contain approximate wavelengths for some of the transitions \citep[e.g.][]{Loidl01,Aringer09}. Smoothing to R $\sim$ 2\,000 allows us to avoid these difficulties. Nevertheless, the residuals show that it remains difficult to fit all the spectral features even at low reso\-lution, and even when a multiplicative polynomial mimicking extinction and flux calibration corrections is allowed to modify the energy distribution. 

For stars from Groups A to C, the effective temperature ranges derived separately from the VIS and NIR wavelength range overlap in general. It is therefore possible to provide a via\-ble
$T_\mathrm{eff}$ estimate for the carbon stars in these groups. 
When temperature differences are obtained between the two wavelength ranges, the VIS-based temperature is warmer than the NIR-based temperature.
It becomes progressively more difficult to obtain reasonable matches of the data for redder observed spectra, in particular for stars from Group D. This is due to a combination of a more complex forest of molecular lines (that is only matched in detail in small parts of the spectra) and to the effects of circumstellar material on the energy distribution (that a simple extinction law cannot reproduce).
Veiling by dust is important in the NIR spectra of Group D. At a lower level, this 
veiling might contribute to explain the differences in temperature seen between
the VIS and NIR in a few of the other stars.

For the spectra of Groups A to C, fitting models constrain $\log(g)$ and $[\mathrm{Fe/H}]$. The favored surface gravity is $\log(g) = 0$, and the favored metallicity $[\mathrm{Fe/H}]=-0.5$. These values are satis\-factory for TP-AGB stars and for a sample consisting mostly of LMC, SMC and Milky Way halo stars. However, the uncertainties are quite large as are result of the small number of va\-lues avai\-lable for these parameters in the model grid and because of the degeneracy between $T_\mathrm{eff}$ and C/O, and between $T_\mathrm{eff}$ and $[\mathrm{Fe/H}]$.

For now, the fitted models show that the features at R $\sim$ 2\,000 do not perfectly represent the data. This justifies the use of empirical spectra as templates for C stars in the Universe. 
At this stage, however, it is not possible to precisely assign an evolutio\-nary stage to individual stars in the collection based on their spectra alone. When using the C-star observations as templates for stellar population models, it will probably be necessary to continue to use averages, as done for instance by \citet{Lancon02} for O-rich and C-rich LPVs. 

The next step will be to use dynamical models, that is to say models that take into account the pulsating stellar interior as well as the development of dust-driven winds \citep[e.g.][]{Eriksson14}. It has been
suggested that a simple dust-envelope model could be used as an intermediate
step before entering the complexity of dynamical models \citep[e.g.][]{Aringer09,Paladini11}. 
With dust radiative transfer codes, we can expect to reproduce the attenuation of the absorption features and also the overall shift of the SED toward redder wavelengths \citep{Nowotny11}. The situation is not that straightforward however. 
First, large-amplitude pulsation and dust production are closely related \citep{Sloan16}.
Se\-cond, the appearance of the 13.7\,$\mu$m C$_2$H$_2$ band in deeply embedded carbon stars points to the fact that acetylene in gas form is producing molecular absorption bands well into, and possibly beyond, the dust-forming region \citep{Matsuura16}. This complicates the situation for the 1.53\,$\mu$m feature, most likely carried by HCN\,+\,C$_2$H$_2$.
Therefore, it seems more promising to directly use the low-resolution version of the pulsating C-star models directly. 
In particular, a study of their mole\-cular stratification
would be useful: the XSL C stars indicate that some CO should be located above
dust emission layers (Paper I), and this hypo\-thesis requires verification. If high-resolution versions of the pulsating model spectra can be computed, we could also more quantitatively explore
the effects of pulsation on the XSL spectra at R $\simeq$ 8\,000.

\begin{acknowledgements}

We thank the referee, Greg Sloan, for his insightful report that helped to improve the quality of this paper.
A.L. and P.P. thank PNCG (Programme National Cosmologie et Galaxies) for support in 2016. 
B.A. was supported by the ERC Consolidator Grant funding scheme ({\em project STARKEY}, G.A. n.~615604, PI~P.~Marigo).

\end{acknowledgements}



\bibliographystyle{aa}  
\bibliography{gonneau_17_biblio} 



\begin{appendix}

\section{Models that fit our observations best}
\label{obs_plots_all}

\begin{figure*}[h]
\centering
\subfloat[VIS]{\includegraphics[trim=30 10 30 65, clip,width=0.45\hsize]{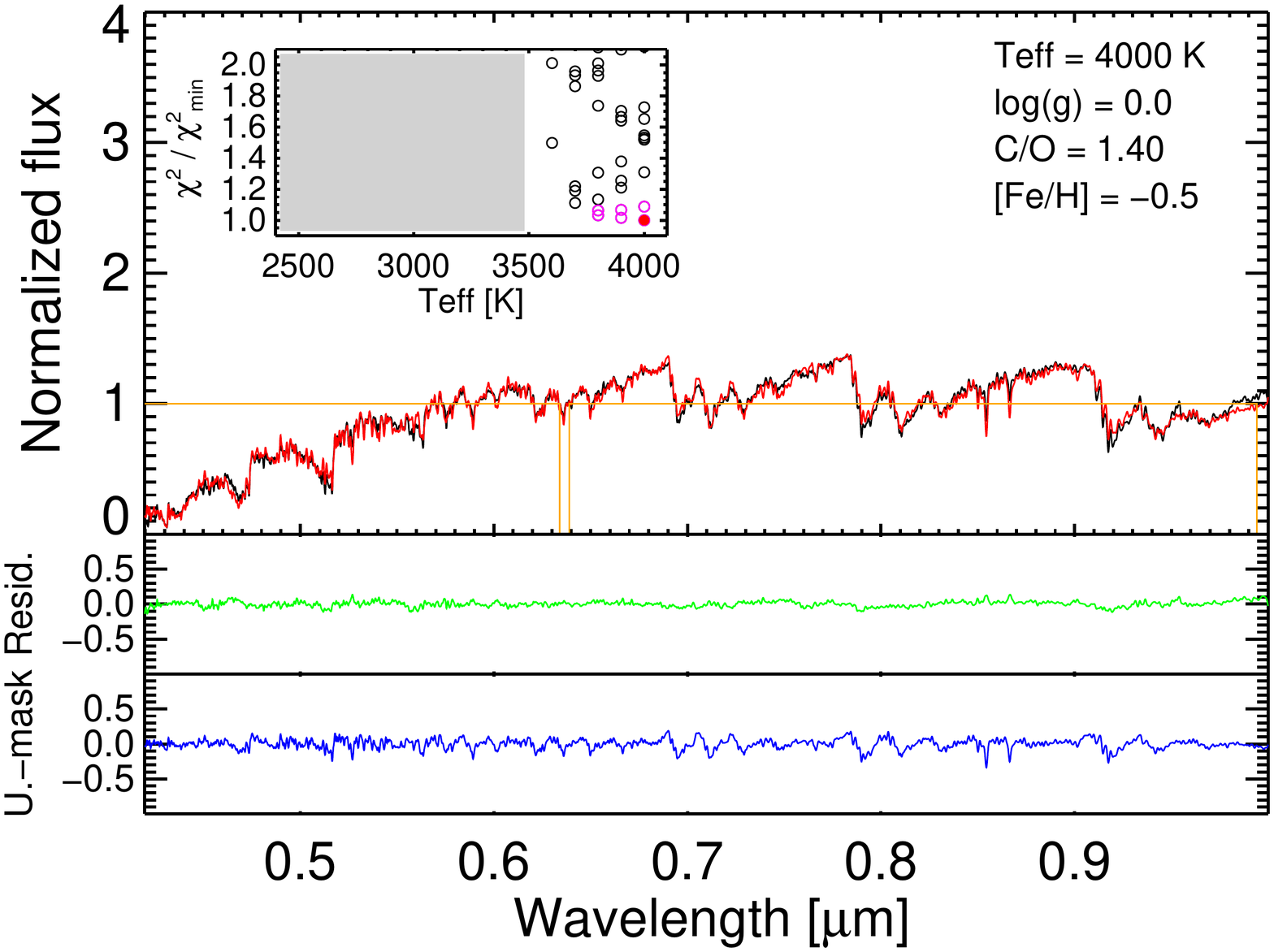}} 
\subfloat[NIR]{\includegraphics[trim=30 10 30 65, clip,width=0.45\hsize]{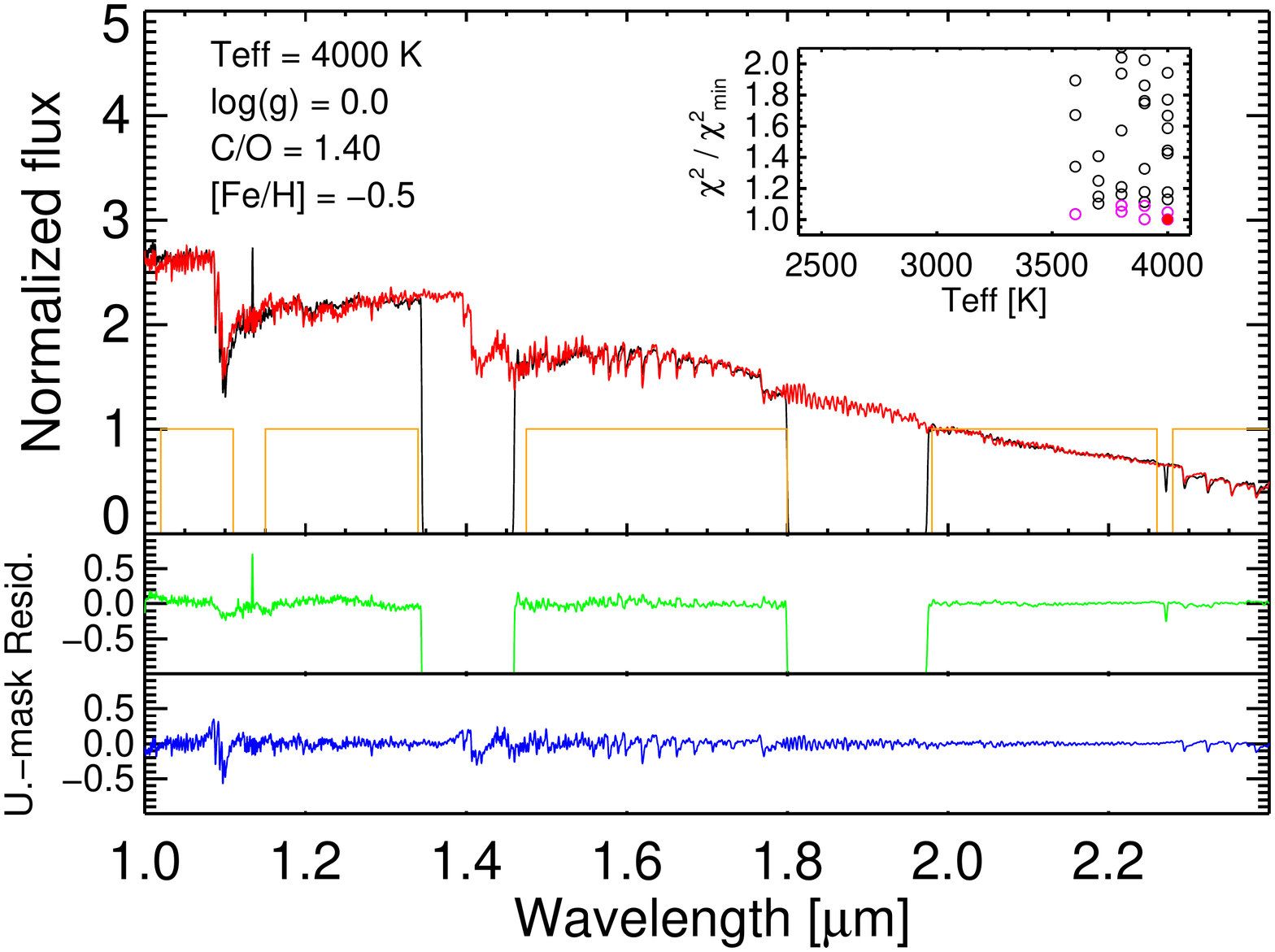}}
\caption{Best-fitting models for Cl* NGC 121 T V8 (Group A). For each observation, the upper panel shows the stellar spectrum as a black curve and the best model times the best-fitting polynomial as a red curve. The weights applied for the $\chi^2$ minimization are displayed in orange. The middle panel shows the residuals (in green), while the bottom panel shows an unsharp-mask filtered version 
of the best-model spectrum (in blue).
In addition, a small inset shows the output $\chi^2 / \chi^2_{min}$ values as a function of the temperatures of the input models. 
The gray-shaded area masks the models excluded from the fitting in order to avoid unphysical dereddening of the models. The red solid point corresponds to the best model, while the magenta points are for the range of models considered similarly acceptable. 
Note that some of the VIS spectra lack data around 0.63\,$\mu$m, e.g., Fig. A4). This results from the bad columns in the VIS arm of the X-Shooter (an artifact restricted to data taken in the ESO semester P84).}
\label{fit_beg_a}
\end{figure*}

Tables~\ref{table_grp_a} to~\ref{table_grp_d} show the range of parameters for the mo\-dels that fit the observed spectra best. 
Figures~\ref{fit_beg_a} to~\ref{fit_end_d} show the best-fitting models for each observation. For each figure, the upper panel shows the observation (black curve), the best model times the polynomial (red curve), and the weights used for the $\chi^2$ minimization (orange). The middle panel shows the residuals (in green), while the bottom panel shows an unsharp-mask filtered version 
of the best-model spectrum (the difference between the synthetic spectrum and a heavily smoothed version thereof, in blue). In addition, a small inset shows the output $\chi^2 / \chi^2_{min}$ values as a function of the temperatures of the input models. 







\begin{table*}
\caption{\label{table_grp_a}Range of parameters for the best-fitting models for the stars from Group A}
\centering
\begin{adjustbox}{width=1\textwidth}
\begin{tabular}{l|c|c|c|c|c|c|c|c|c|c|c|c|c|c|c|c|c}
\cline{3-18}
\multicolumn{2}{l}{} & \multicolumn{4}{|c|}{$T_{eff}$} & \multicolumn{4}{|c|}{log($g$)} & \multicolumn{4}{|c|}{C/O} & \multicolumn{4}{|c}{[Fe/H]} \\ 
\hline

Name & Range & Best & Wei & Min & Max & Best & Wei & Min & Max & Best & Wei & Min & Max & Best & Wei & Min & Max \\

\hline

Cl* NGC 121 T V8 & V & 4000 & 3900 & 3800 & 4000 &        0.0 &        0.0 &
       0.0 &        0.0 &       1.40 &       1.34 &       1.05 &       2.00 &
      -0.5 &       -0.5 &       -0.5 &       -0.5 \\
(Figure A.1) & N & 4000 & 3857 & 3600 & 4000 &        0.0 &        0.6 &        0.0 &
       2.0 &       1.40 &       1.12 &       1.01 &       1.40 &       -0.5 &
      -0.4 &       -0.5 &        0.0 \\
\hline
SHV 0517337-725738 & V & 3600 & 3632 & 3500 & 3800 &        0.0 &        0.3 &
       0.0 &        2.0 &       1.10 &       1.17 &       1.05 &       1.40 &
      -0.5 &       -0.4 &       -0.5 &        0.0 \\
(Figure A.2) & N & 4000 & 3861 & 3700 & 4000 &        0.0 &        0.0 &        0.0 &
       0.0 &       1.40 &       1.22 &       1.10 &       1.40 &       -0.5 &
      -0.5 &       -0.5 &       -0.5 \\
\hline
SHV 0518161-683543 & V & 3600 & 3649 & 3600 & 3700 &        0.0 &        0.0 &
       0.0 &        0.0 &       1.40 &       1.47 &       1.10 &       2.00 &
      -0.5 &       -0.5 &       -0.5 &       -0.5 \\
(Figure A.3)& N & 3600 & 3662 & 3500 & 3900 &        0.0 &        0.2 &        0.0 &
       2.0 &       1.05 &       1.11 &       1.01 &       1.40 &       -0.5 &
      -0.4 &       -0.5 &        0.0 \\
\hline

\end{tabular}
\end{adjustbox}

\tablefoot{
The column ``Range'' indicates the wavelength range used for the fitting:
V=visible, N=near-infrared. \\
The columns ``Best'' indicate the parameters of the best-fitting model. \\
The columns ``Wei'' indicate the weighted values for each parameter as calculated in Eq.~\ref{eq_wei}. \\
The columns ``Min'' and ``Max'' give the range of values for each parameter.\\
}

\end{table*}

\begin{table*}
\caption{\label{table_grp_b}Range of parameters for the best-fitting models for the stars from Group B}
\centering
\begin{adjustbox}{width=1\textwidth}
\begin{tabular}{l|c|c|c|c|c|c|c|c|c|c|c|c|c|c|c|c|c}
\cline{3-18}
\multicolumn{2}{l}{} & \multicolumn{4}{|c|}{$T_{eff}$} & \multicolumn{4}{|c|}{log($g$)} & \multicolumn{4}{|c|}{C/O} & \multicolumn{4}{|c}{[Fe/H]} \\ 
\hline

Name & Range & Best & Wei & Min & Max & Best & Wei & Min & Max & Best & Wei & Min & Max & Best & Wei & Min & Max \\

\hline

2MASS J00571648-7310527 & V & 3700 & 3699 & 3600 & 3800 &        0.0 &
       0.0 &        0.0 &        0.0 &       2.00 &       1.80 &       1.40 &
      2.00 &       -0.5 &       -0.5 &       -0.5 &       -0.5 \\
(Figure A.4) & N & 3600 & 3488 & 3300 & 3700 &        0.0 &        0.4 &        0.0 &
       2.0 &       1.40 &       1.28 &       1.05 &       2.00 &       -0.5 &
      -0.4 &       -0.5 &        0.0 \\
\hline
2MASS J01003150-7307237 & V & 3700 & 3649 & 3500 & 3800 &        0.0 &
       0.0 &        0.0 &        0.0 &       1.40 &       1.39 &       1.05 &
      2.00 &       -0.5 &       -0.5 &       -0.5 &       -0.5 \\
(Figure A.5) & N & 3800 & 3585 & 3400 & 3800 &        0.0 &        0.0 &        0.0 &
       0.0 &       1.40 &       1.18 &       1.05 &       1.40 &       -0.5 &
      -0.5 &       -0.5 &       -0.5 \\
\hline
2MASS J00563906-7304529 & V & 3200 & 3133 & 2800 & 3400 &        0.0 &
      -0.1 &       -0.4 &        0.0 &       1.40 &       1.46 &       1.10 &
      2.00 &        0.0 &       -0.2 &       -0.5 &        0.0 \\
(Figure A.6) & N & 3200 & 3190 & 3000 & 3400 &        0.0 &        0.0 &        0.0 &
       0.0 &       2.00 &       1.73 &       1.40 &       2.00 &       -0.5 &
      -0.3 &       -0.5 &        0.0 \\
\hline
2MASS J00530765-7307477 & V & 3600 & 3699 & 3600 & 3800 &        0.0 &
       0.0 &        0.0 &        0.0 &       2.00 &       2.00 &       2.00 &
      2.00 &       -0.5 &       -0.3 &       -0.5 &        0.0 \\
(Figure A.7) & N & 3600 & 3501 & 3400 & 3600 &        0.0 &        0.0 &        0.0 &
       0.0 &       2.00 &       1.60 &       1.40 &       2.00 &       -0.5 &
      -0.5 &       -0.5 &       -0.5 \\
\hline
2MASS J00493262-7317523 & V & 3500 & 3549 & 3500 & 3600 &        0.0 &
       0.0 &        0.0 &        0.0 &       1.40 &       1.70 &       1.40 &
      2.00 &       -0.5 &       -0.5 &       -0.5 &       -0.5 \\
(Figure A.8) & N & 3500 & 3477 & 3300 & 3700 &        0.0 &        0.7 &        0.0 &
       2.0 &       1.40 &       1.35 &       1.05 &       2.00 &       -0.5 &
      -0.3 &       -0.5 &        0.0 \\
\hline
2MASS J00490032-7322238 & V & 3500 & 3499 & 3400 & 3600 &        0.0 &
       0.0 &        0.0 &        0.0 &       2.00 &       1.80 &       1.40 &
      2.00 &       -0.5 &       -0.5 &       -0.5 &       -0.5 \\
(Figure A.9) & N & 3400 & 3416 & 3200 & 3600 &        0.0 &        0.0 &        0.0 &
       0.0 &       1.40 &       1.55 &       1.10 &       2.00 &       -0.5 &
      -0.5 &       -0.5 &       -0.5 \\
\hline
2MASS J00571214-7307045 & V & 3500 & 3400 & 3300 & 3500 &        0.0 &
       0.0 &        0.0 &        0.0 &       2.00 &       1.70 &       1.40 &
      2.00 &       -0.5 &       -0.5 &       -0.5 &       -0.5 \\
(Figure A.10) & N & 3400 & 3416 & 3200 & 3600 &        0.0 &        0.0 &        0.0 &
       0.0 &       1.40 &       1.55 &       1.10 &       2.00 &       -0.5 &
      -0.5 &       -0.5 &       -0.5 \\
\hline

\end{tabular}
\end{adjustbox}
\end{table*}

\begin{table*}
\caption{\label{table_grp_c}Range of parameters for the best-fitting models for the stars from Group C}
\centering
\begin{adjustbox}{width=1\textwidth}
\begin{tabular}{l|c|c|c|c|c|c|c|c|c|c|c|c|c|c|c|c|c}
\cline{3-18}
\multicolumn{2}{l}{} & \multicolumn{4}{|c|}{$T_{eff}$} & \multicolumn{4}{|c|}{log($g$)} & \multicolumn{4}{|c|}{C/O} & \multicolumn{4}{|c}{[Fe/H]} \\ 
\hline

Name & Range & Best & Wei & Min & Max & Best & Wei & Min & Max & Best & Wei & Min & Max & Best & Wei & Min & Max \\

\hline

2MASS J00570070-7307505 & V & 3500 & 3440 & 3300 & 3600 &        0.0 &
       0.0 &        0.0 &        0.0 &       2.00 &       1.76 &       1.40 &
      2.00 &       -0.5 &       -0.5 &       -0.5 &       -0.5 \\
(Figure A.11) & N & 3400 & 3340 & 3200 & 3500 &        0.0 &        0.0 &        0.0 &
       0.0 &       2.00 &       1.76 &       1.40 &       2.00 &       -0.5 &
      -0.5 &       -0.5 &       -0.5 \\
\hline
[W65] c2 & V & 3200 & 3270 & 3100 & 3400 &       -0.4 &       -0.2 &       -0.4
&        0.0 &       1.01 &       1.04 &       1.01 &       1.05 &        0.0 &
      -0.1 &       -0.5 &        0.0 \\
(Figure A.12) & N & 3400 & 3423 & 3300 & 3500 &        0.0 &        0.0 &        0.0 &
       0.0 &       1.05 &       1.06 &       1.05 &       1.10 &       -0.5 &
      -0.5 &       -0.5 &       -0.5 \\
\hline
2MASS J00564478-7314347 & V & 3600 & 3576 & 3500 & 3700 &        0.0 &
       0.0 &        0.0 &        0.0 &       2.00 &       1.85 &       1.40 &
      2.00 &       -0.5 &       -0.5 &       -0.5 &       -0.5 \\
(Figure A.13) & N & 3500 & 3475 & 3400 & 3600 &        0.0 &        0.0 &        0.0 &
       0.0 &       2.00 &       1.85 &       1.40 &       2.00 &       -0.5 &
      -0.5 &       -0.5 &       -0.5 \\
\hline
2MASS J00542265-7301057 & V & 3500 & 3527 & 3400 & 3700 &        0.0 &
       0.0 &        0.0 &        0.0 &       2.00 &       1.83 &       1.40 &
      2.00 &       -0.5 &       -0.4 &       -0.5 &        0.0 \\
(Figure A.14) & N & 3400 & 3376 & 3300 & 3500 &        0.0 &        0.0 &        0.0 &
       0.0 &       2.00 &       1.85 &       1.40 &       2.00 &       -0.5 &
      -0.5 &       -0.5 &       -0.5 \\
\hline
Cl* NGC 419 LE 27 & V & 3600 & 3442 & 3200 & 3700 &        0.0 &        0.0 &
       0.0 &        0.0 &       2.00 &       1.47 &       1.05 &       2.00 &
      -0.5 &       -0.5 &       -0.5 &       -0.5 \\
(Figure A.15) & N & 3300 & 3300 & 3100 & 3500 &        0.0 &        0.0 &        0.0 &
       0.0 &       2.00 &       1.78 &       1.40 &       2.00 &       -0.5 &
      -0.4 &       -0.5 &        0.0 \\
\hline
IRAS 09484-6242 & V & 3100 & 2967 & 2700 & 3200 &        0.0 &        0.0 &
       0.0 &        0.0 &       2.00 &       1.67 &       1.40 &       2.00 &
      -0.5 &       -0.5 &       -0.5 &       -0.5 \\
(Figure A.16) & N & 3100 & 3132 & 3000 & 3300 &        0.0 &        0.0 &        0.0 &
       0.0 &       2.00 &       1.93 &       1.40 &       2.00 &       -0.5 &
      -0.2 &       -0.5 &        0.0 \\
\hline
Cl* NGC 419 LE 35 & V & 3800 & 3616 & 3400 & 3900 &        0.0 &        0.5 &
       0.0 &        2.0 &       2.00 &       1.36 &       1.05 &       2.00 &
      -0.5 &       -0.4 &       -0.5 &        0.0 \\
(Figure A.17) & N & 3400 & 3329 & 3200 & 3500 &        0.0 &        0.0 &        0.0 &
       0.0 &       2.00 &       1.83 &       1.40 &       2.00 &       -0.5 &
      -0.4 &       -0.5 &        0.0 \\
\hline
2MASS J00553091-7310186 & V & 3600 & 3470 & 3300 & 3700 &        0.0 &
       0.0 &        0.0 &        0.0 &       2.00 &       1.55 &       1.05 &
      2.00 &       -0.5 &       -0.5 &       -0.5 &       -0.5 \\
(Figure A.18) & N & 3300 & 3333 & 3200 & 3500 &        0.0 &        0.0 &        0.0 &
       0.0 &       2.00 &       1.90 &       1.40 &       2.00 &       -0.5 &
      -0.4 &       -0.5 &        0.0 \\
\hline
SHV 0520427-693637 & V & 3500 & 3426 & 3300 & 3600 &        0.0 &        0.0 &
       0.0 &        0.0 &       2.00 &       1.51 &       1.05 &       2.00 &
      -0.5 &       -0.5 &       -0.5 &       -0.5 \\
(Figure A.19) & N & 3500 & 3301 & 3100 & 3500 &        0.0 &        0.6 &        0.0 &
       2.0 &       2.00 &       1.49 &       1.10 &       2.00 &       -0.5 &
      -0.4 &       -0.5 &        0.0 \\
\hline
SHV 0504353-712622 & V & 3400 & 3351 & 3200 & 3500 &        0.0 &        0.0 &
       0.0 &        0.0 &       2.00 &       1.70 &       1.40 &       2.00 &
      -0.5 &       -0.5 &       -0.5 &       -0.5 \\
(Figure A.20) & N & 3500 & 3341 & 3200 & 3500 &        0.0 &        0.0 &        0.0 &
       0.0 &       2.00 &       1.76 &       1.40 &       2.00 &       -0.5 &
      -0.5 &       -0.5 &       -0.5 \\
\hline
[ABC89] Pup 42 & V & 3500 & 3339 & 3100 & 3600 &        0.0 &        0.2 &
       0.0 &        2.0 &       2.00 &       1.37 &       1.05 &       2.00 &
      -0.5 &       -0.5 &       -0.5 &        0.0 \\
(Figure A.21) & N & 3300 & 3320 & 3100 & 3500 &        0.0 &        0.0 &        0.0 &
       0.0 &       2.00 &       1.82 &       1.40 &       2.00 &       -0.5 &
      -0.4 &       -0.5 &        0.0 \\
\hline
[ABC89] Cir 18 & V & 3500 & 3320 & 3100 & 3600 &        0.0 &        0.3 &
       0.0 &        2.0 &       2.00 &       1.39 &       1.05 &       2.00 &
      -0.5 &       -0.4 &       -0.5 &        0.0 \\
(Figure A.22) & N & 3500 & 3329 & 3200 & 3500 &        0.0 &        0.3 &        0.0 &
       2.0 &       2.00 &       1.62 &       1.10 &       2.00 &       -0.5 &
      -0.4 &       -0.5 &        0.0 \\
\hline
[ABC89] Cir 18 & V & 3500 & 3381 & 3100 & 3700 &        0.0 &        0.5 &
       0.0 &        2.0 &       2.00 &       1.37 &       1.05 &       2.00 &
      -0.5 &       -0.4 &       -0.5 &        0.0 \\
(Figure A.23) & N & 3300 & 3249 & 3100 & 3400 &        0.0 &        0.0 &        0.0 &
       0.0 &       2.00 &       1.85 &       1.40 &       2.00 &       -0.5 &
      -0.4 &       -0.5 &        0.0 \\
\hline

\end{tabular}
\end{adjustbox}
\end{table*}

\begin{table*}
\caption{\label{table_grp_d}Range of parameters for the best-fitting models for the stars from Group D. 
\textit{Warning:} Numerous spectra in this group cannot be fitted well with static
models (see text), and the best-fitted parameters listed here for completeness
are to be considered with extreme caution.}
\centering
\begin{adjustbox}{width=1\textwidth}
\begin{tabular}{l|c|c|c|c|c|c|c|c|c|c|c|c|c|c|c|c|c}
\cline{3-18}
\multicolumn{2}{l}{} & \multicolumn{4}{|c|}{$T_{eff}$} & \multicolumn{4}{|c|}{log($g$)} & \multicolumn{4}{|c|}{C/O} & \multicolumn{4}{|c}{[Fe/H]} \\ 
\hline

Name & Range & Best & Wei & Min & Max & Best & Wei & Min & Max & Best & Wei & Min & Max & Best & Wei & Min & Max \\

\hline

SHV 0500412-684054 & V & 3600 & 3516 & 3300 & 3700 &        0.0 &        0.0 &
       0.0 &        0.0 &       2.00 &       1.49 &       1.05 &       2.00 &
      -0.5 &       -0.5 &       -0.5 &       -0.5 \\
(Figure A.25) & N & 3300 & 3349 & 2700 & 3800 &        2.0 &        1.3 &        0.0 &
       2.0 &       1.05 &       1.30 &       1.05 &       2.00 &        0.0 &
      -0.2 &       -0.5 &        0.0 \\
\hline
SHV 0502469-692418 & V & 3300 & 3385 & 3200 & 3600 &        0.0 &        0.0 &
       0.0 &        0.0 &       1.05 &       1.25 &       1.05 &       2.00 &
      -0.5 &       -0.5 &       -0.5 &       -0.5 \\
(Figure A.26) & N & 2700 & 3270 & 2600 & 3900 &        2.0 &        1.2 &        0.0 &
       2.0 &       1.05 &       1.28 &       1.01 &       2.00 &        0.0 &
      -0.2 &       -0.5 &        0.0 \\
\hline
SHV 0520505-705019 & V & 3700 & 3699 & 3600 & 3800 &        0.0 &        0.0 &
       0.0 &        0.0 &       2.00 &       2.00 &       2.00 &       2.00 &
      -0.5 &       -0.5 &       -0.5 &       -0.5 \\
(Figure A.27) & N & 3600 & 3456 & 3100 & 3800 &        0.0 &        1.0 &        0.0 &
       2.0 &       2.00 &       1.40 &       1.05 &       2.00 &       -0.5 &
      -0.2 &       -0.5 &        0.0 \\
\hline
SHV 0518222-750327 & V & 3600 & 3600 & 3500 & 3700 &        0.0 &        0.0 &
       0.0 &        0.0 &       2.00 &       1.70 &       1.40 &       2.00 &
      -0.5 &       -0.5 &       -0.5 &       -0.5 \\
(Figure A.28) & N & 3700 & 3585 & 3200 & 4000 &        0.0 &        0.7 &        0.0 &
       2.0 &       2.00 &       1.49 &       1.05 &       2.00 &       -0.5 &
      -0.3 &       -0.5 &        0.0 \\
\hline
SHV 0527072-701238 & V & 3500 & 3656 & 3500 & 3900 &        2.0 &        0.6 &
       0.0 &        2.0 &       1.05 &       1.38 &       1.05 &       2.00 &
       0.0 &       -0.4 &       -0.5 &        0.0 \\
(Figure A.29) & N & 3400 & 3476 & 3000 & 3900 &        2.0 &        1.1 &        0.0 &
       2.0 &       1.05 &       1.40 &       1.05 &       2.00 &        0.0 &
      -0.2 &       -0.5 &        0.0 \\
\hline
SHV 0525478-690944 & V & 3800 & 3836 & 3600 & 4000 &        2.0 &        1.5 &
       0.0 &        2.0 &       1.05 &       1.20 &       1.01 &       2.00 &
       0.0 &       -0.1 &       -0.5 &        0.0 \\
(Figure A.30) & N & 4000 & 3802 & 3600 & 4000 &        0.0 &        0.8 &        0.0 &
       2.0 &       2.00 &       1.63 &       1.05 &       2.00 &       -0.5 &
      -0.3 &       -0.5 &        0.0 \\
\hline
SHV 0536139-701604 & V & 3500 & 3613 & 3400 & 3900 &        2.0 &        0.4 &
       0.0 &        2.0 &       1.05 &       1.40 &       1.05 &       2.00 &
       0.0 &       -0.4 &       -0.5 &        0.0 \\
(Figure A.31) & N & 3500 & 3393 & 3100 & 3700 &        0.0 &        0.6 &        0.0 &
       2.0 &       2.00 &       1.57 &       1.05 &       2.00 &       -0.5 &
      -0.3 &       -0.5 &        0.0 \\
\hline
SHV 0528537-695119 & V & 4000 & 3862 & 3700 & 4000 &        0.0 &        0.5 &
       0.0 &        2.0 &       2.00 &       1.38 &       1.05 &       2.00 &
      -0.5 &       -0.4 &       -0.5 &        0.0 \\
(Figure A.32) & N & 3900 & 3745 & 3500 & 4000 &        0.0 &        0.8 &        0.0 &
       2.0 &       2.00 &       1.60 &       1.05 &       2.00 &       -0.5 &
      -0.3 &       -0.5 &        0.0 \\
\hline

\end{tabular}
\end{adjustbox}

\tablefoot{
The column ``Range'' indicates the wavelength range used for the fitting:
V=visible, N=near-infrared. \\
The columns ``Best'' indicate the parameters of the best-fitting model. \\
The columns ``Wei'' indicate the weighted values for each parameter as calculated in Eq.~\ref{eq_wei}. \\
The columns ``Min'' and ``Max'' give the range of values for each parameter.\\
}

\end{table*}




\begin{figure*}[h]
\centering
\subfloat[VIS]{\includegraphics[trim=30 10 30 65, clip,width=0.45\hsize]{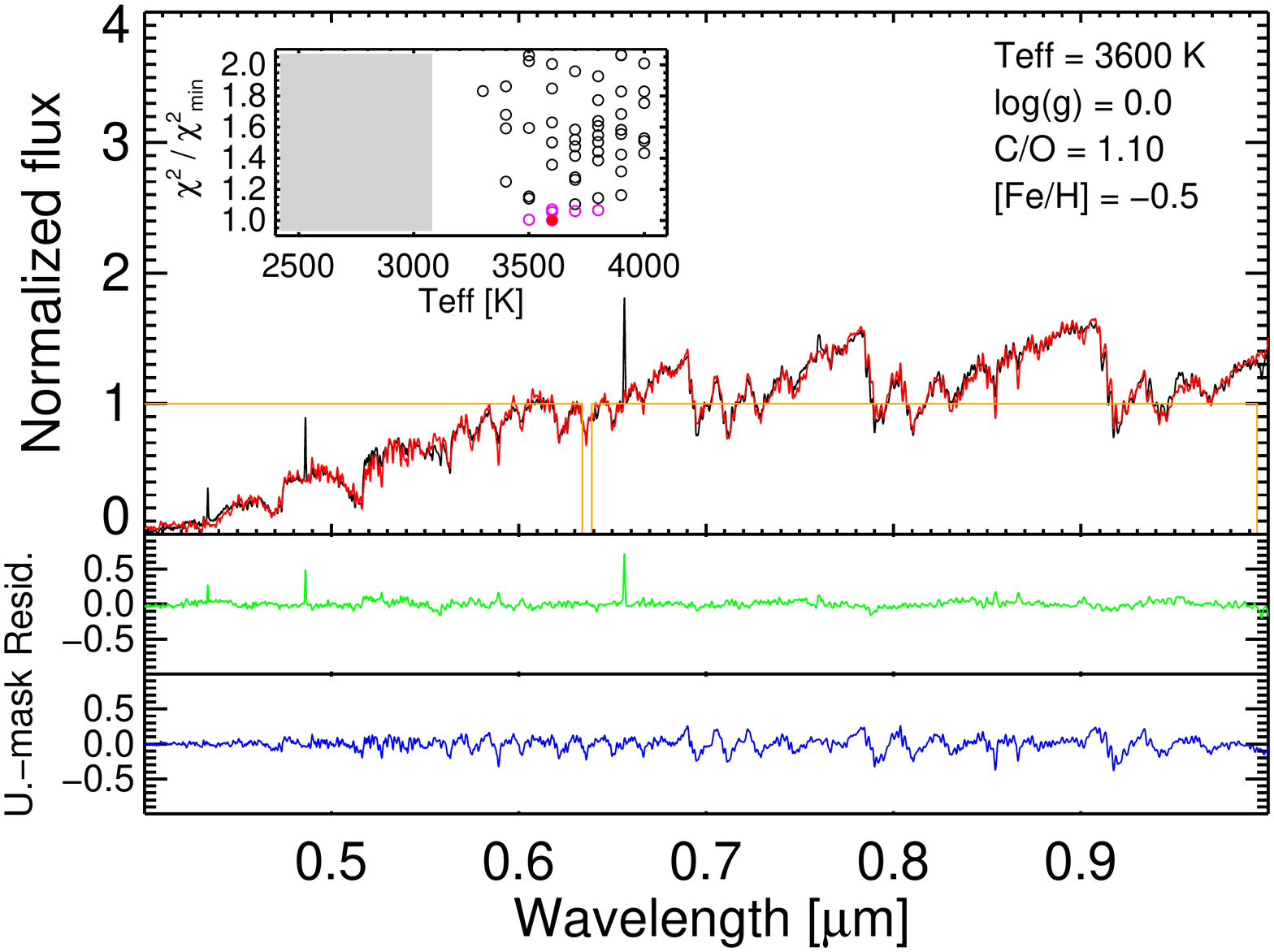}} 
\subfloat[NIR]{\includegraphics[trim=30 10 30 65, clip,width=0.45\hsize]{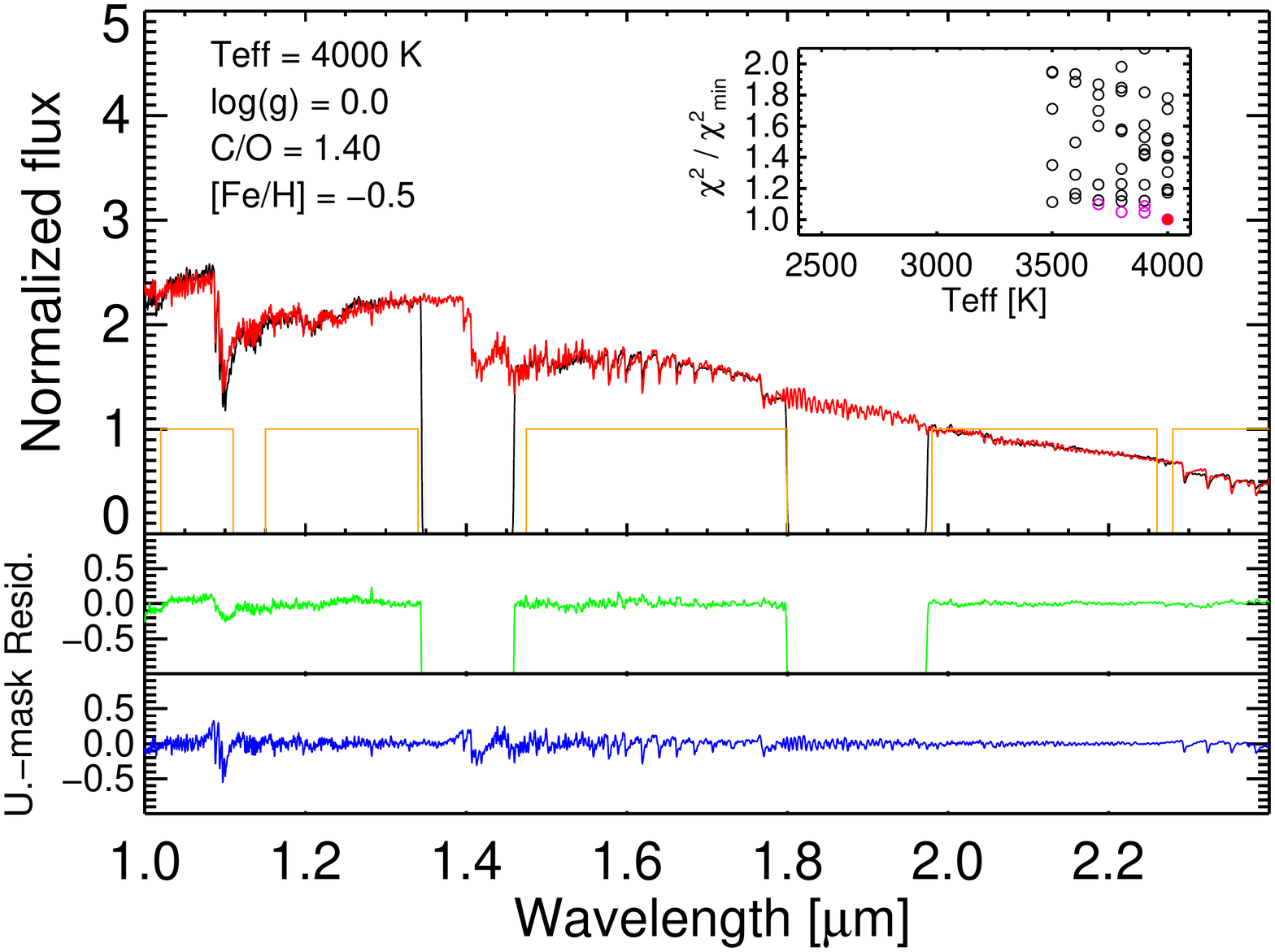}}
\caption{Best-fitting models for SHV 0517337-725738 (Group A). Same legend as for Fig.~\ref{fit_beg_a}.}
\end{figure*}


\begin{figure*}[h]
\centering
\subfloat[VIS]{\includegraphics[trim=30 10 30 65, clip,width=0.45\hsize]{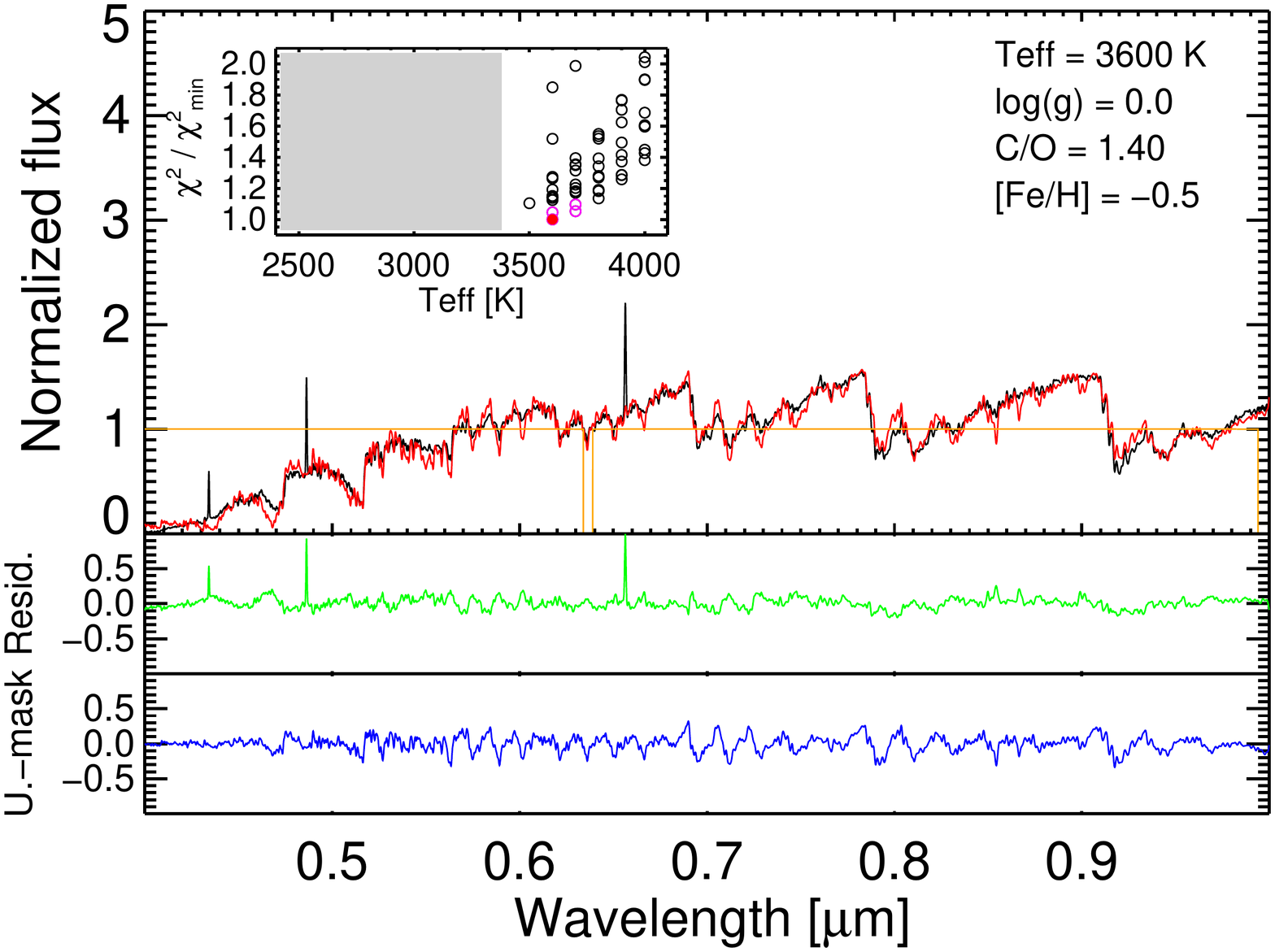}} 
\subfloat[NIR]{\includegraphics[trim=30 10 30 65, clip,width=0.45\hsize]{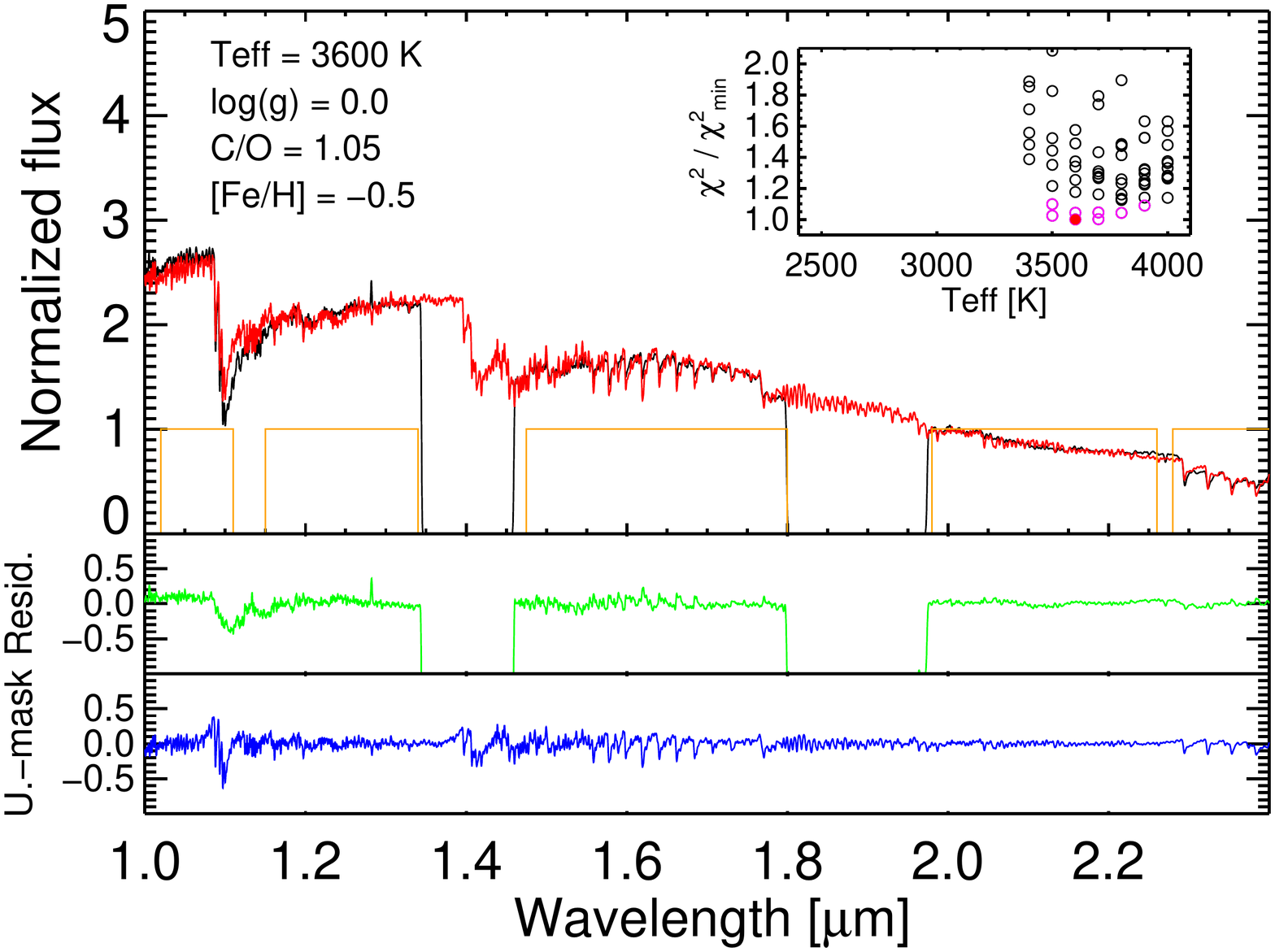}}
\caption{Best-fitting models for SHV 0518161-683543 (Group A). Same legend as for Fig.~\ref{fit_beg_a}.}
\label{fit_end_a}
\end{figure*}


\begin{figure*}[h] 
\centering
\subfloat[VIS]{\includegraphics[trim=30 10 30 65, clip,width=0.45\hsize]{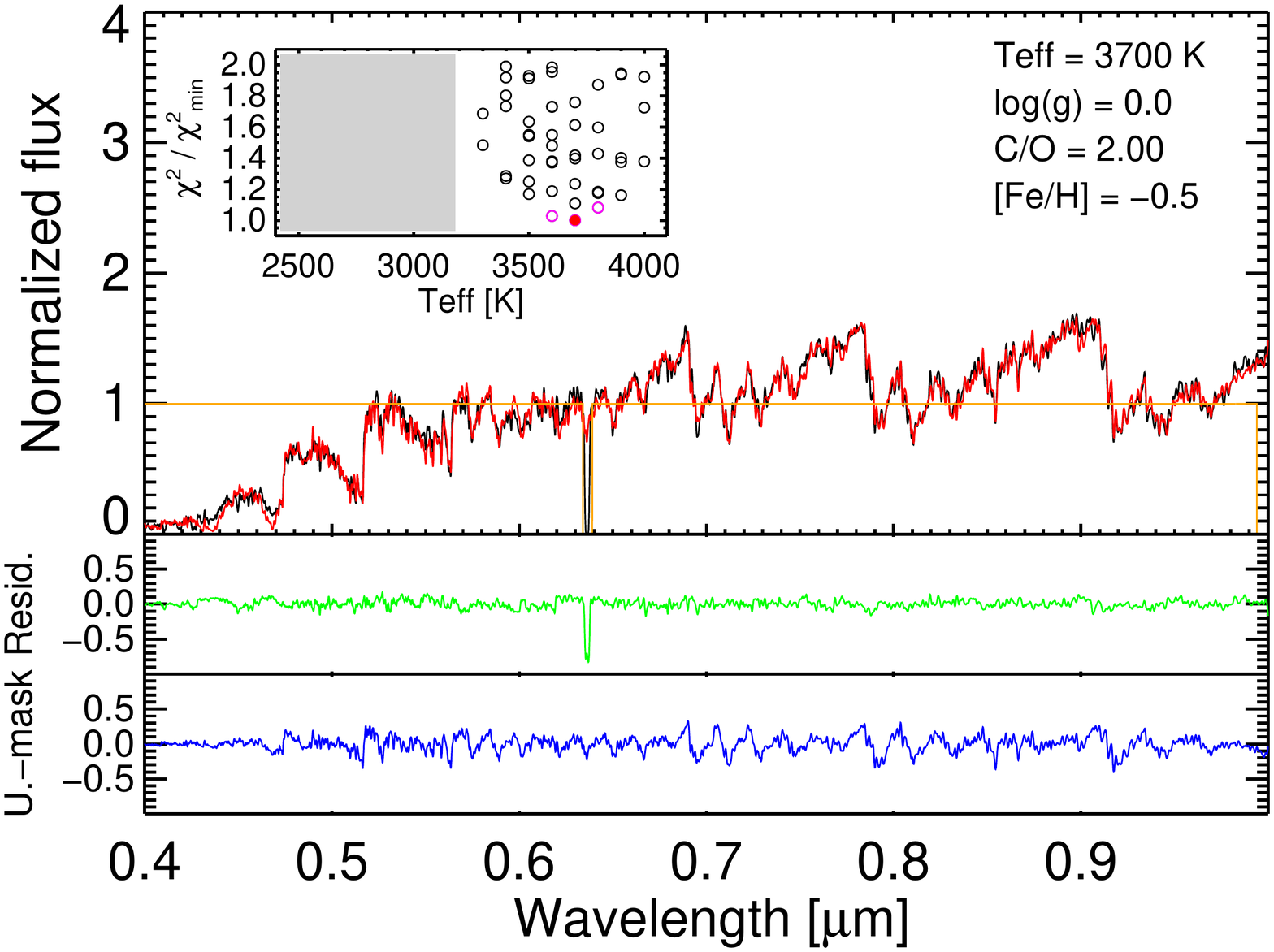}} 
\subfloat[NIR]{\includegraphics[trim=30 10 30 65, clip,width=0.45\hsize]{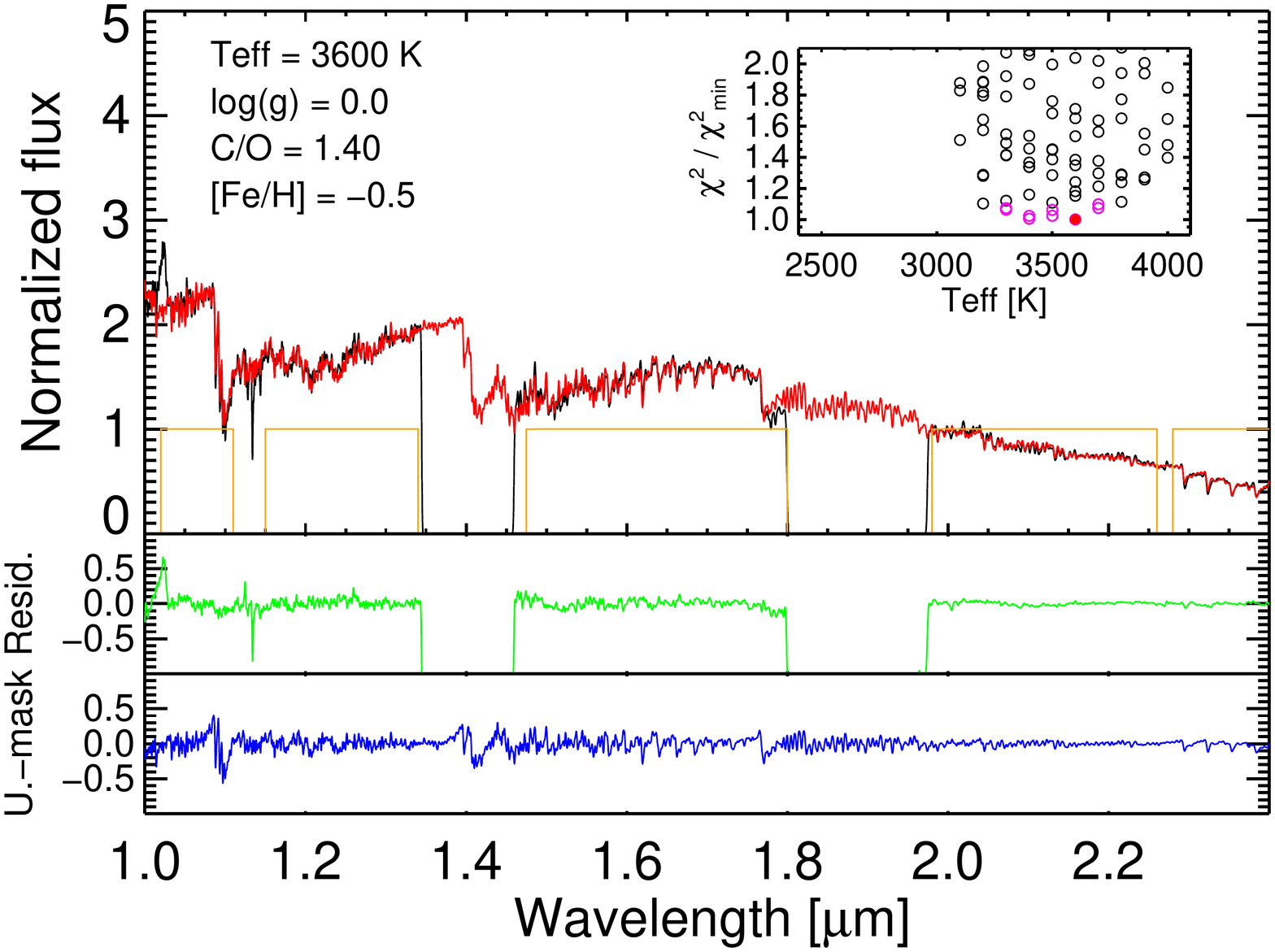}}
\caption{Best-fitting models for 2MASS J00571648-7310527 (Group B). Same legend as for Fig.~\ref{fit_beg_a}.}
\label{fit_beg_b}
\end{figure*}


\begin{figure*}[h] 
\centering
\subfloat[VIS]{\includegraphics[trim=30 10 30 65, clip,width=0.45\hsize]{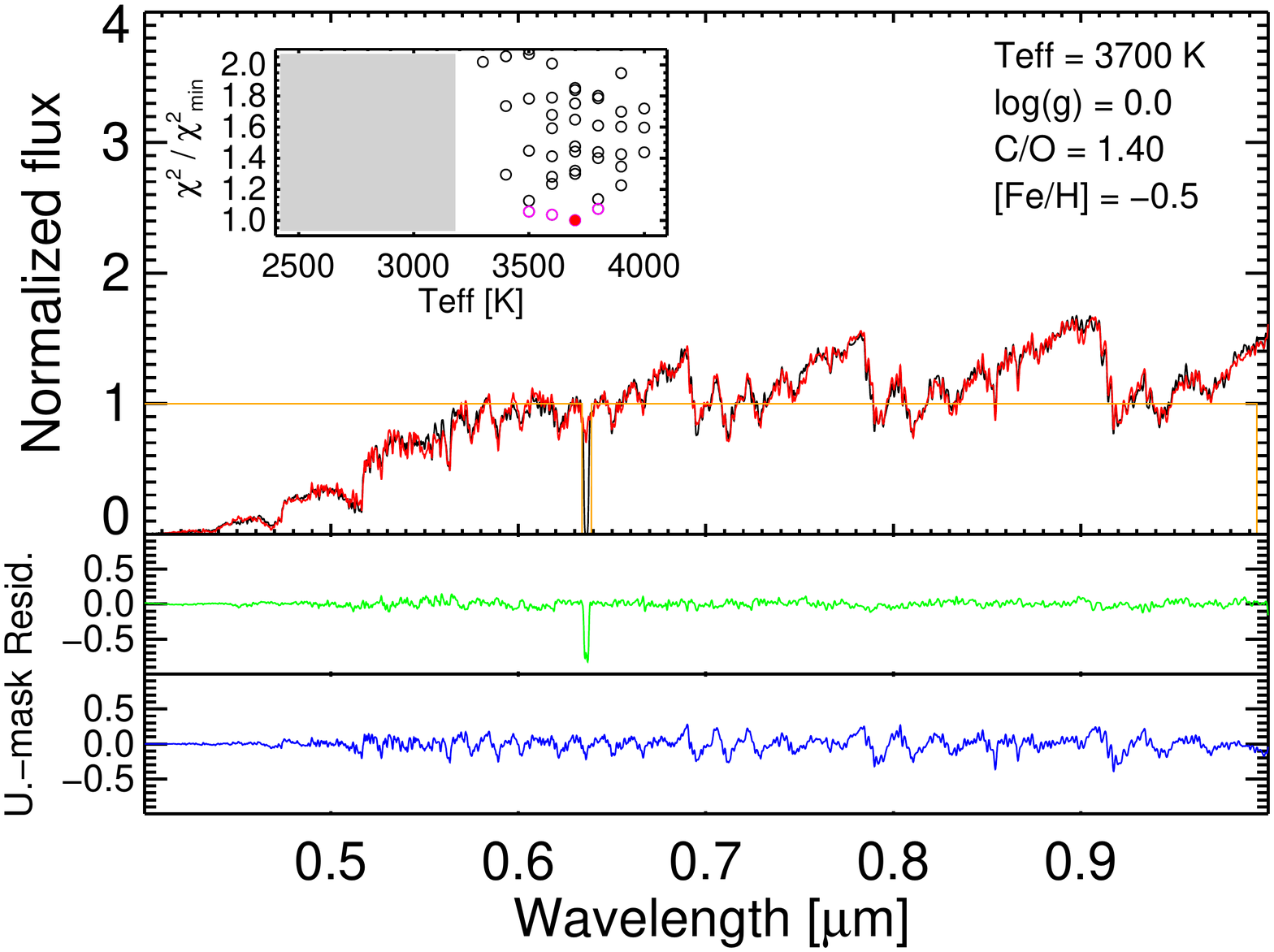}} 
\subfloat[NIR]{\includegraphics[trim=30 10 30 65, clip,width=0.45\hsize]{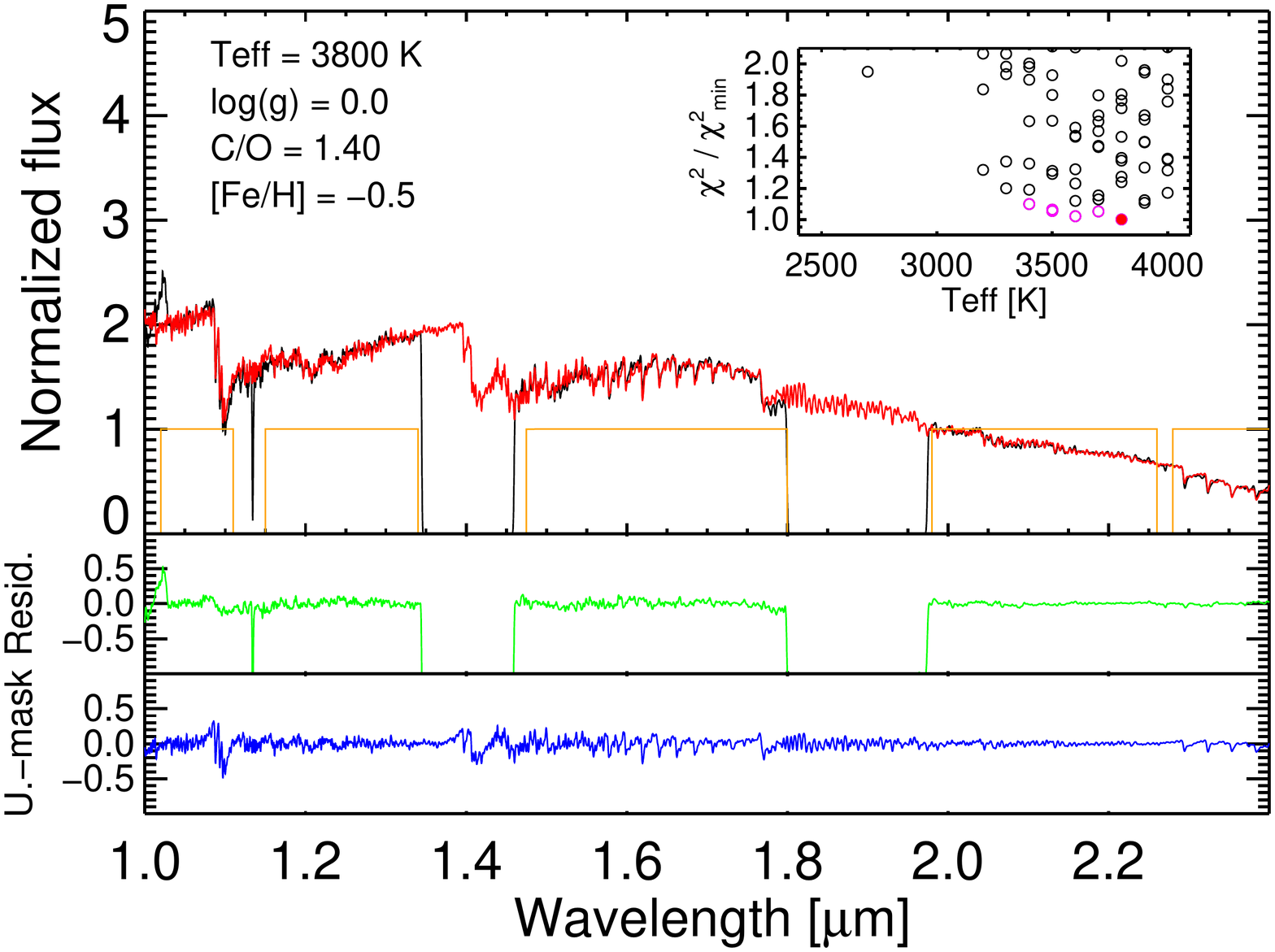}}
\caption{Best-fitting models for 2MASS J01003150-7307237 (Group B). Same legend as for Fig.~\ref{fit_beg_a}.}
\end{figure*}


\begin{figure*}[h] 
\centering
\subfloat[VIS]{\includegraphics[trim=30 10 30 65, clip,width=0.45\hsize]{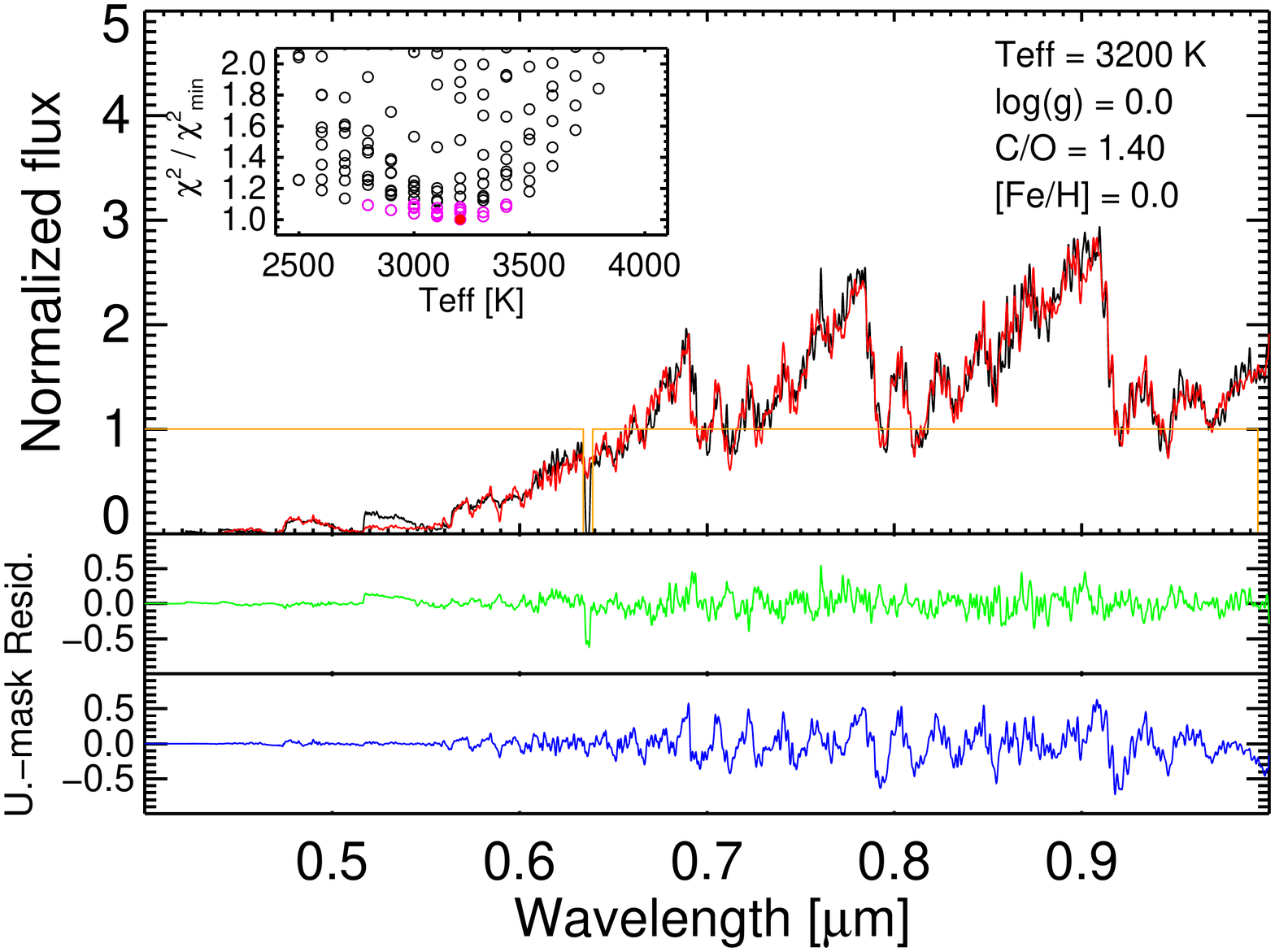}} 
\subfloat[NIR]{\includegraphics[trim=30 10 30 65, clip,width=0.45\hsize]{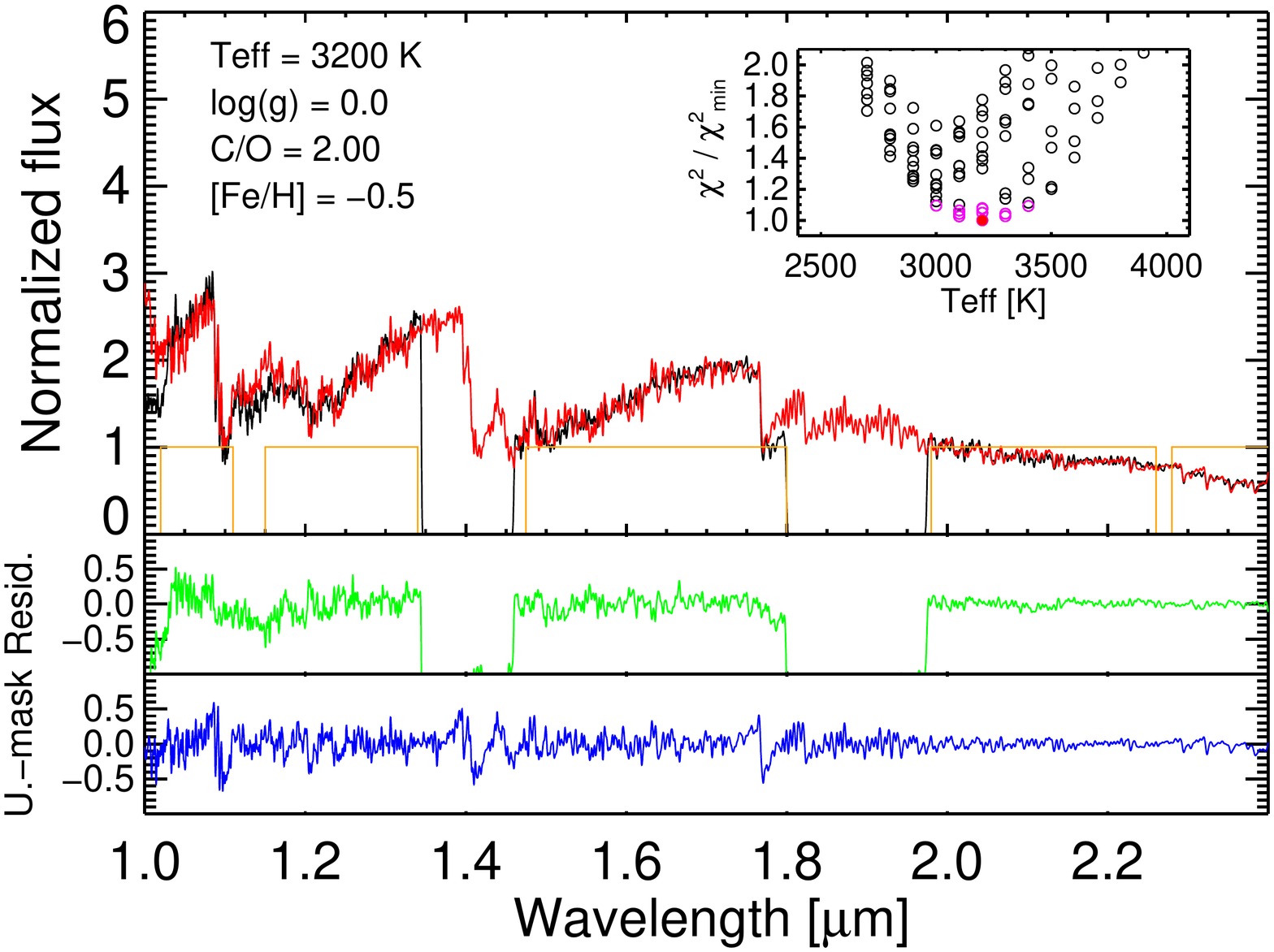}}
\caption{Best-fitting models for 2MASS J00563906-7304529 (Group B). Same legend as for Fig.~\ref{fit_beg_a}.}
\label{ex_grp_b_temp}
\end{figure*}



\begin{figure*}[h] 
\centering
\subfloat[VIS]{\includegraphics[trim=30 10 30 65, clip,width=0.45\hsize]{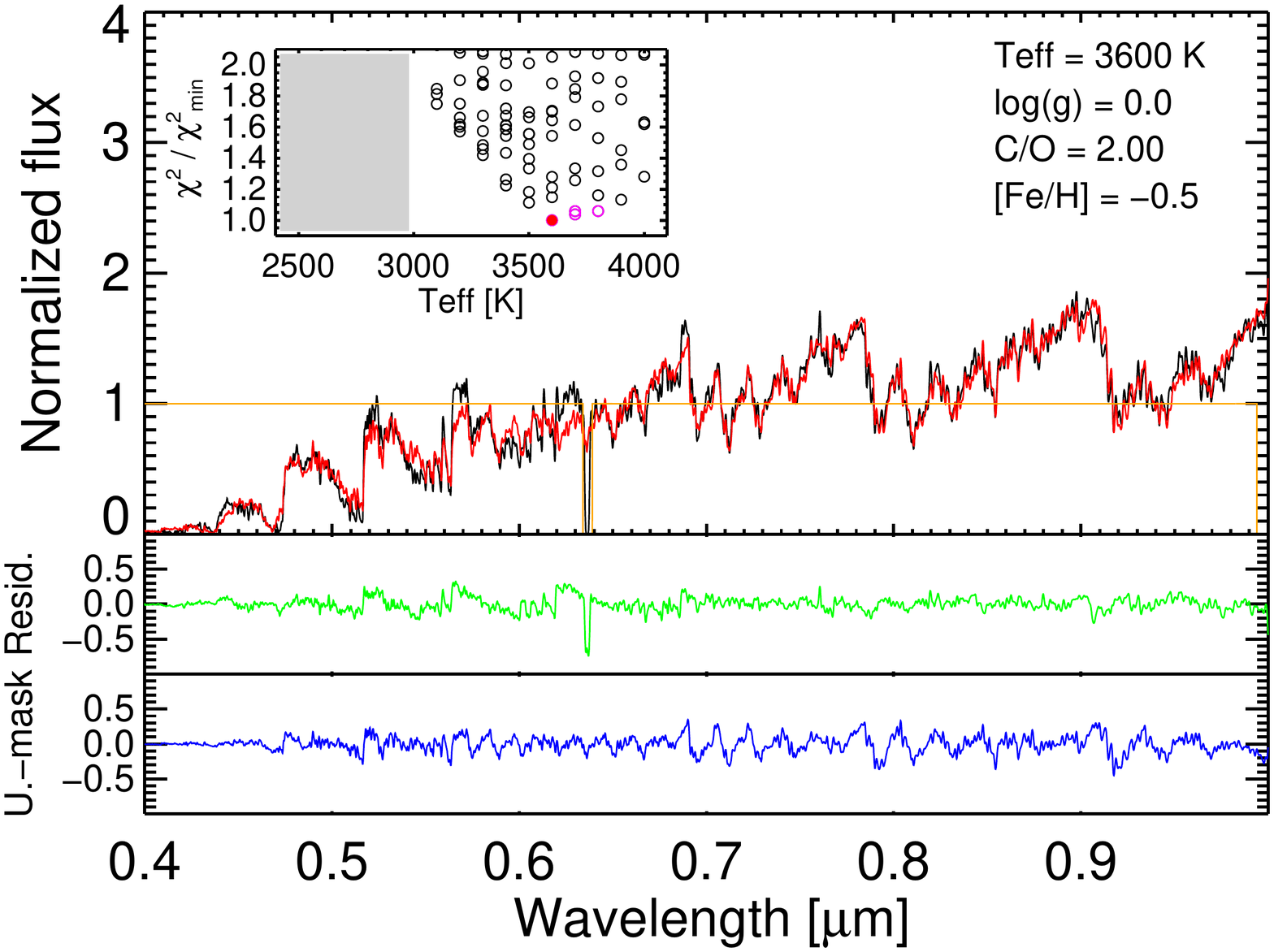}} 
\subfloat[NIR]{\includegraphics[trim=30 10 30 65, clip,width=0.45\hsize]{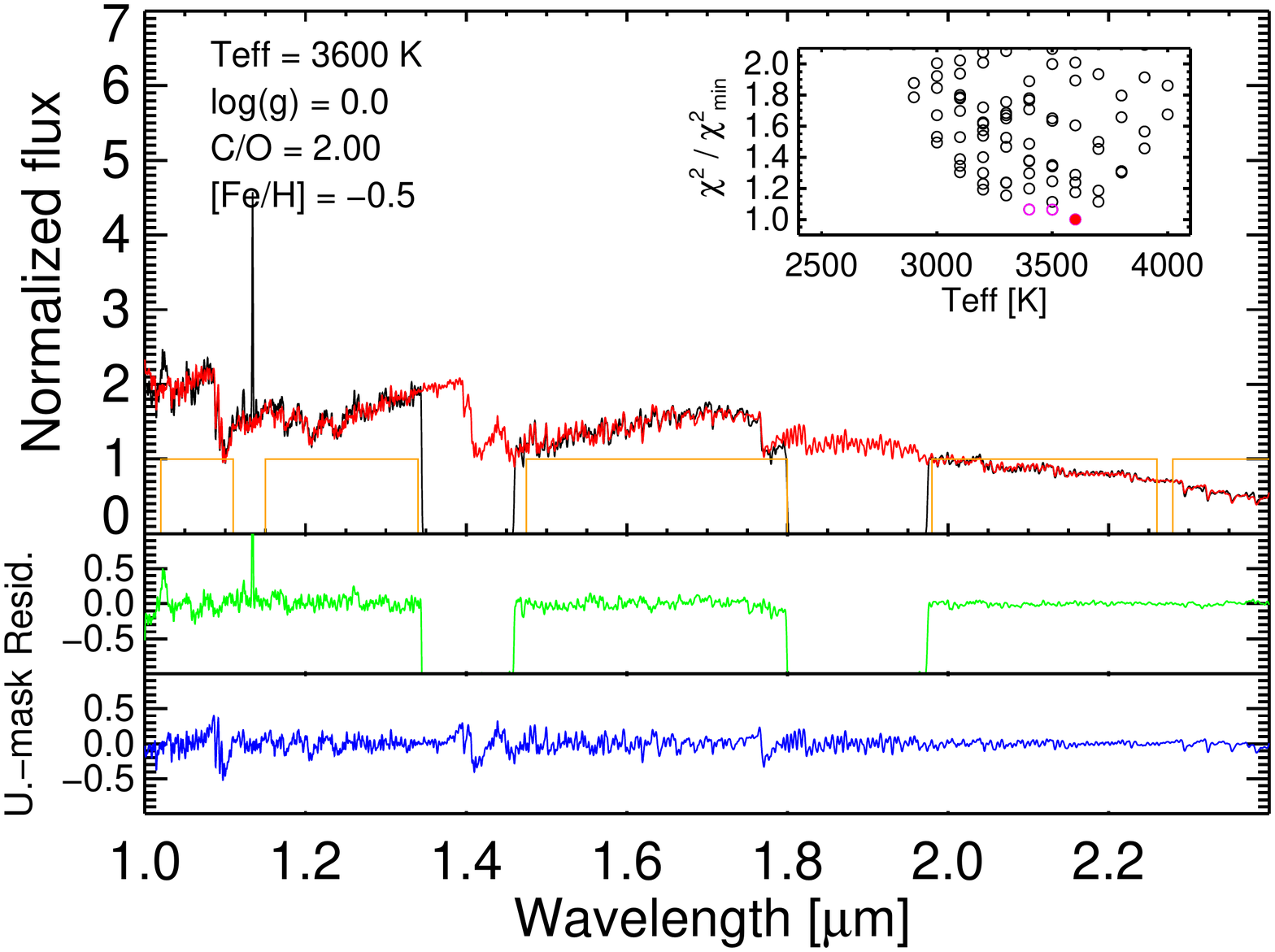}}
\caption{Best-fitting models for 2MASS J00530765-7307477 (Group B). Same legend as for Fig.~\ref{fit_beg_a}.}
\end{figure*}


\begin{figure*}[h]
 \centering
\subfloat[VIS]{\includegraphics[trim=30 10 30 65, clip,width=0.45\hsize]{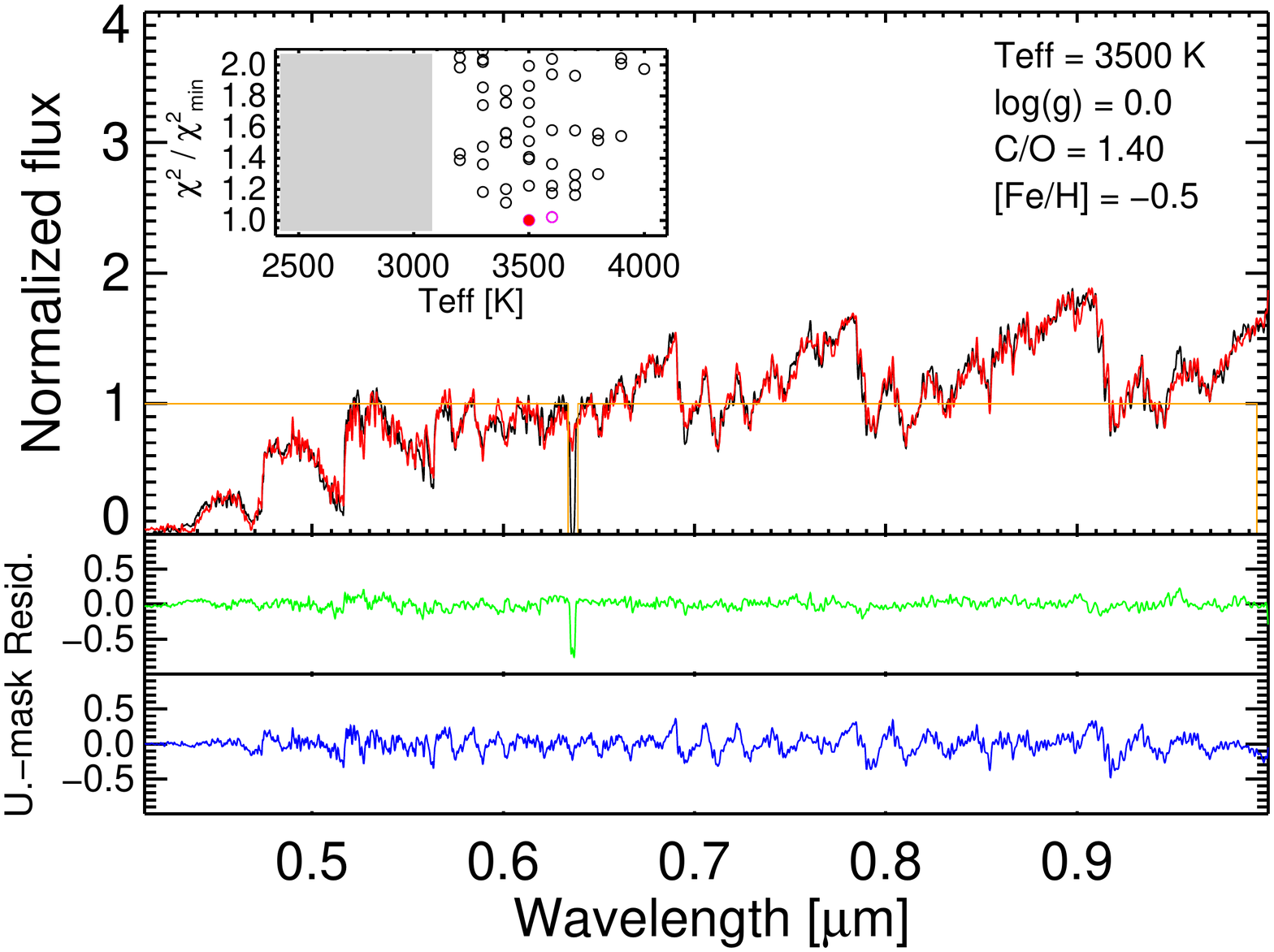}} 
\subfloat[NIR]{\includegraphics[trim=30 10 30 65, clip,width=0.45\hsize]{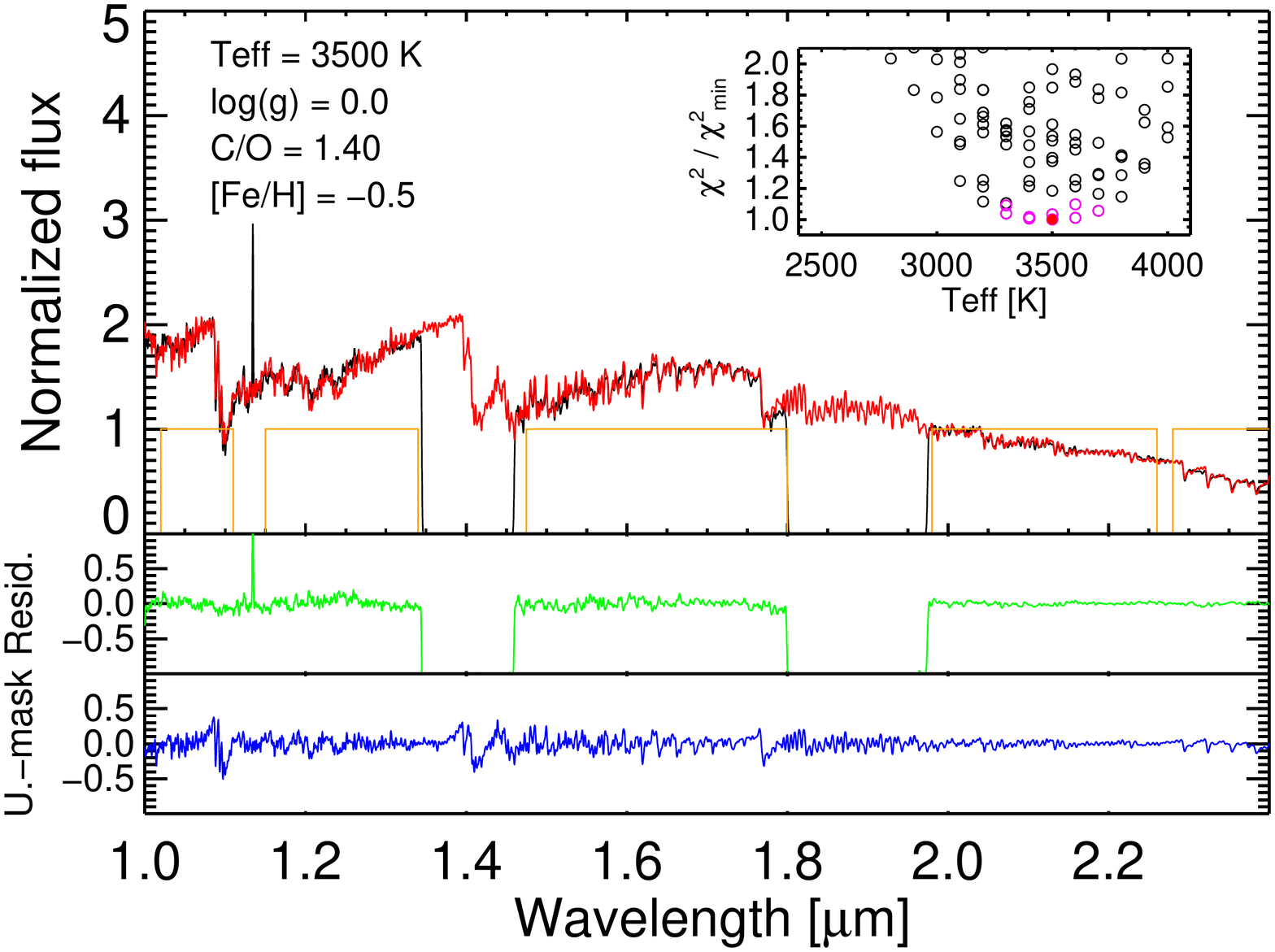}}
\caption{Best-fitting models for 2MASS J00493262-7317523 (Group B). Same legend as for Fig.~\ref{fit_beg_a}.}
\end{figure*}


\begin{figure*}[h] 
\centering
\subfloat[VIS]{\includegraphics[trim=30 10 30 65, clip,width=0.45\hsize]{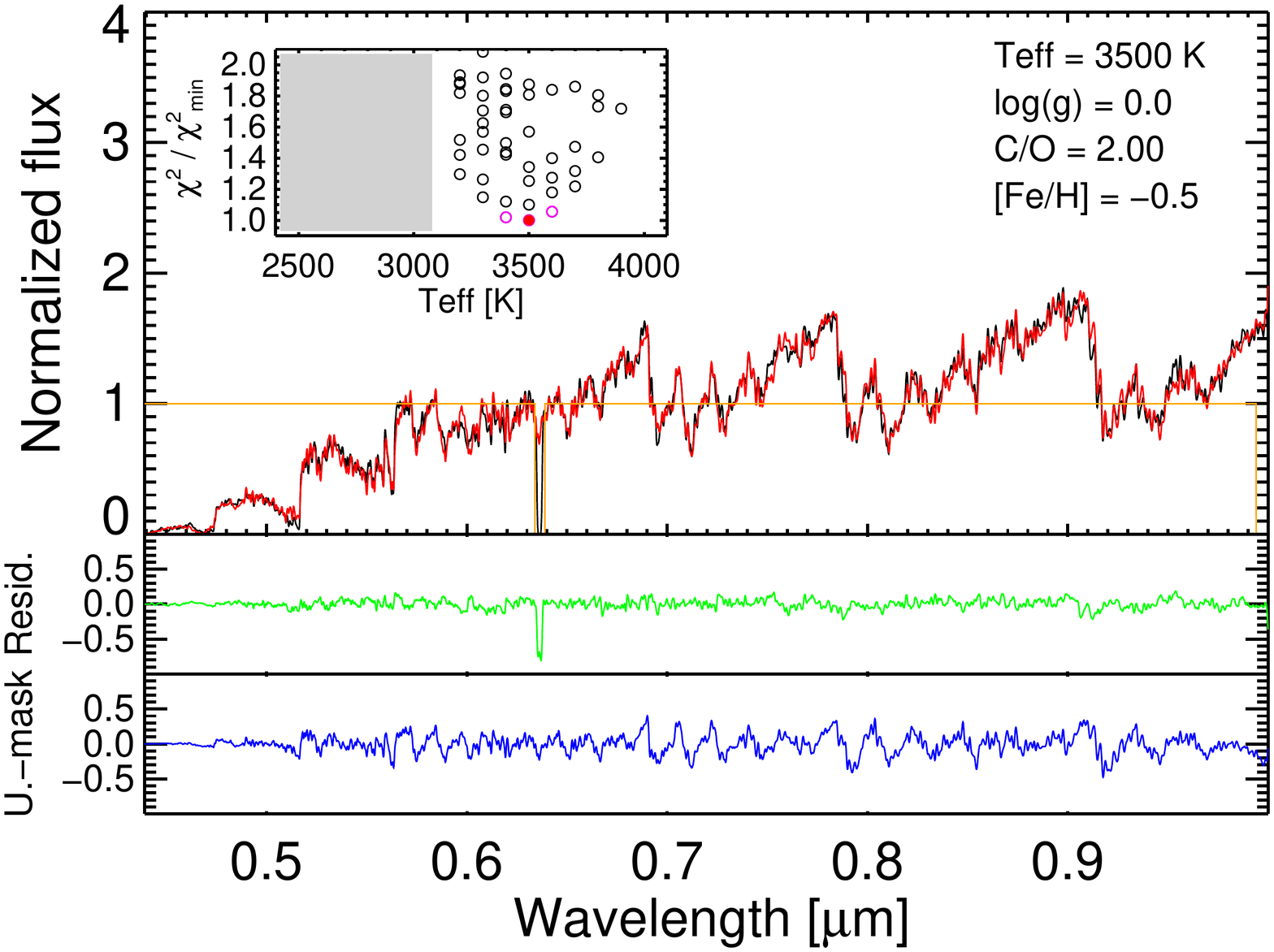}} 
\subfloat[NIR]{\includegraphics[trim=30 10 30 65, clip,width=0.45\hsize]{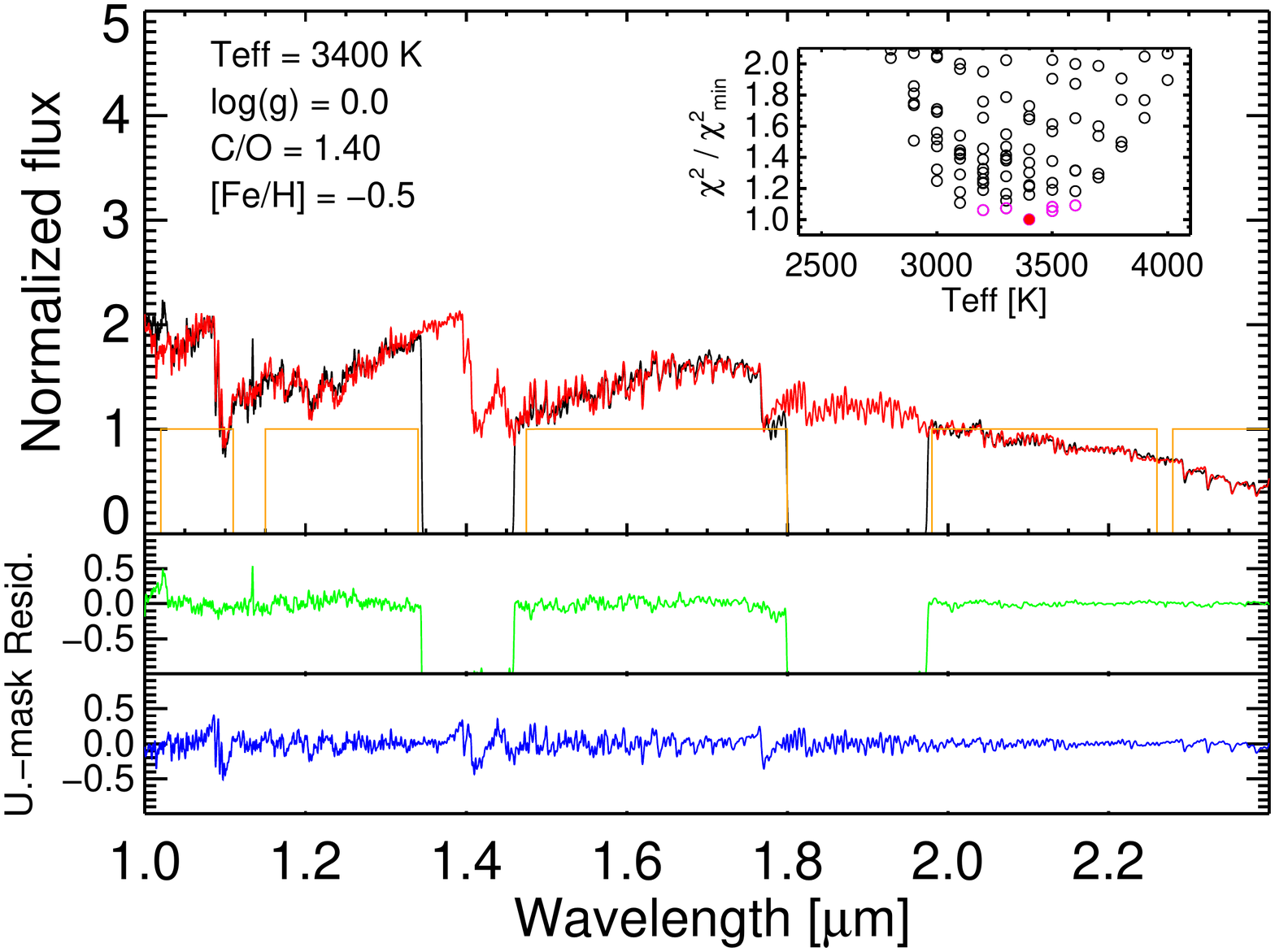}}
\caption{Best-fitting models for 2MASS J00490032-7322238 (Group B). Same legend as for Fig.~\ref{fit_beg_a}.}
\end{figure*}


\begin{figure*}[h] 
\centering
\subfloat[VIS]{\includegraphics[trim=30 10 30 65, clip,width=0.45\hsize]{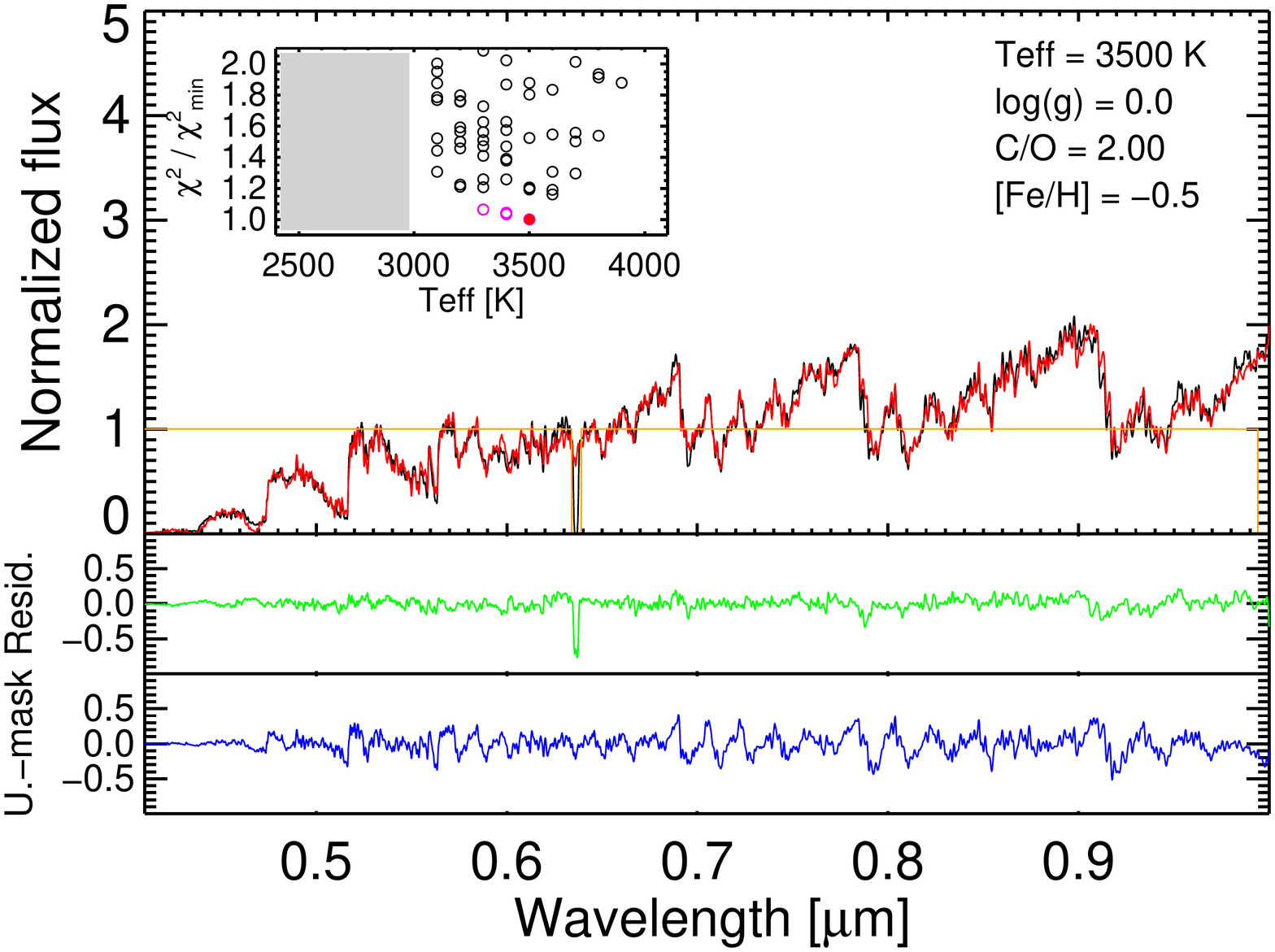}} 
\subfloat[NIR]{\includegraphics[trim=30 10 30 65, clip,width=0.45\hsize]{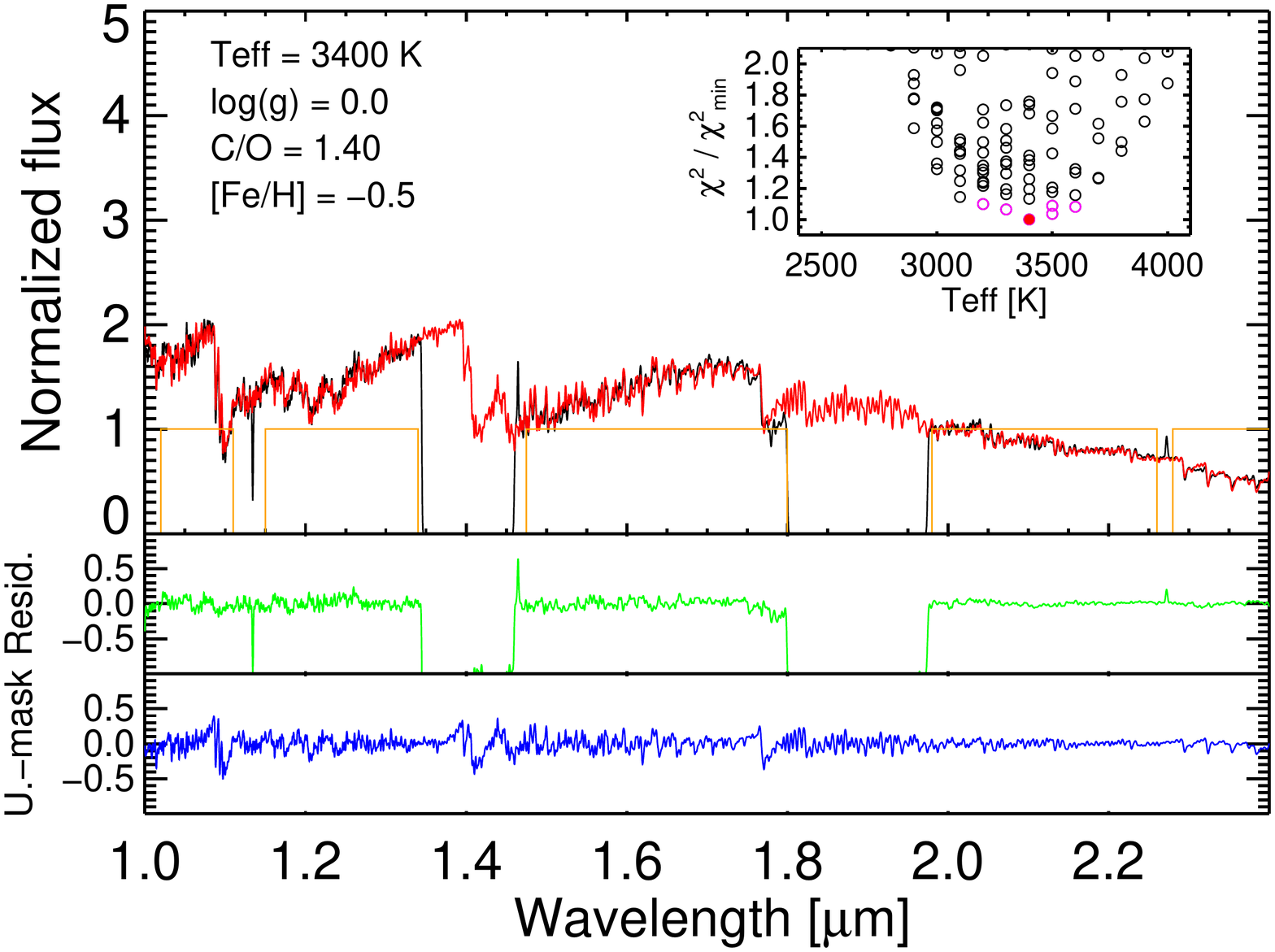}}
\caption{Best-fitting models for 2MASS J00571214-7307045 (Group B). Same legend as for Fig.~\ref{fit_beg_a}.}
\label{fit_end_b}
\end{figure*}


\begin{figure*}[h] 
\centering
\subfloat[VIS]{\includegraphics[trim=30 10 30 65, clip,width=0.45\hsize]{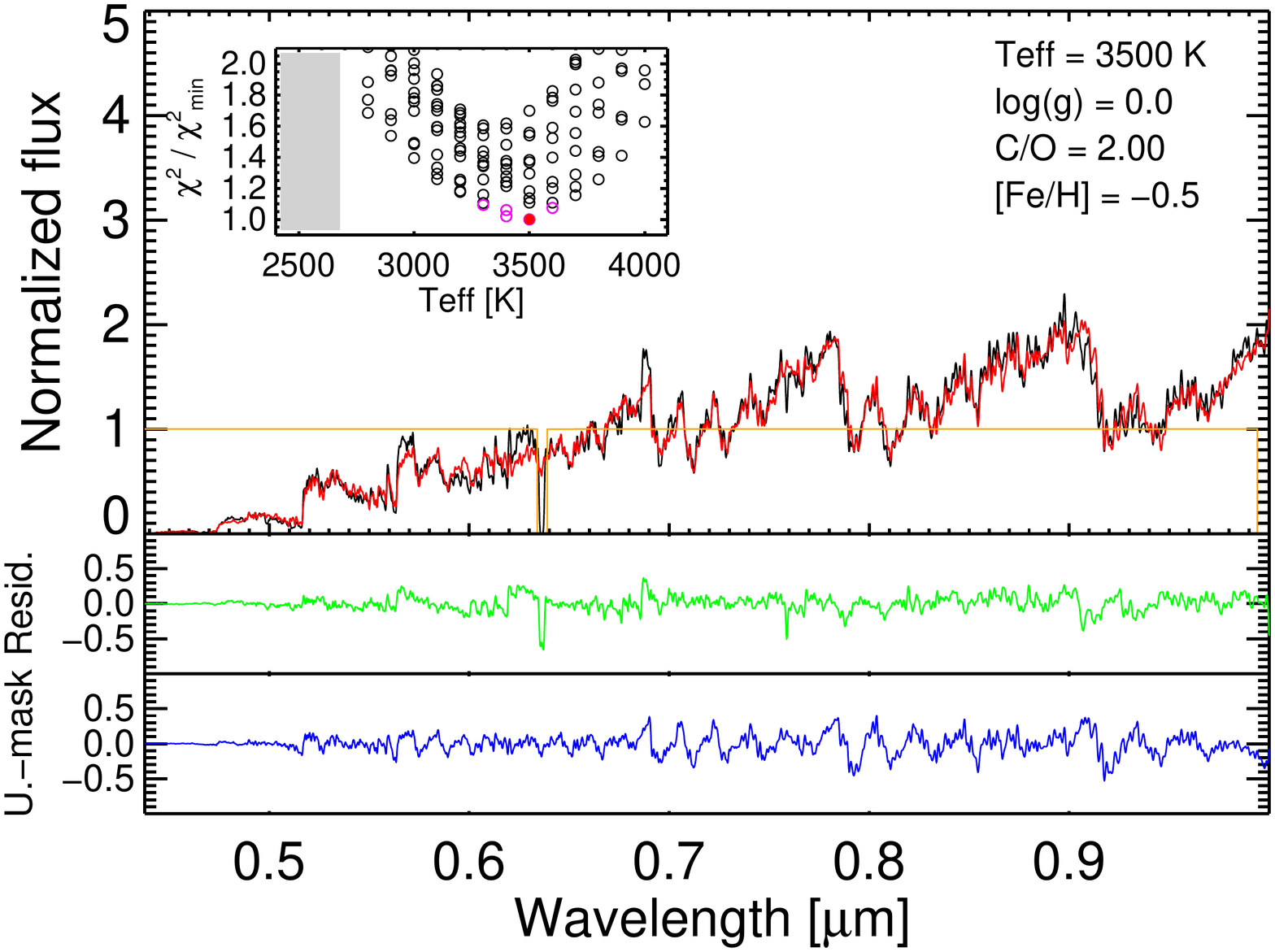}} 
\subfloat[NIR]{\includegraphics[trim=30 10 30 65, clip,width=0.45\hsize]{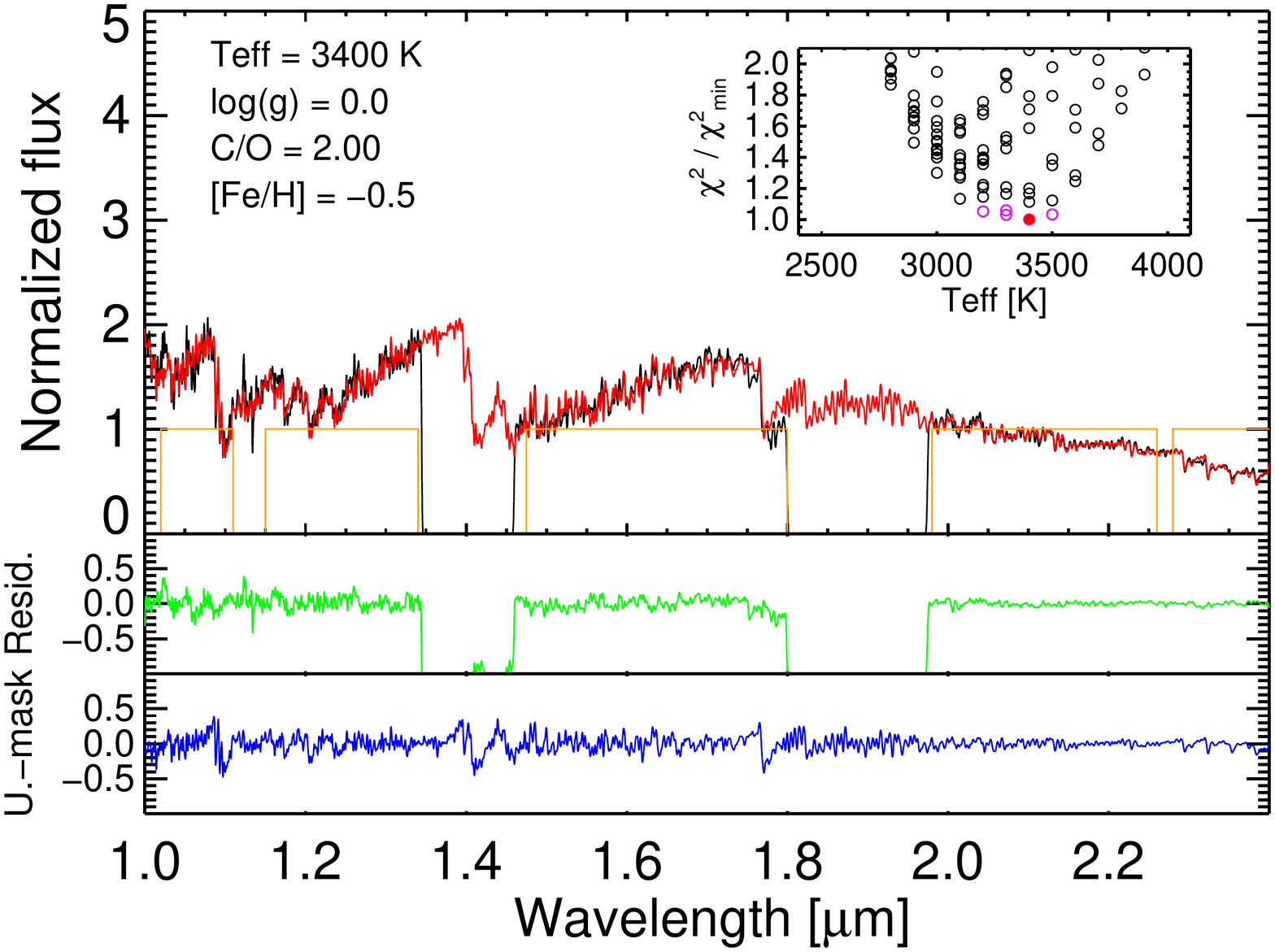}}
\caption{Best-fitting models for 2MASS J00570070-7307505 (Group C). Same legend as for Fig.~\ref{fit_beg_a}.}
\label{fit_beg_c}
\end{figure*}


\begin{figure*}[h] 
\centering
\subfloat[VIS]{\includegraphics[trim=30 10 30 65, clip,width=0.45\hsize]{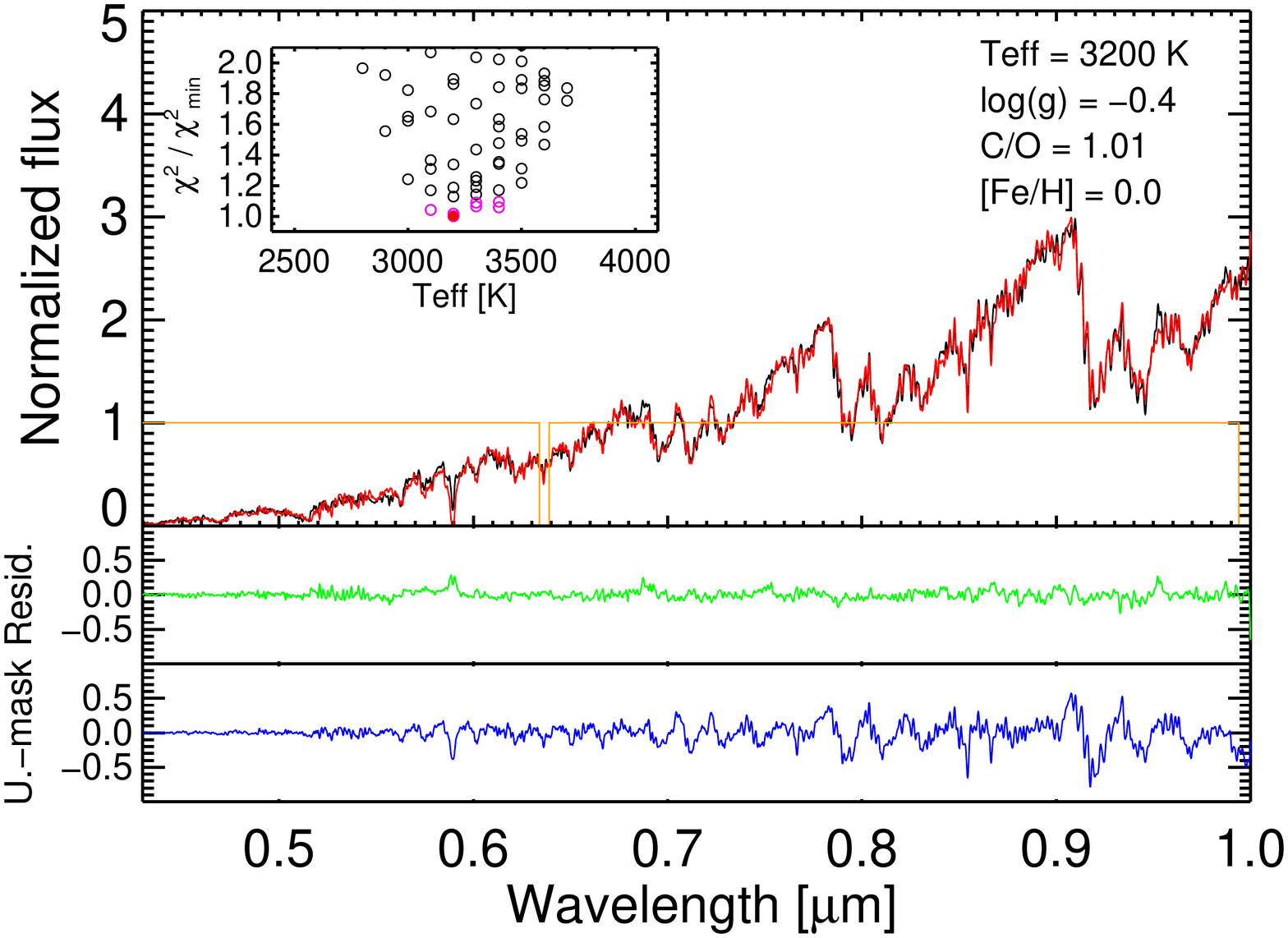}} 
\subfloat[NIR]{\includegraphics[trim=30 10 30 65, clip,width=0.45\hsize]{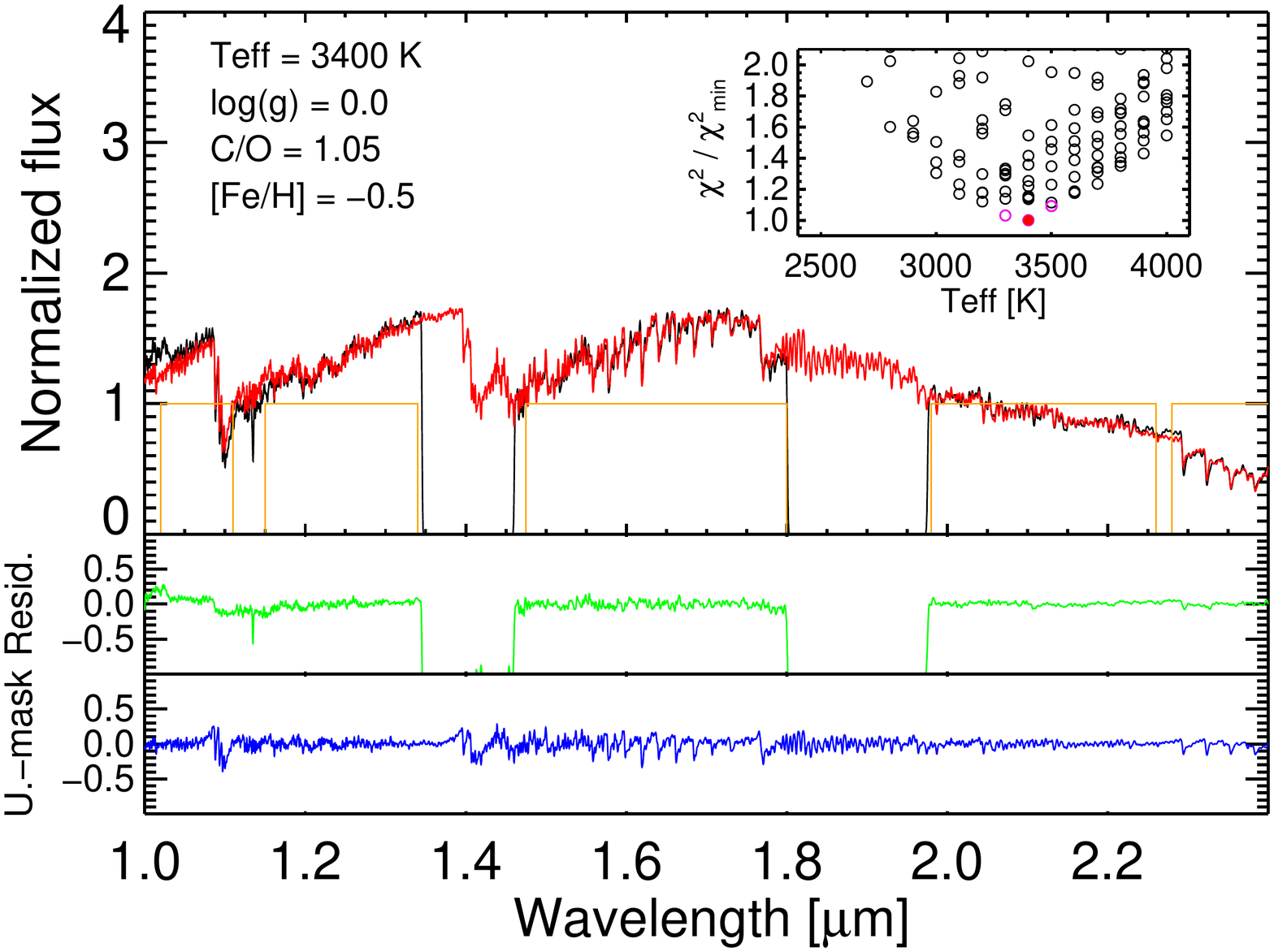}}
\caption{Best-fitting models for {[}W65] c2 (Group C). Same legend as for Fig.~\ref{fit_beg_a}.}
\end{figure*}


\begin{figure*}[h] 
\centering
\subfloat[VIS]{\includegraphics[trim=30 10 30 65, clip,width=0.45\hsize]{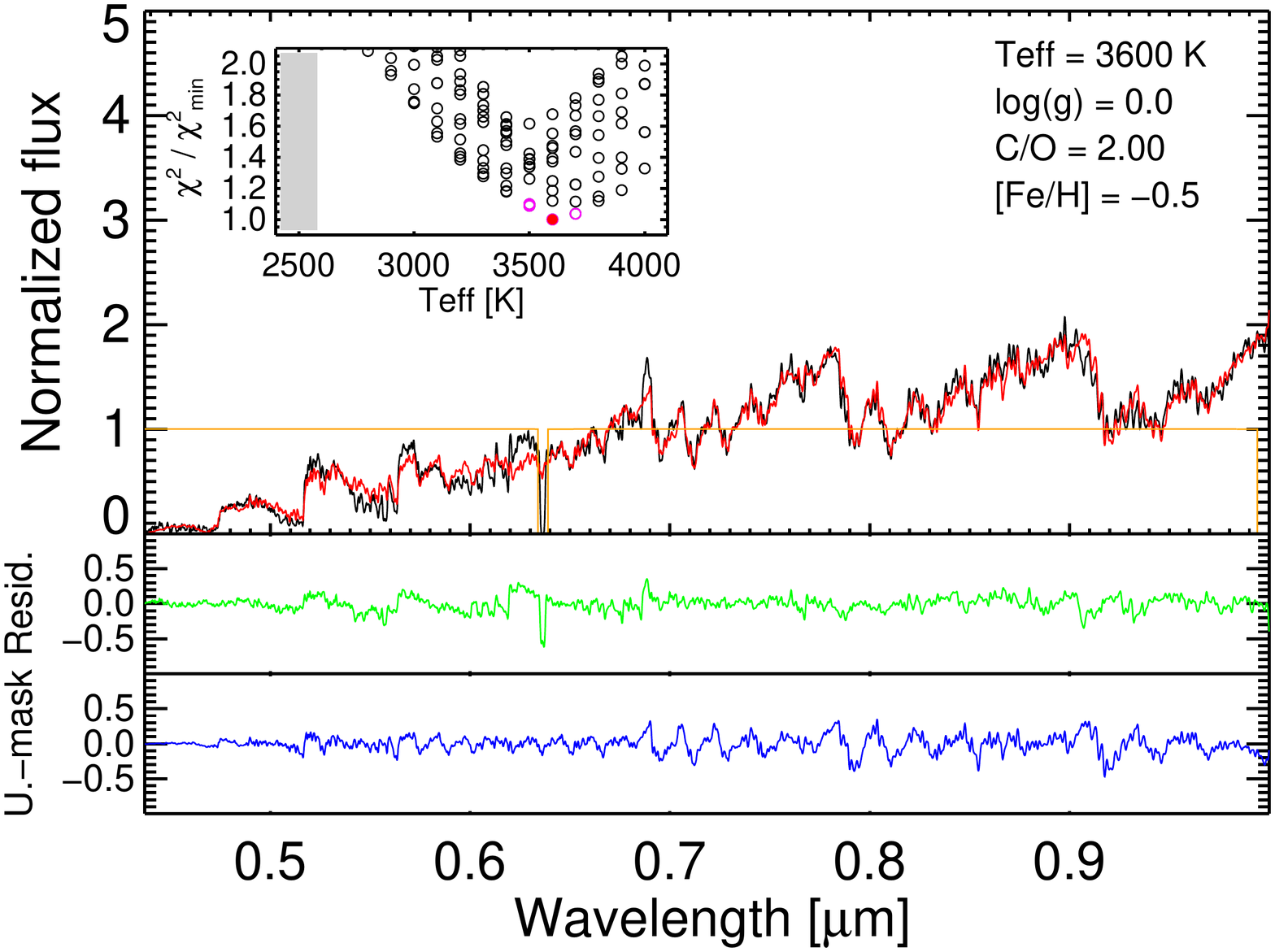}} 
\subfloat[NIR]{\includegraphics[trim=30 10 30 65, clip,width=0.45\hsize]{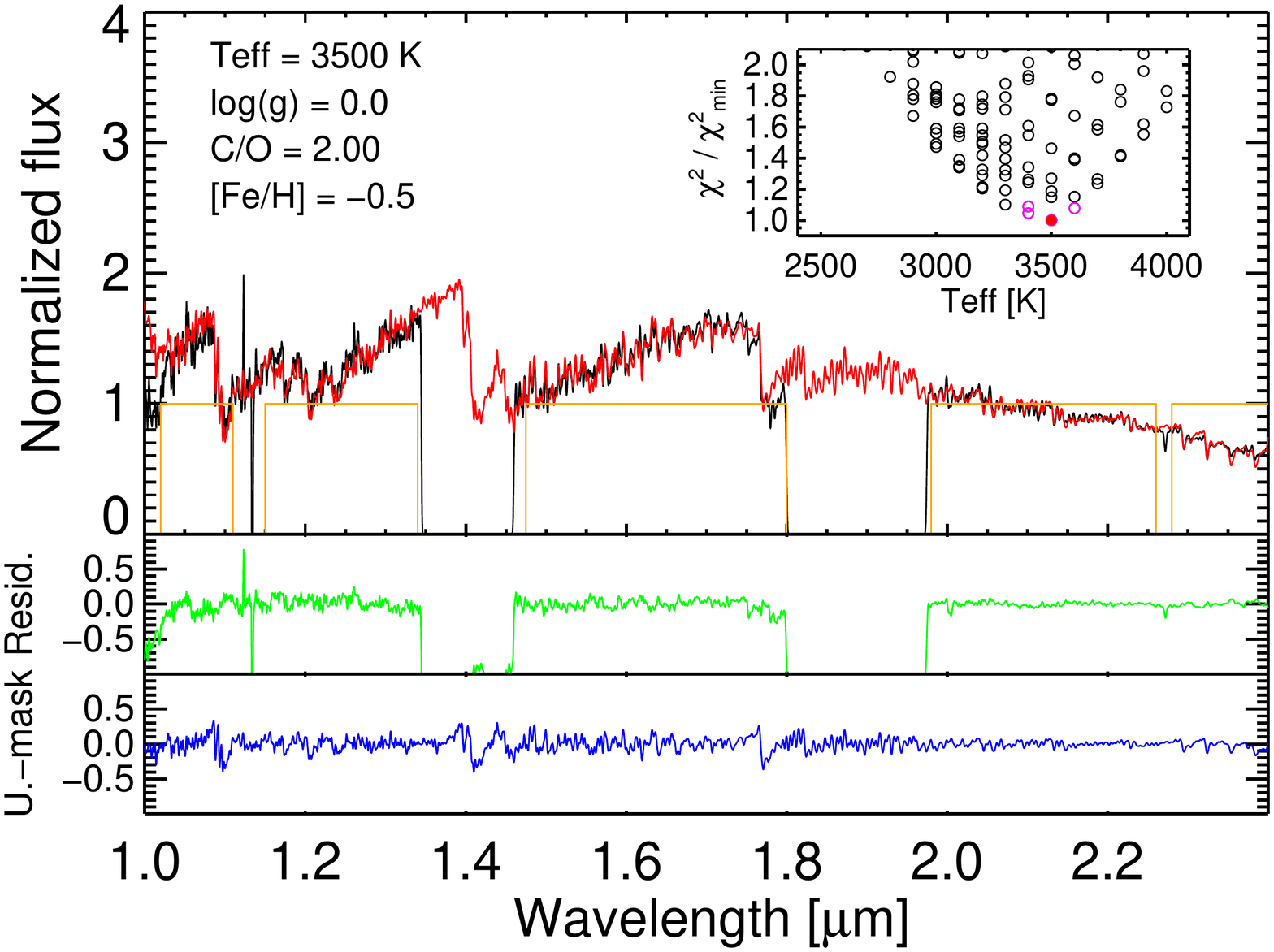}}
\caption{Best-fitting models for 2MASS J00564478-7314347 (Group C). Same legend as for Fig.~\ref{fit_beg_a}.}
\end{figure*}


\begin{figure*}[h] 
\centering
\subfloat[VIS]{\includegraphics[trim=30 10 30 65, clip,width=0.45\hsize]{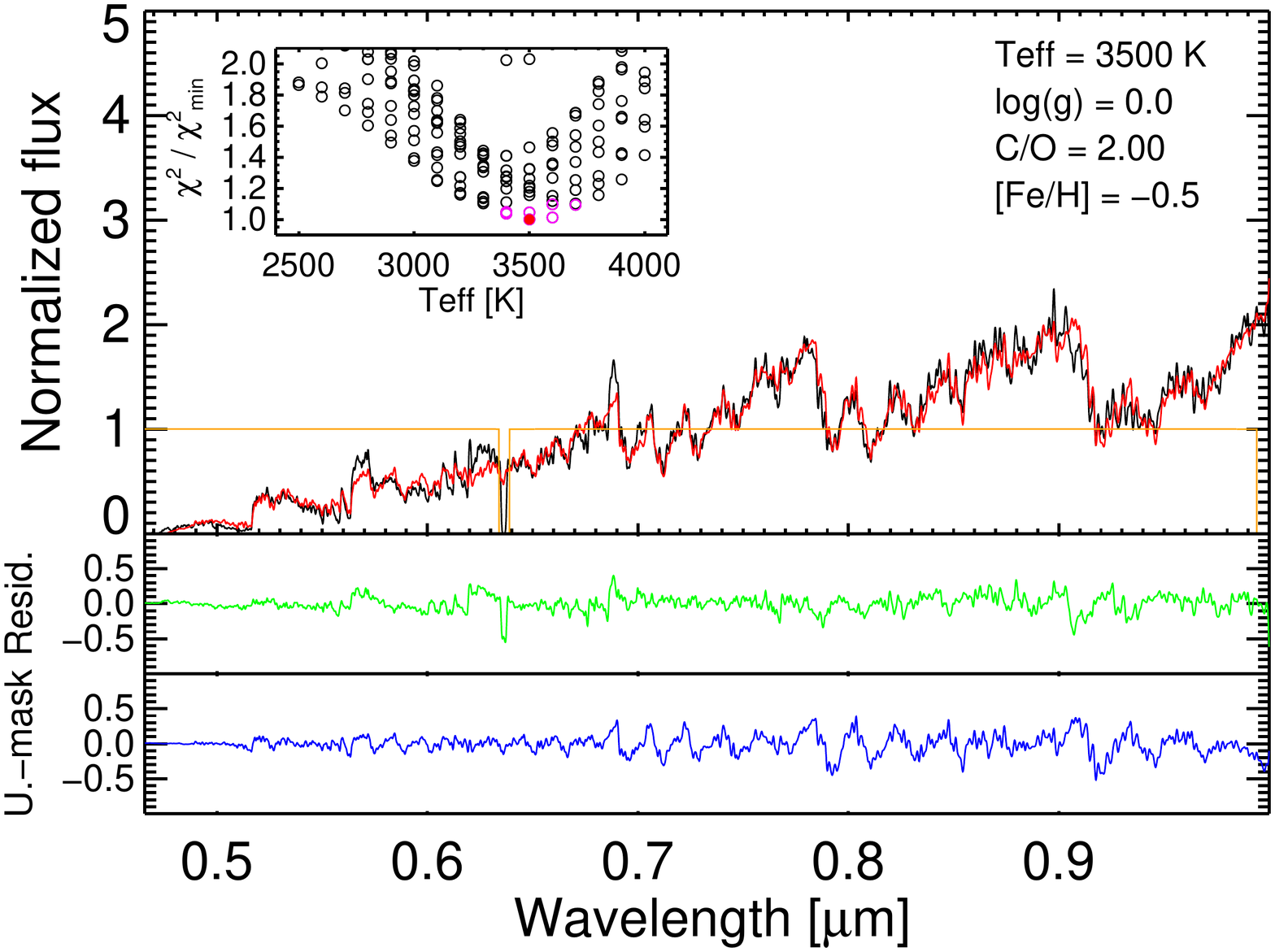}} 
\subfloat[NIR]{\includegraphics[trim=30 10 30 65, clip,width=0.45\hsize]{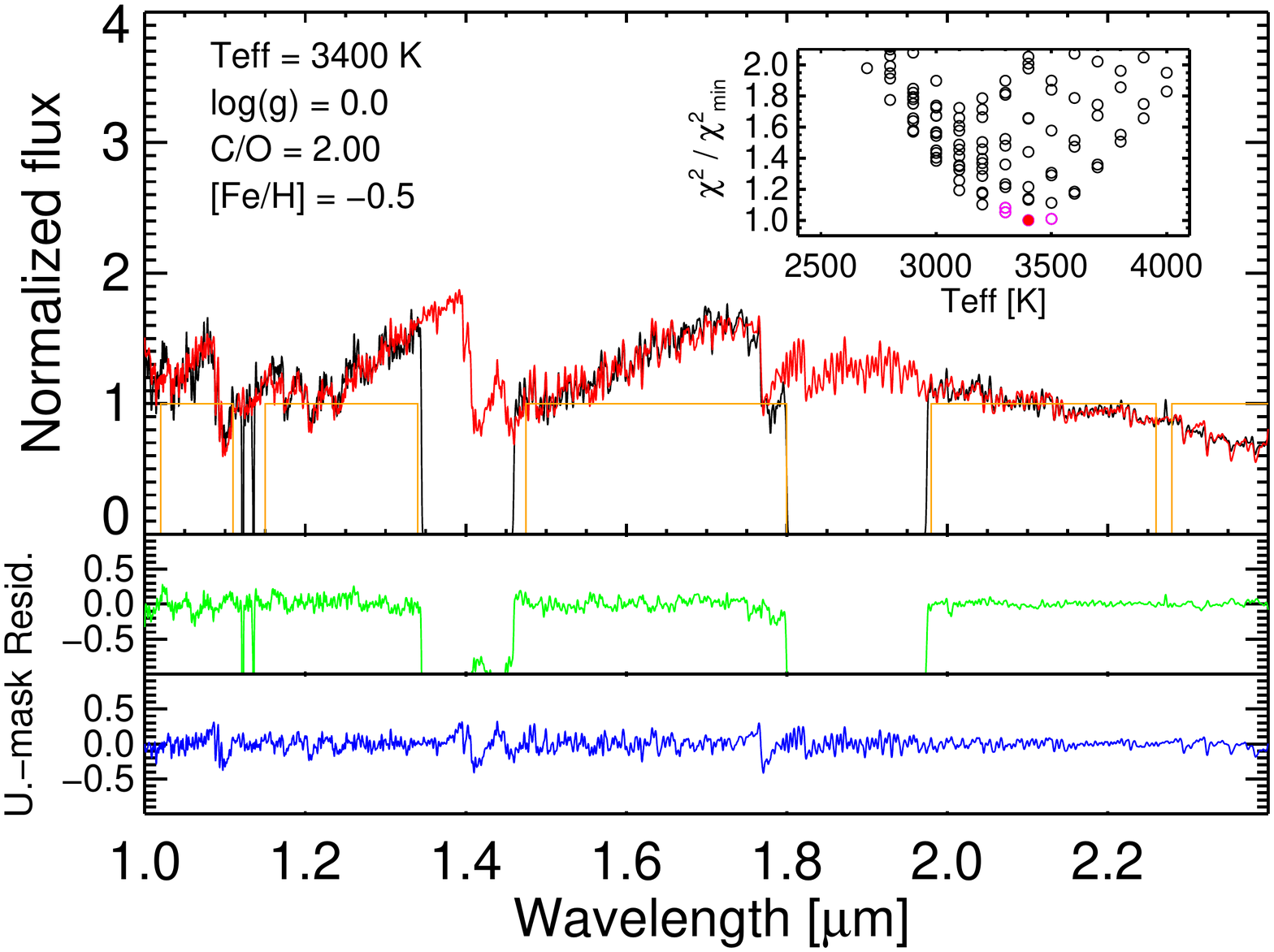}}
\caption{Best-fitting models for 2MASS J00542265-7301057 (Group C). Same legend as for Fig.~\ref{fit_beg_a}.}
\end{figure*}


\begin{figure*}[h] 
\centering
\subfloat[VIS]{\includegraphics[trim=30 10 30 65, clip,width=0.45\hsize]{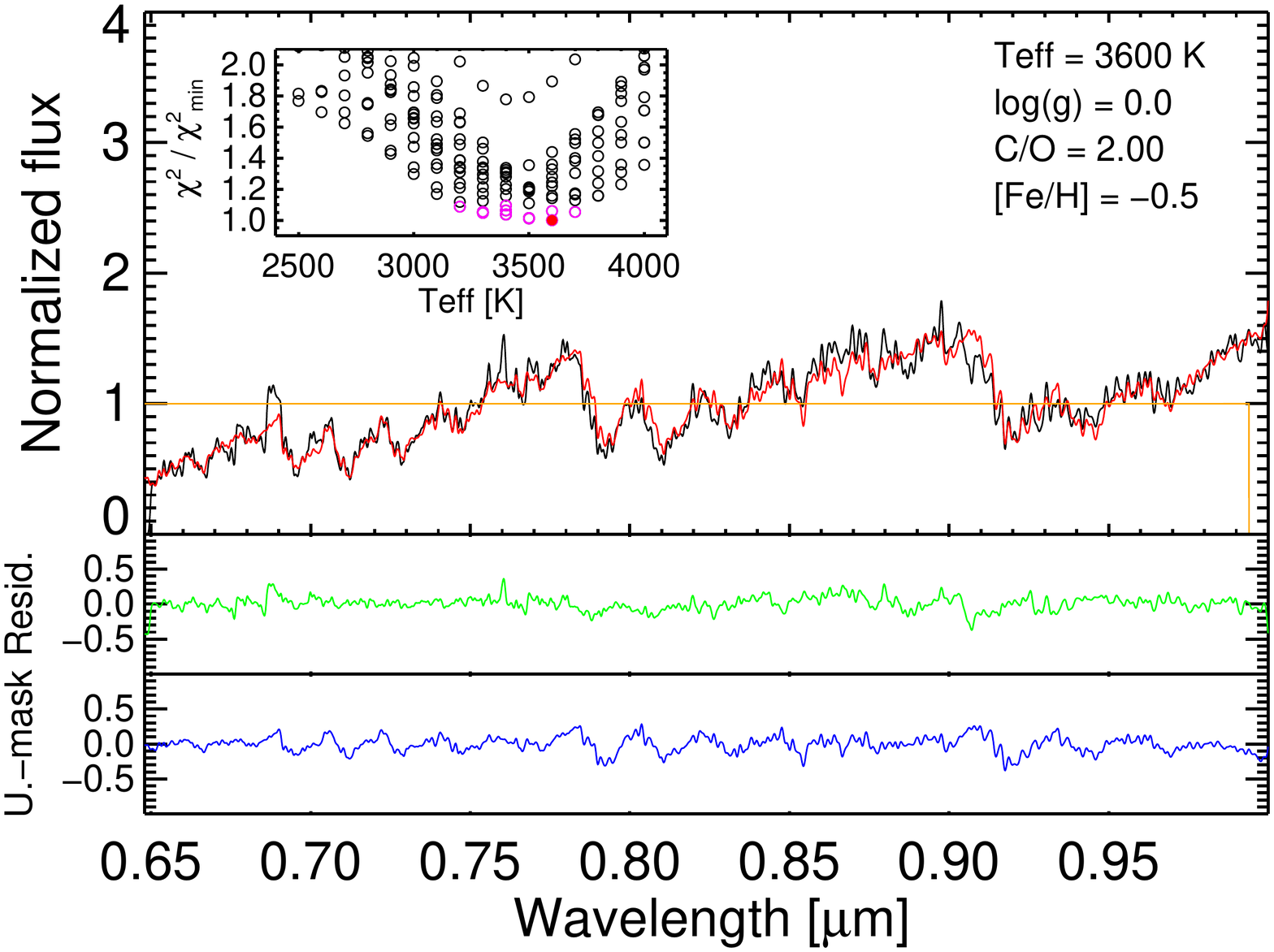}} 
\subfloat[NIR]{\includegraphics[trim=30 10 30 65, clip,width=0.45\hsize]{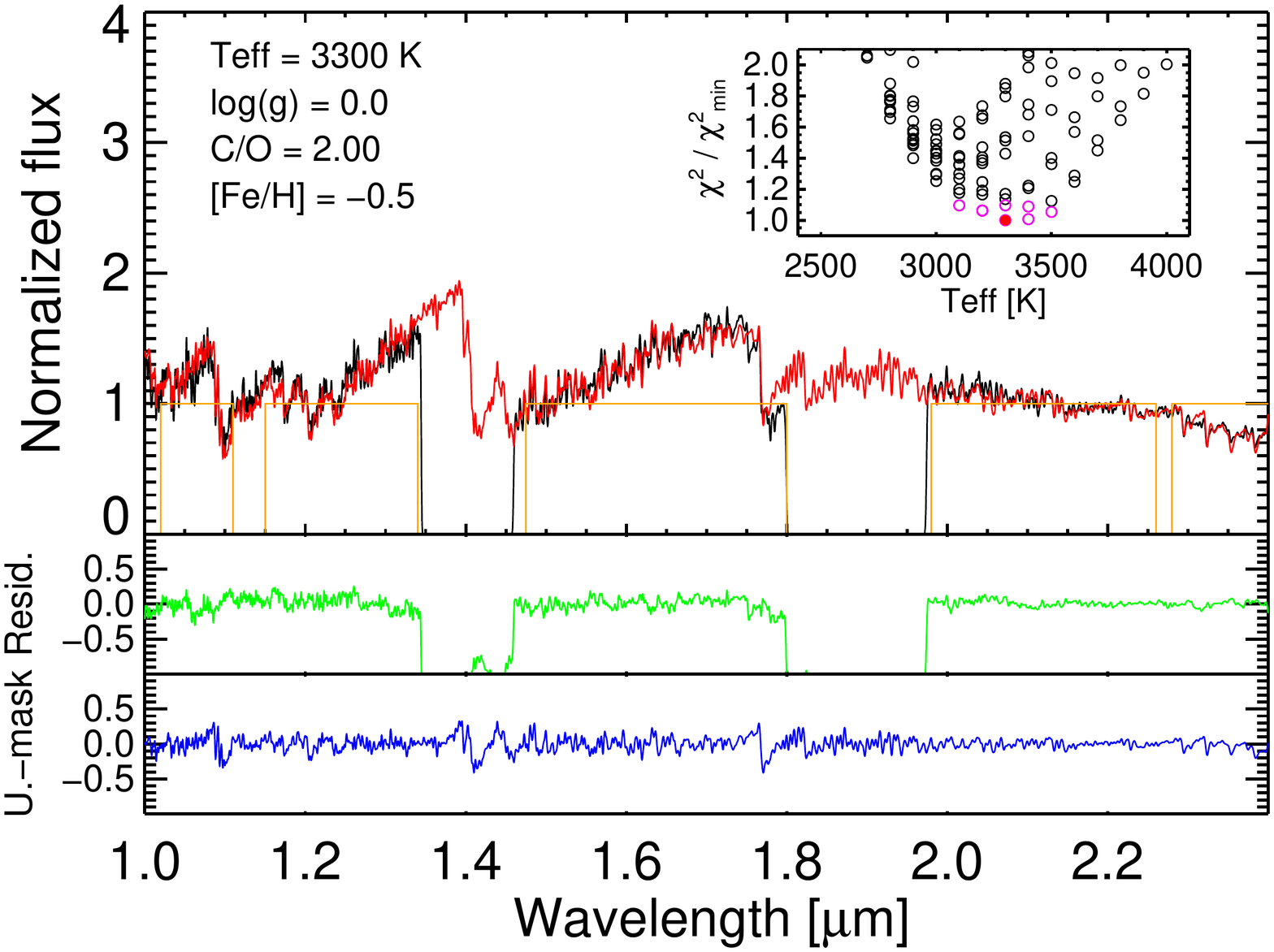}}
\caption{Best-fitting models for Cl* NGC 419 LE 27 (Group C). Same legend as for Fig.~\ref{fit_beg_a}.}
\label{ex_grp_c_temp1}
\end{figure*}


\begin{figure*}[h] 
\centering
\subfloat[VIS]{\includegraphics[trim=30 10 30 65, clip,width=0.45\hsize]{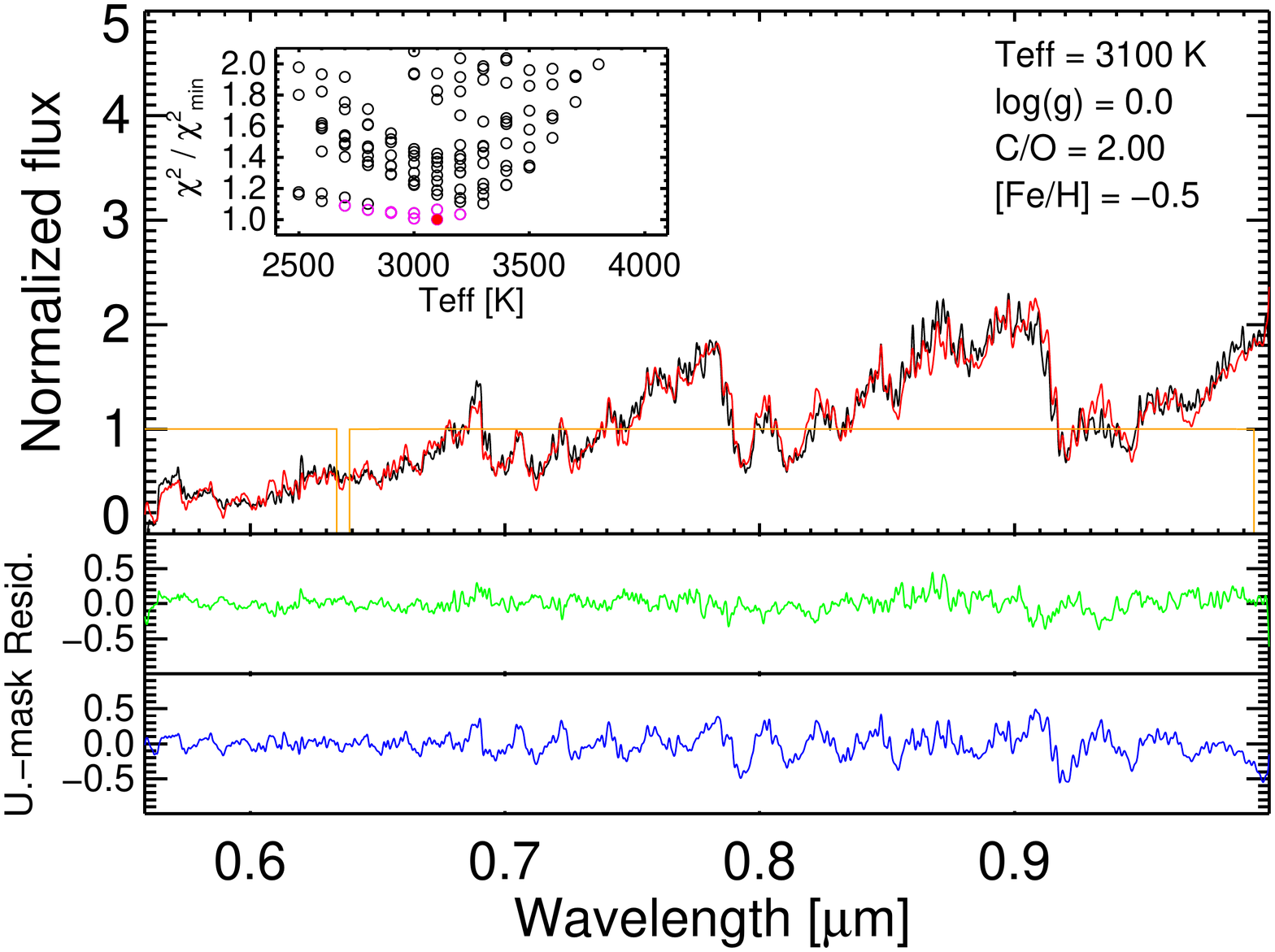}} 
\subfloat[NIR]{\includegraphics[trim=30 10 30 65, clip,width=0.45\hsize]{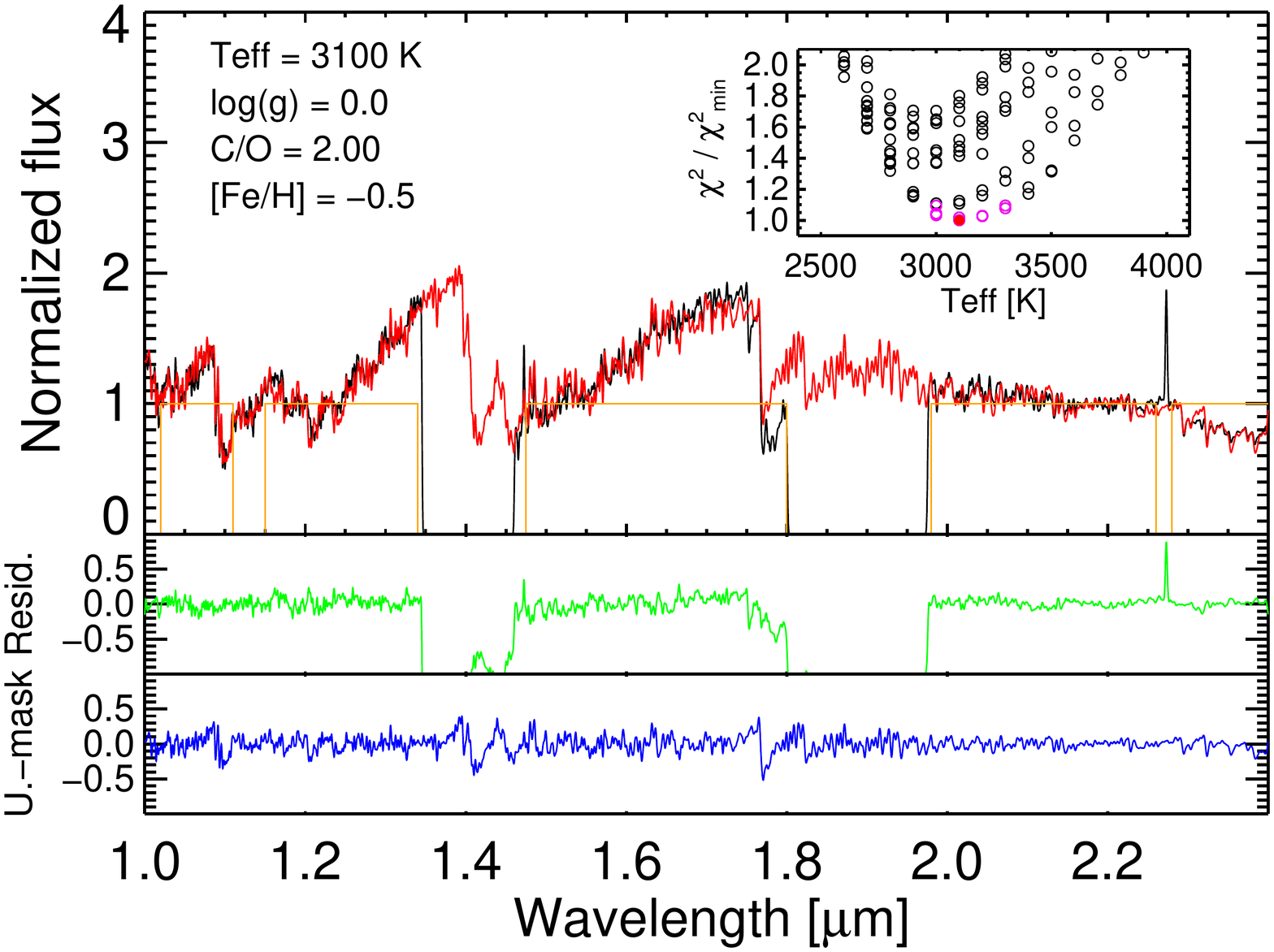}}
\caption{Best-fitting models for IRAS 09484-6242 (Group C). Same legend as for Fig.~\ref{fit_beg_a}.}
\end{figure*}


\begin{figure*}[h] 
\centering
\subfloat[VIS]{\includegraphics[trim=30 10 30 65, clip,width=0.45\hsize]{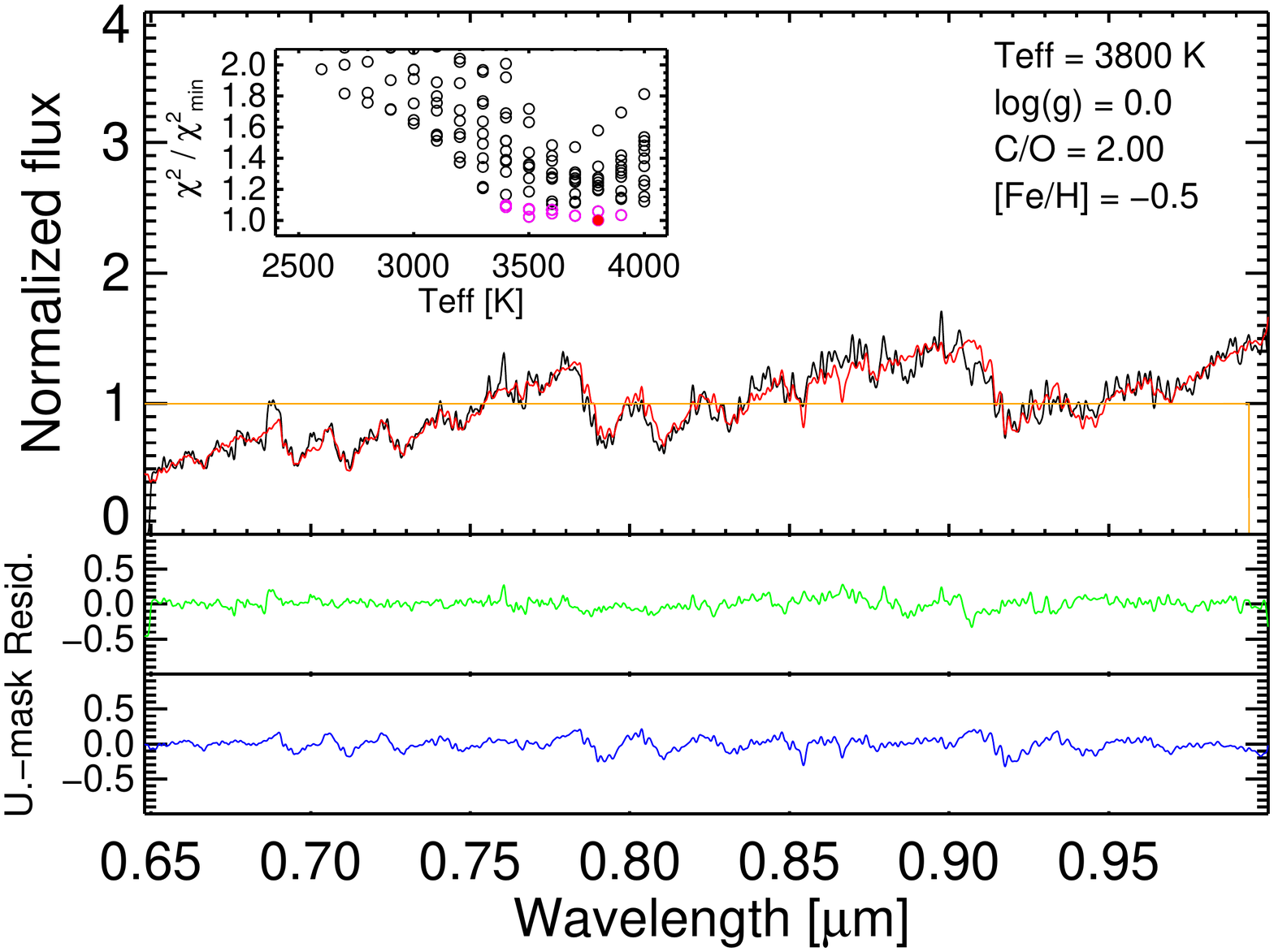}} 
\subfloat[NIR]{\includegraphics[trim=30 10 30 65, clip,width=0.45\hsize]{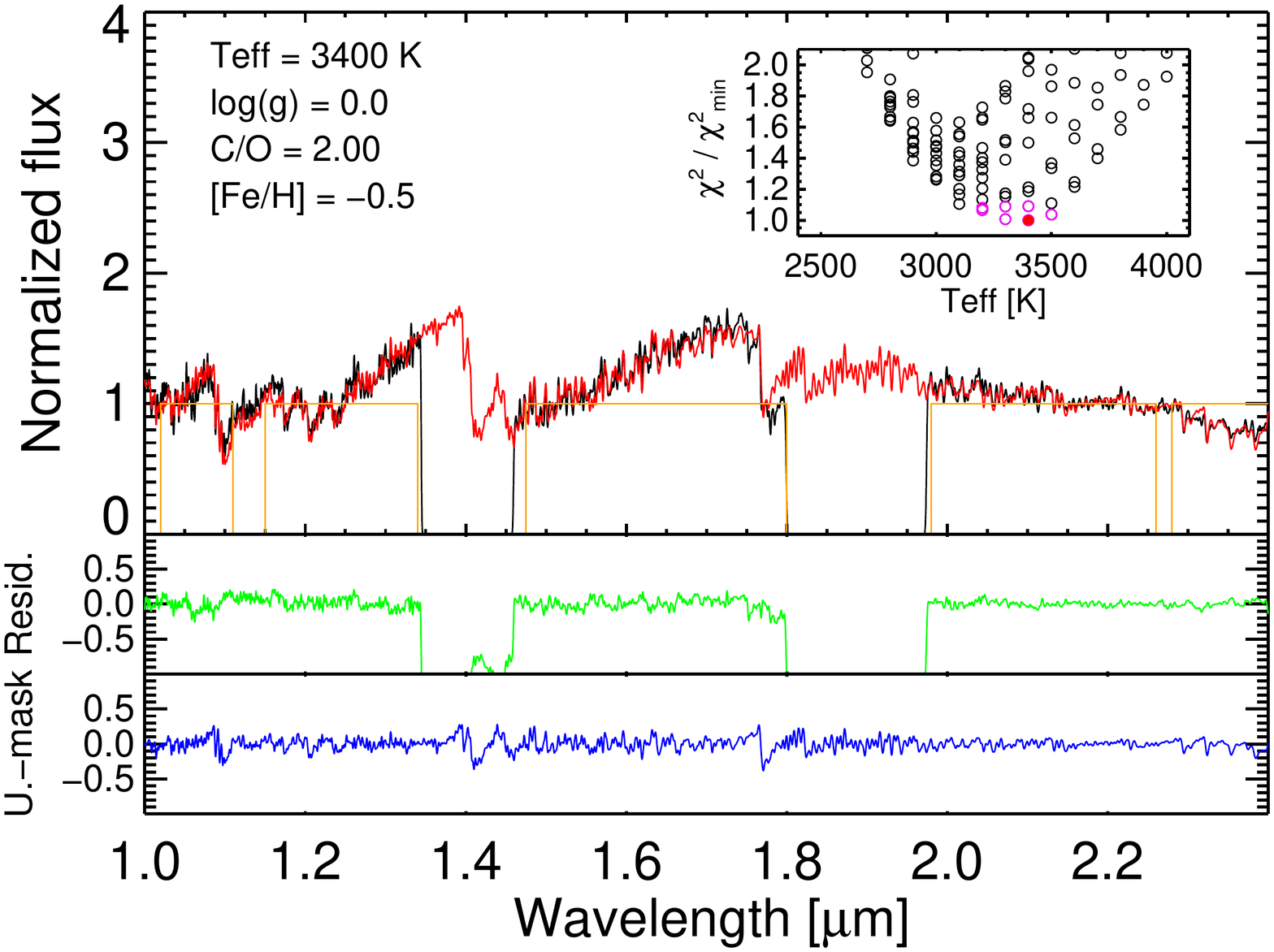}}
\caption{Best-fitting models for Cl* NGC 419 LE 35 (Group C). Same legend as for Fig.~\ref{fit_beg_a}.}
\label{ex_grp_c_temp2}
\end{figure*}


\begin{figure*}[h] 
\centering
\subfloat[VIS]{\includegraphics[trim=30 10 30 65, clip,width=0.45\hsize]{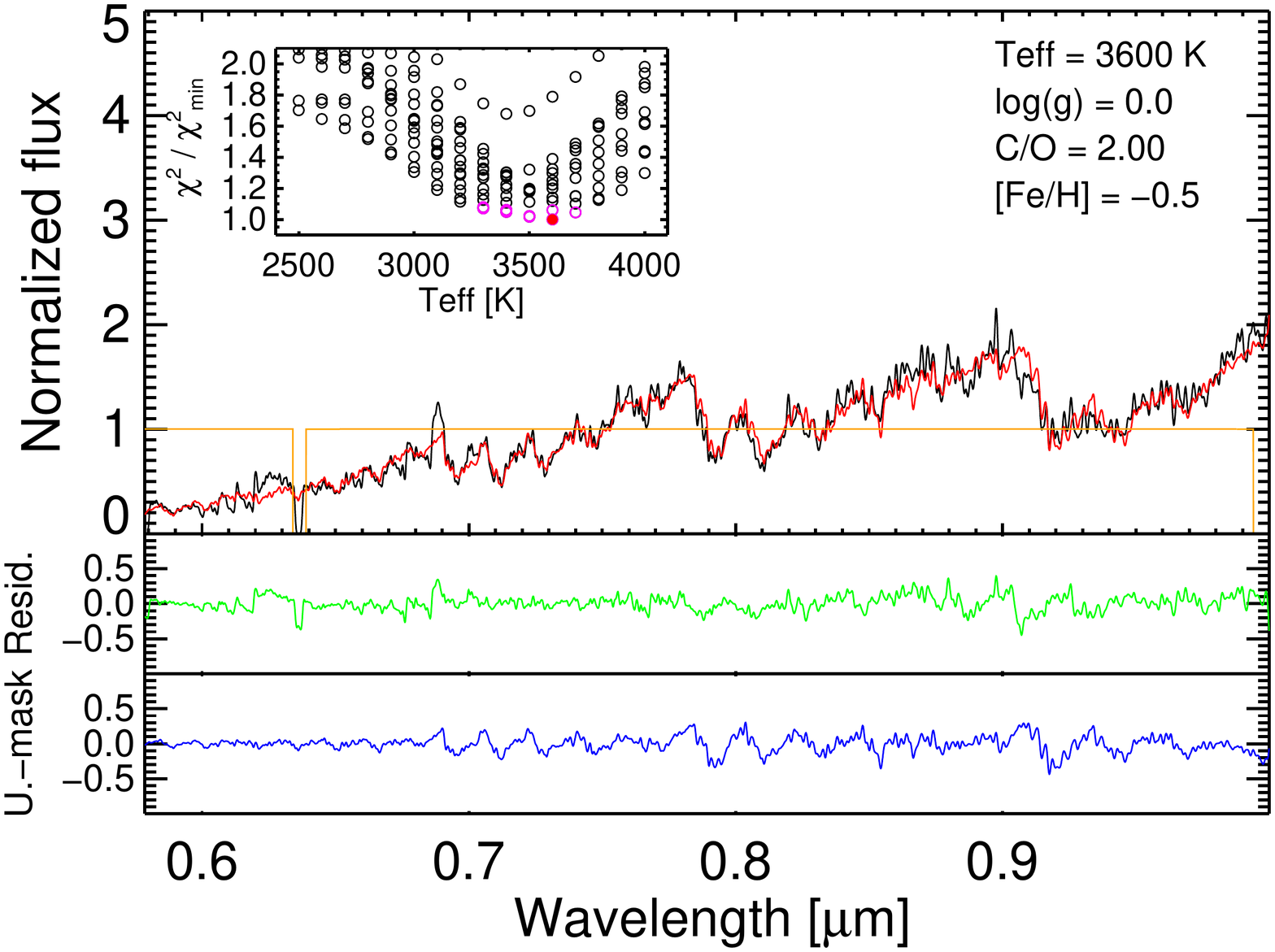}} 
\subfloat[NIR]{\includegraphics[trim=30 10 30 65, clip,width=0.45\hsize]{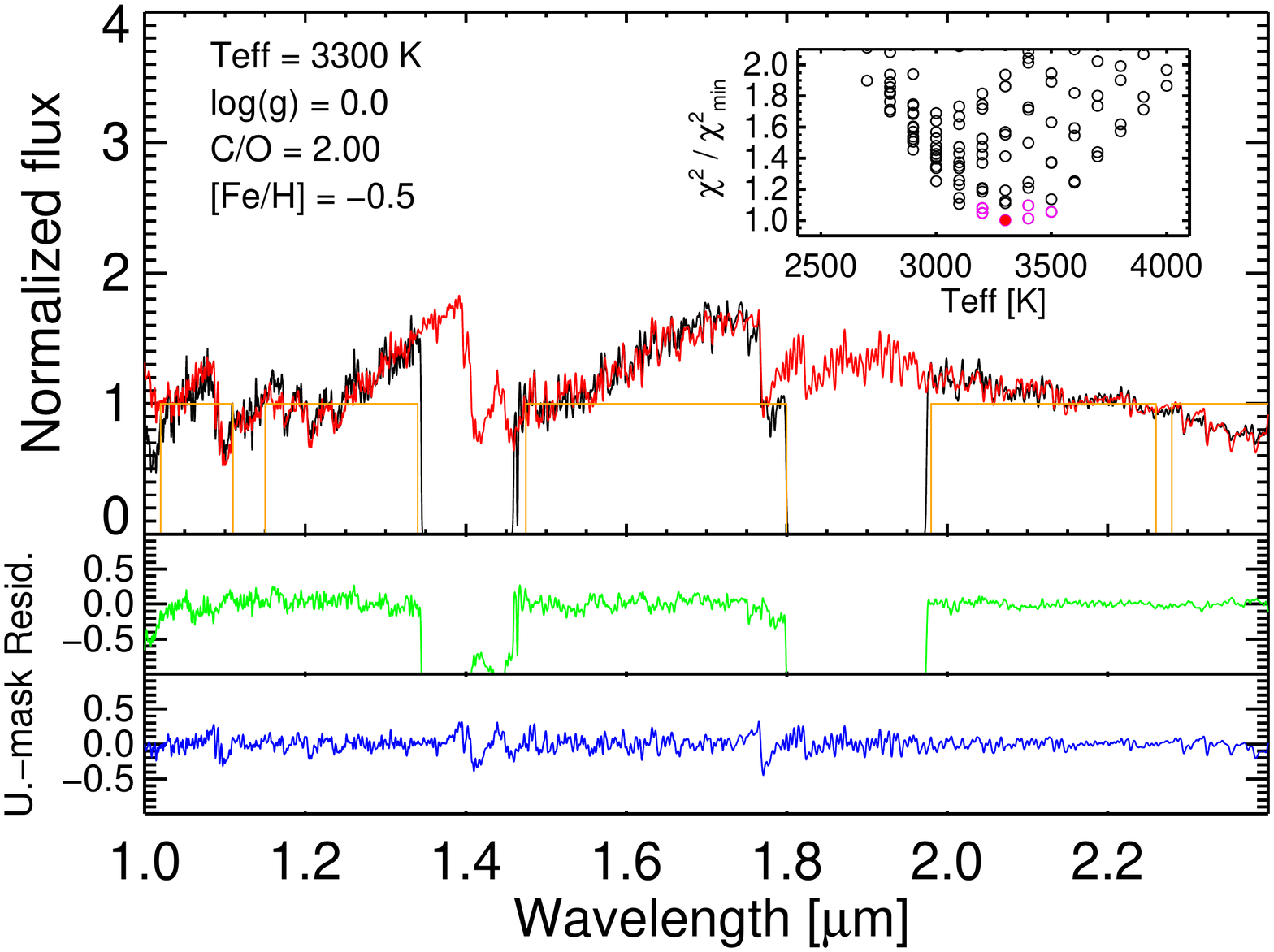}}
\caption{Best-fitting models for 2MASS J00553091-7310186 (Group C). Same legend as for Fig.~\ref{fit_beg_a}.}
\label{ex_grp_c_temp3}
\end{figure*}


\begin{figure*}[h] 
\centering
\subfloat[VIS]{\includegraphics[trim=30 10 30 65, clip,width=0.45\hsize]{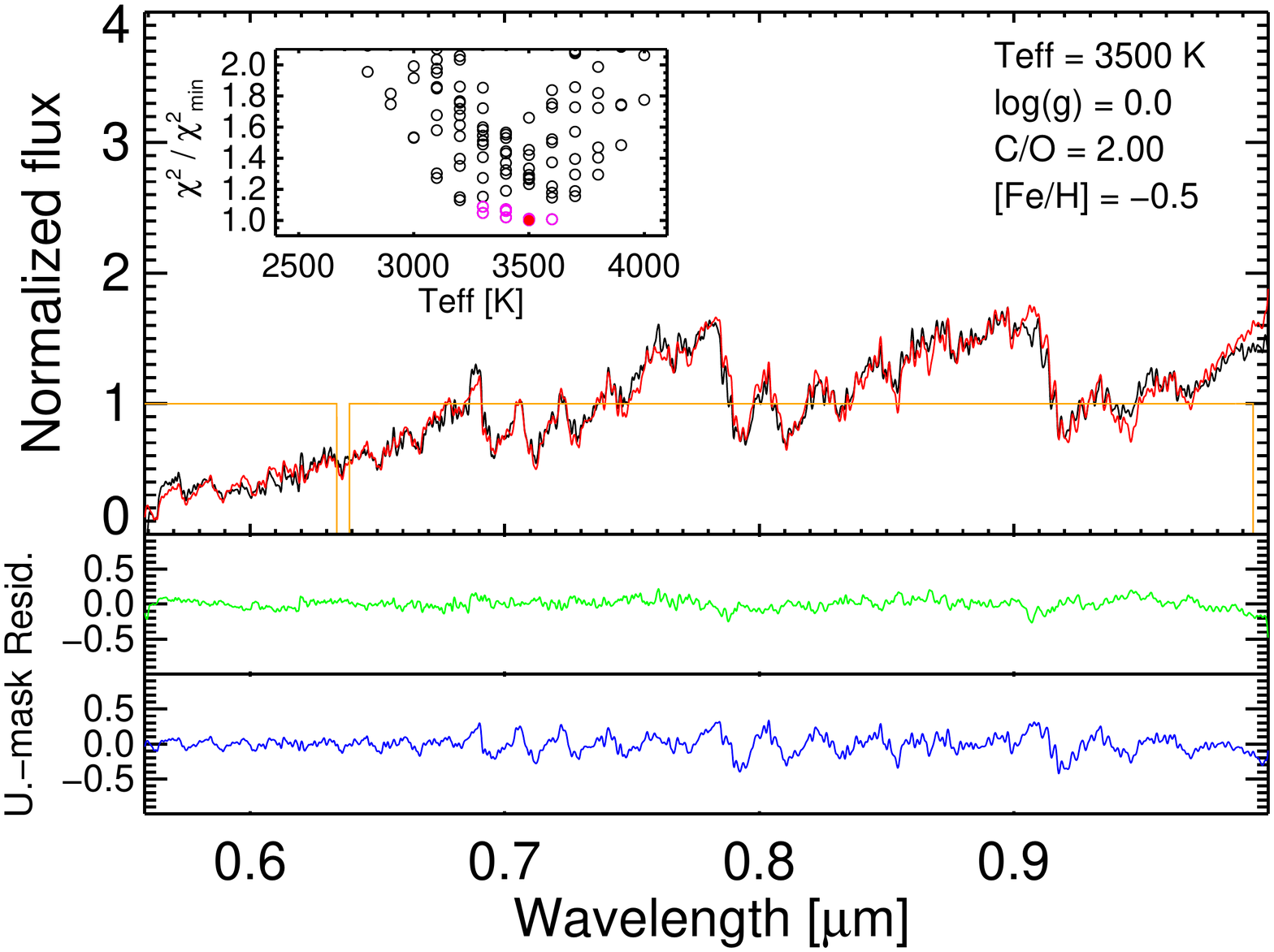}} 
\subfloat[NIR]{\includegraphics[trim=30 10 30 65, clip,width=0.45\hsize]{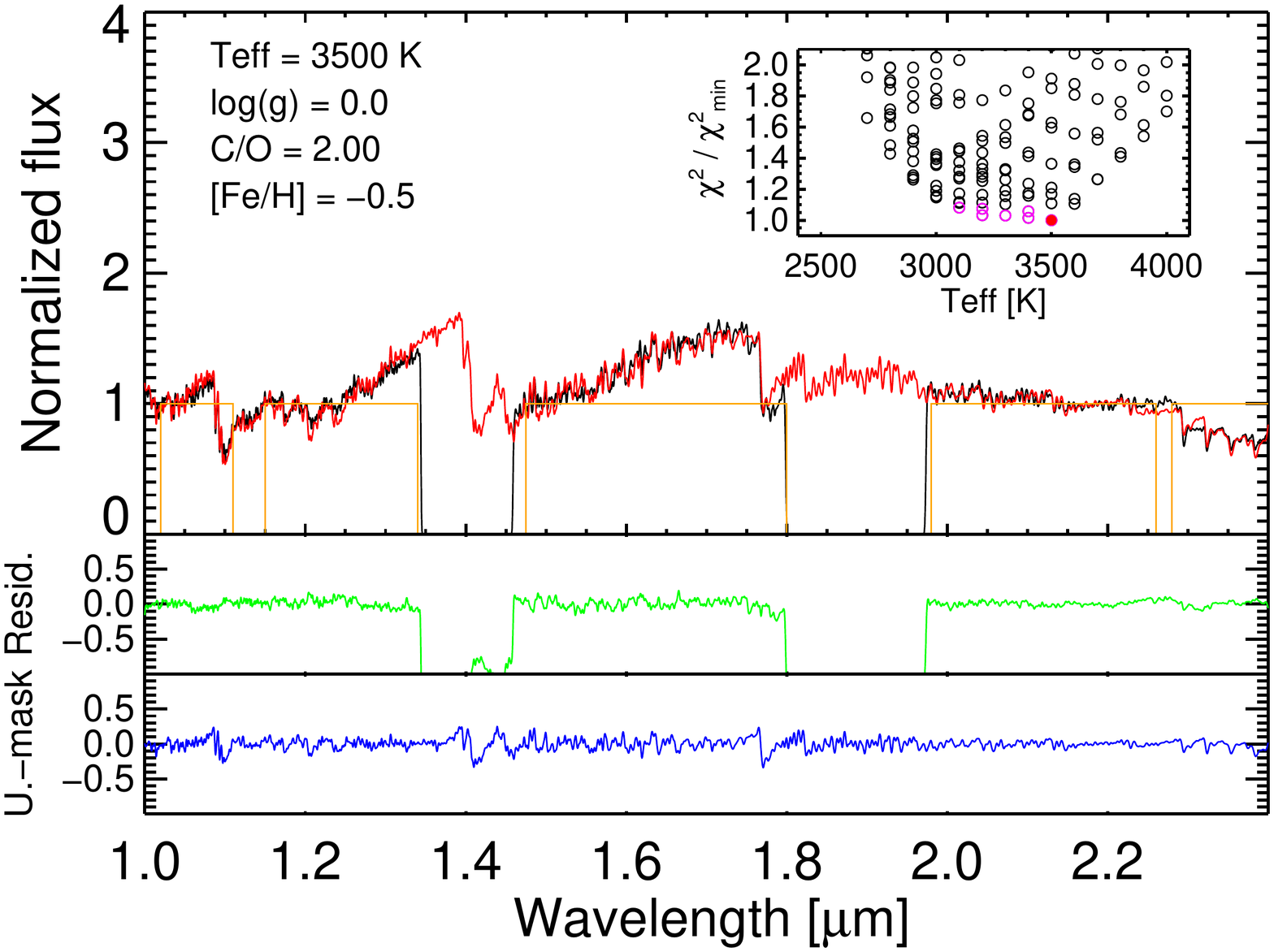}}
\caption{Best-fitting models for SHV 0520427-693637 (Group C). Same legend as for Fig.~\ref{fit_beg_a}.}
\end{figure*}



\begin{figure*}[h] 
\centering
\subfloat[VIS]{\includegraphics[trim=30 10 30 65, clip,width=0.45\hsize]{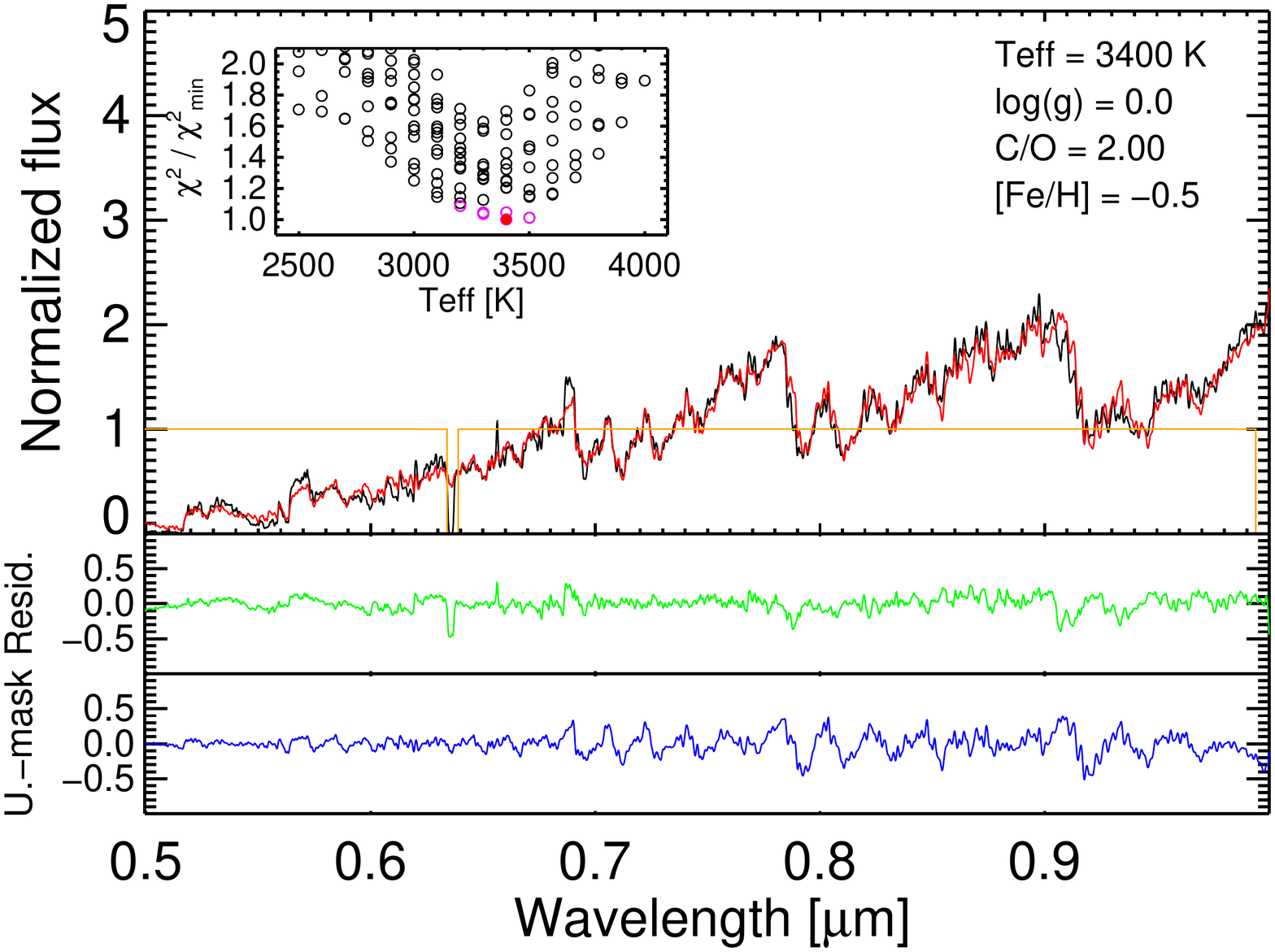}} 
\subfloat[NIR]{\includegraphics[trim=30 10 30 65, clip,width=0.45\hsize]{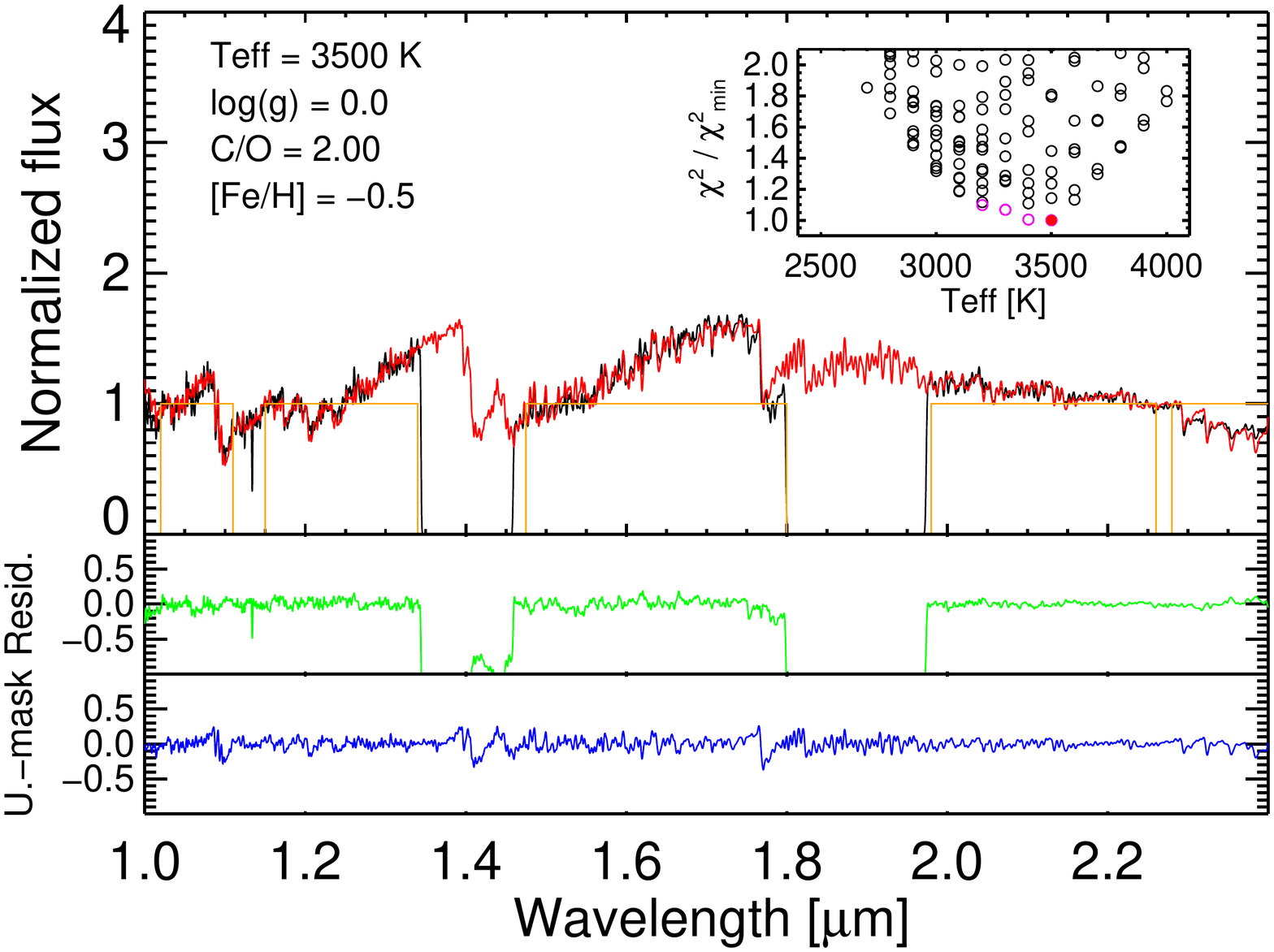}}
\caption{Best-fitting models for SHV 0504353-712622 (Group C). Same legend as for Fig.~\ref{fit_beg_a}.}
\end{figure*}


\begin{figure*}[h] 
\centering
\subfloat[VIS]{\includegraphics[trim=30 10 30 65, clip,width=0.45\hsize]{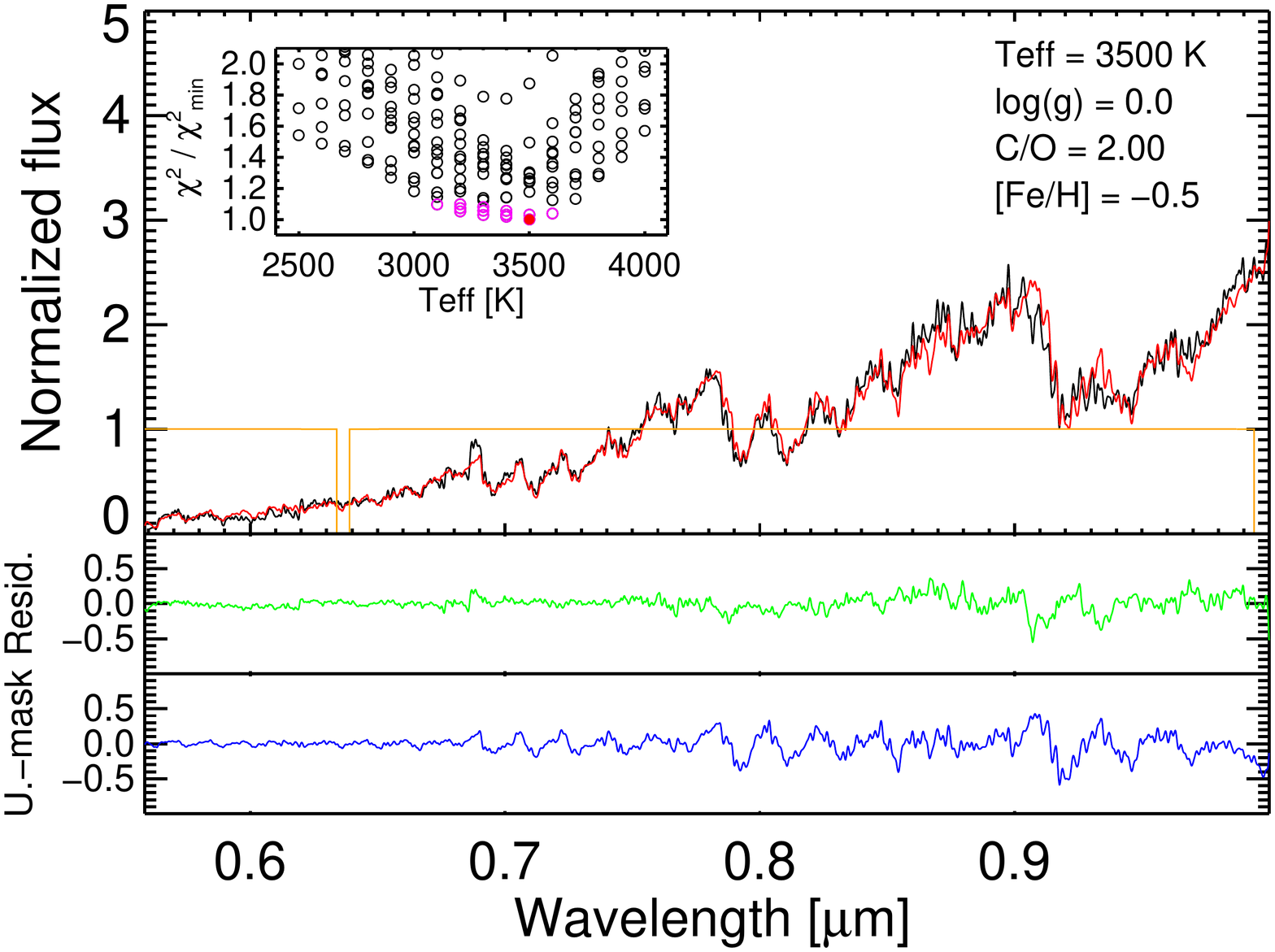}} 
\subfloat[NIR]{\includegraphics[trim=30 10 30 65, clip,width=0.45\hsize]{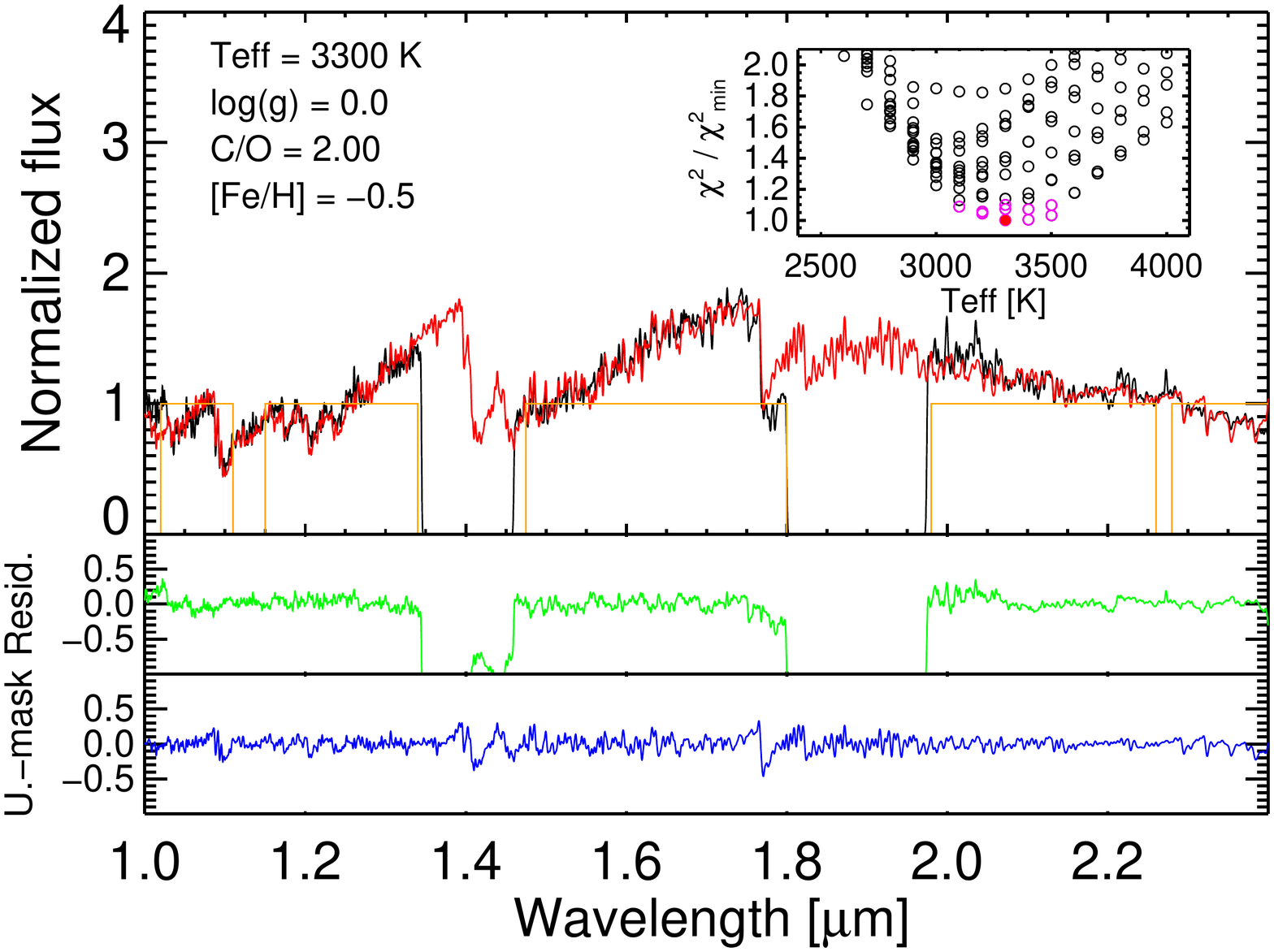}}
\caption{Best-fitting models for {[}ABC89] Pup 42 (Group C). Same legend as for Fig.~\ref{fit_beg_a}.}
\end{figure*}


\begin{figure*}[h] 
\centering
\subfloat[VIS]{\includegraphics[trim=30 10 30 65, clip,width=0.45\hsize]{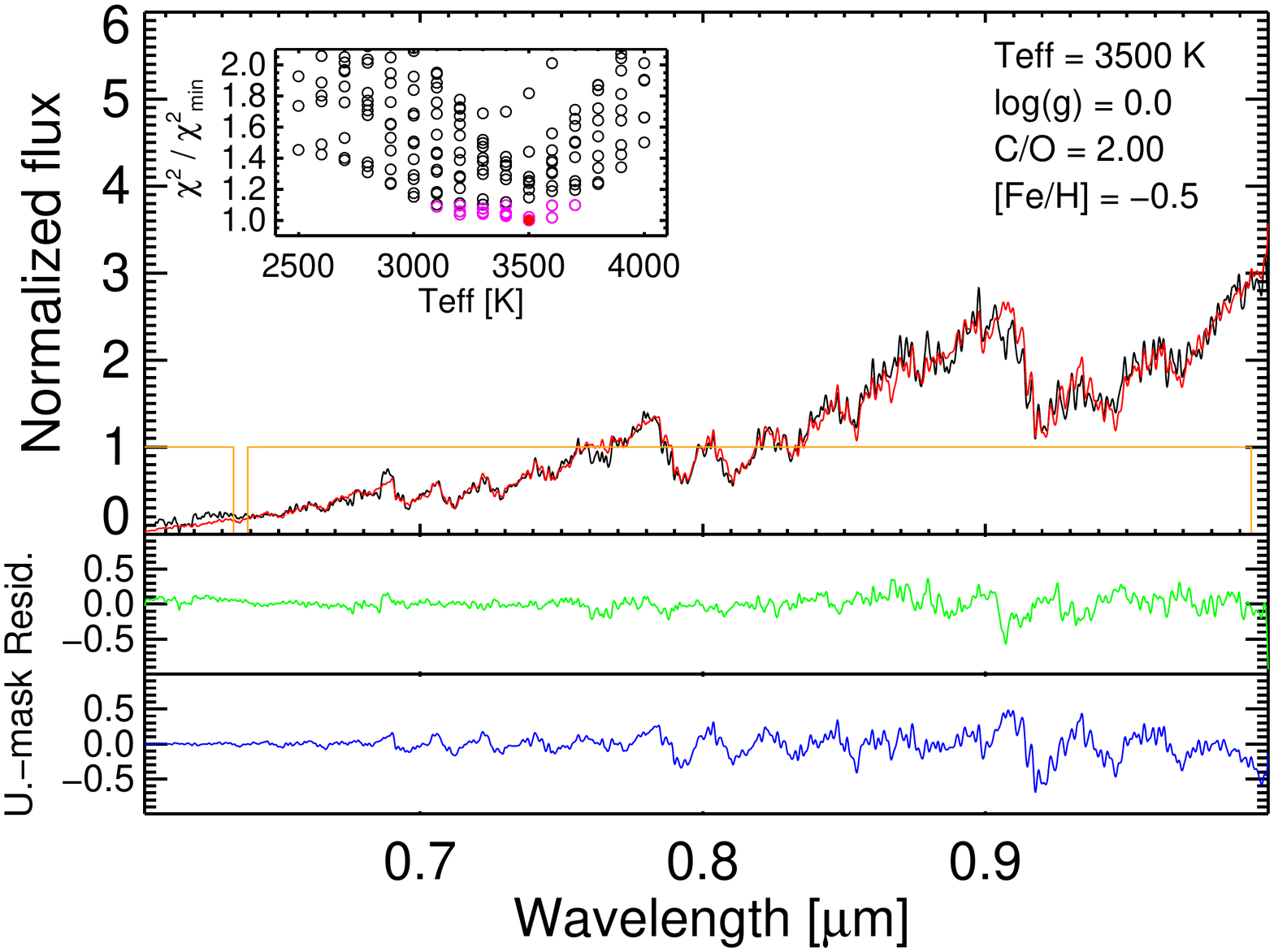}}
\subfloat[NIR]{\includegraphics[trim=30 10 30 65, clip,width=0.45\hsize]{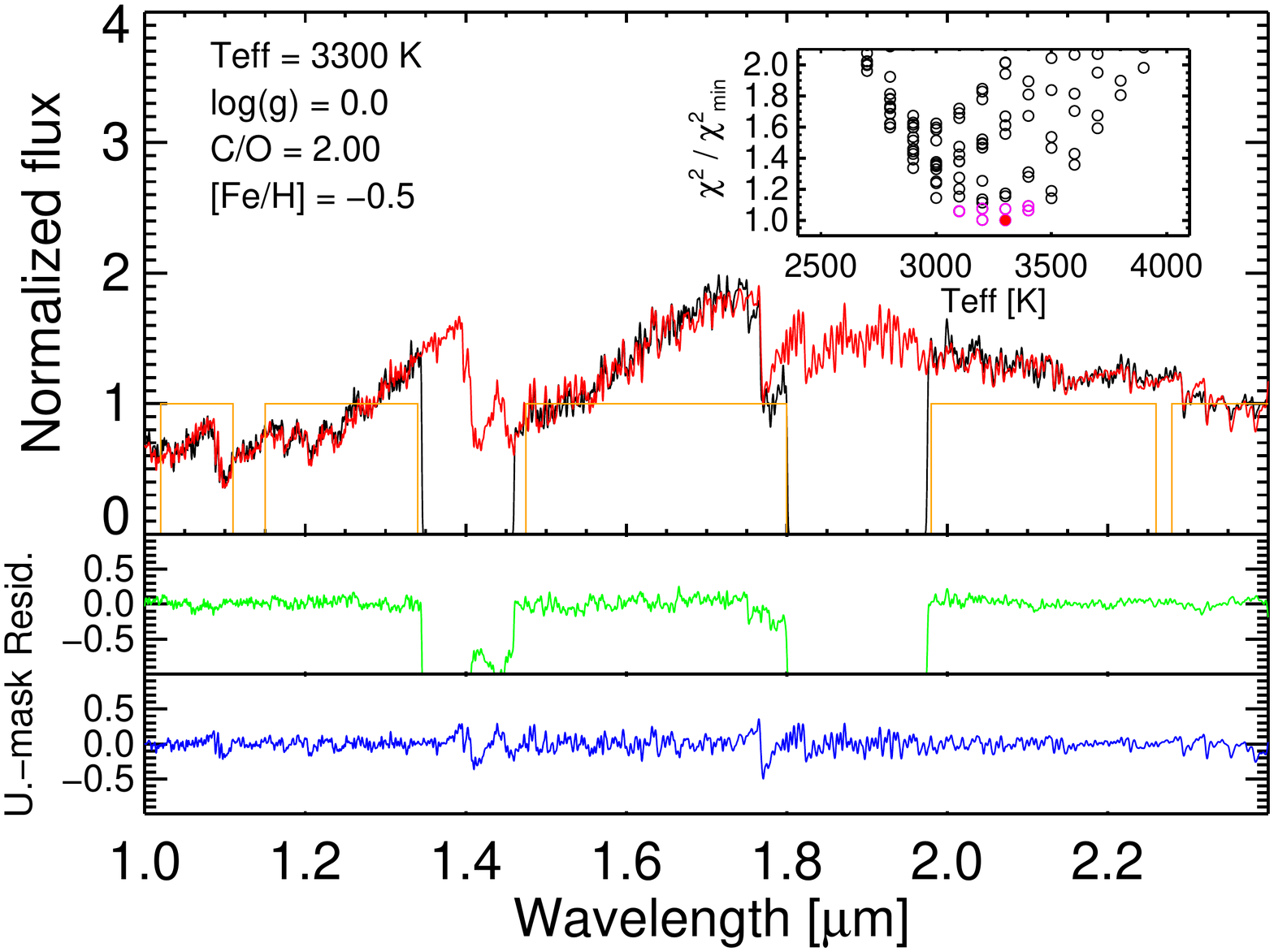}}
\caption{Best-fitting models for {[}ABC89] Cir 18 (Group C). Same legend as for Fig.~\ref{fit_beg_a}.}
\end{figure*}


\begin{figure*}[h] 
\centering
\subfloat[VIS]{\includegraphics[trim=30 10 30 65, clip,width=0.45\hsize]{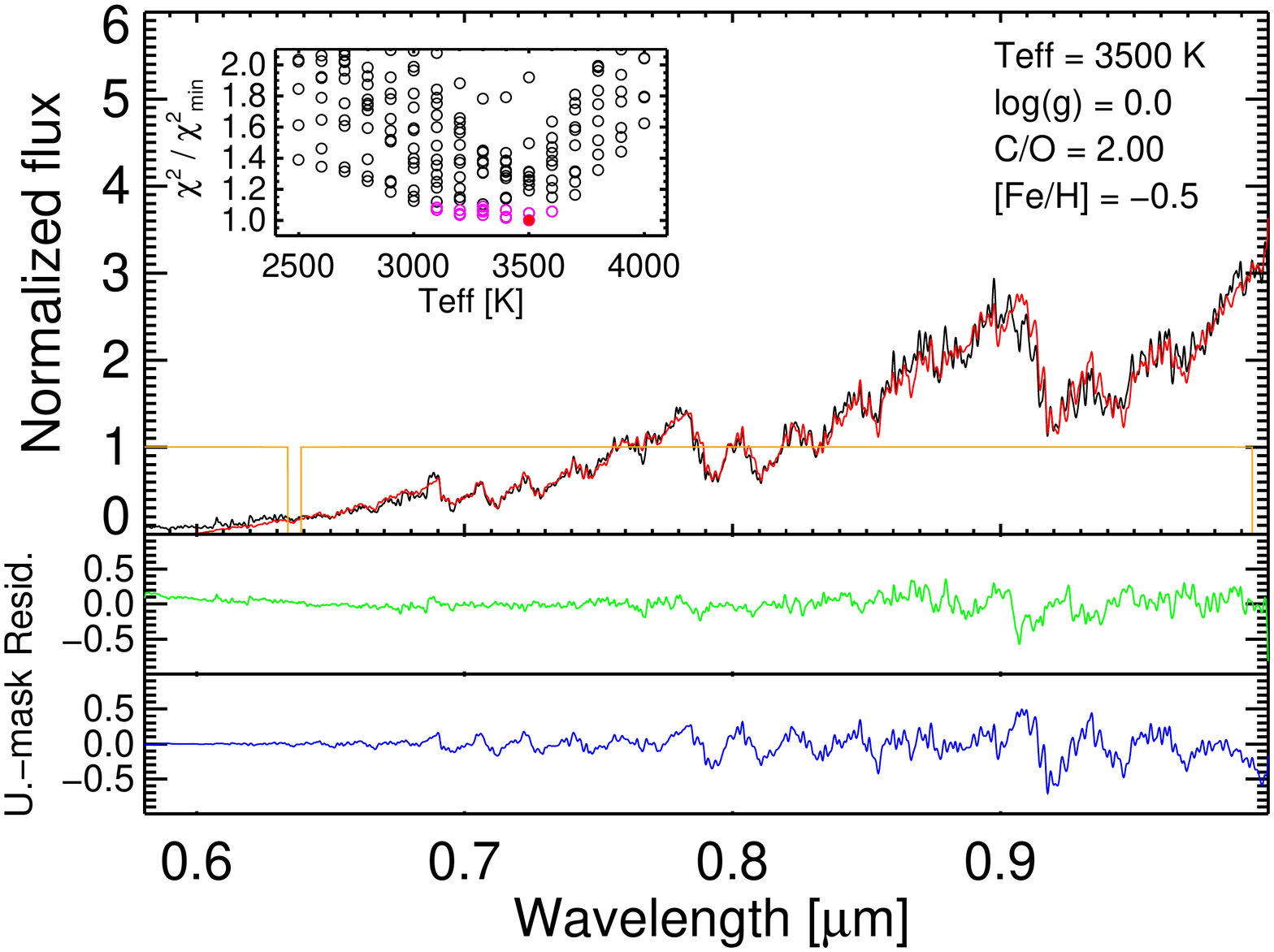}} 
\subfloat[NIR]{\includegraphics[trim=30 10 30 65, clip,width=0.45\hsize]{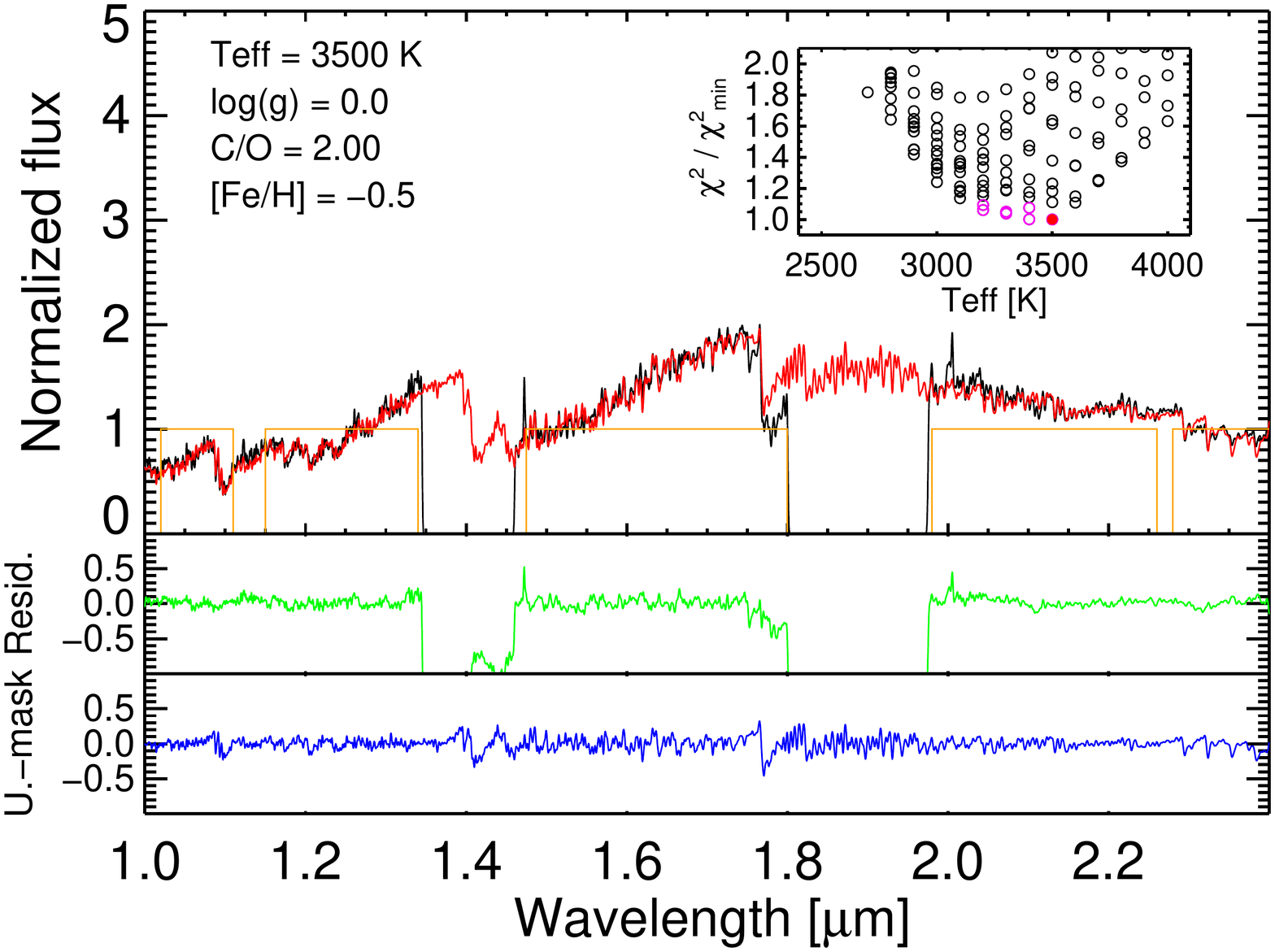}}
\caption{Best-fitting models for {[}ABC89] Cir 18 (Group C). Same legend as for Fig.~\ref{fit_beg_a}.}
\label{fit_end_c}
\end{figure*}


\begin{figure*}[h] 
\centering
\subfloat[VIS]{\includegraphics[trim=40 10 20 20, clip,width=0.45\hsize]{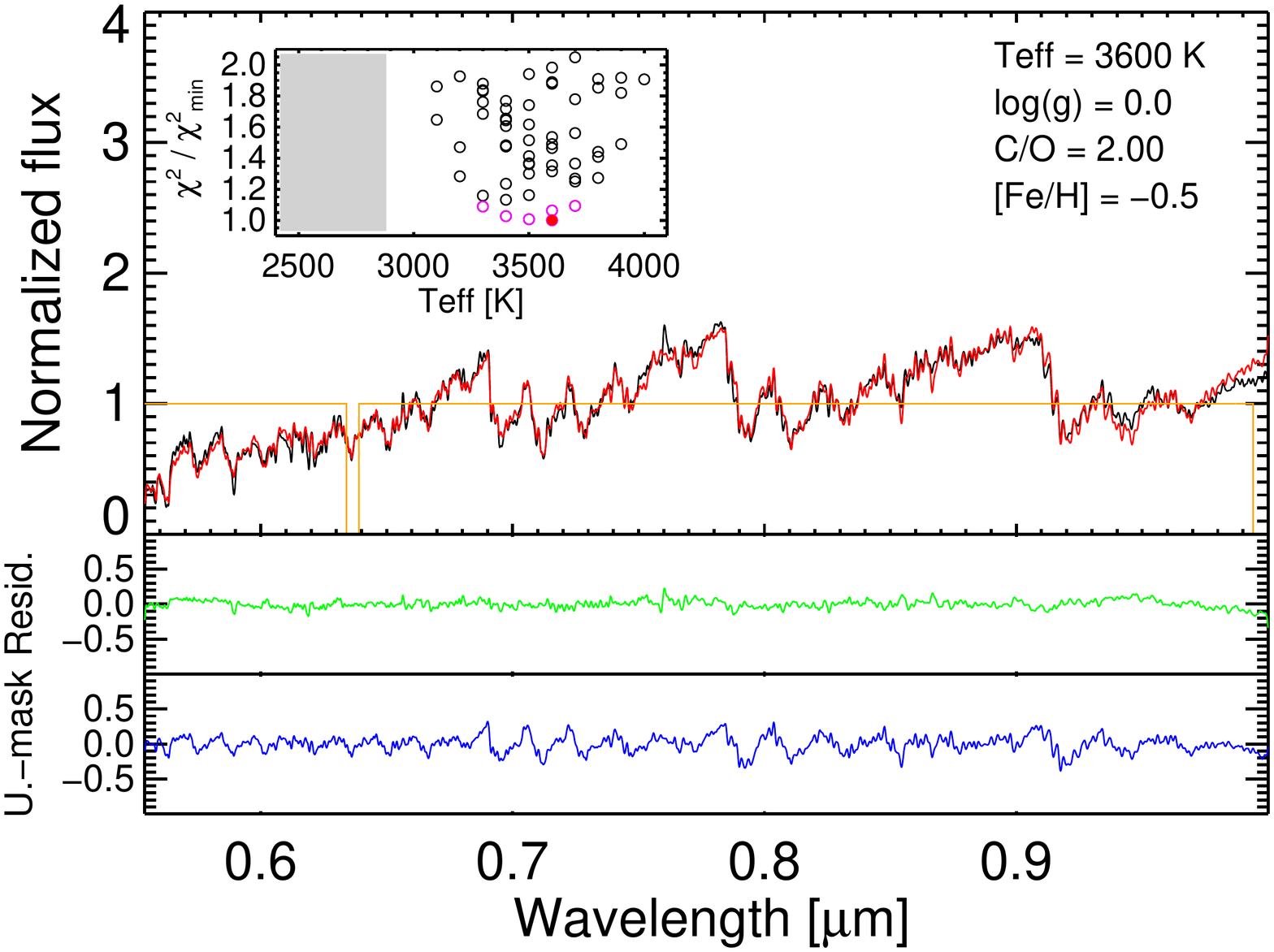}} 
\subfloat[NIR]{\includegraphics[trim=30 10 30 65, clip,width=0.45\hsize]{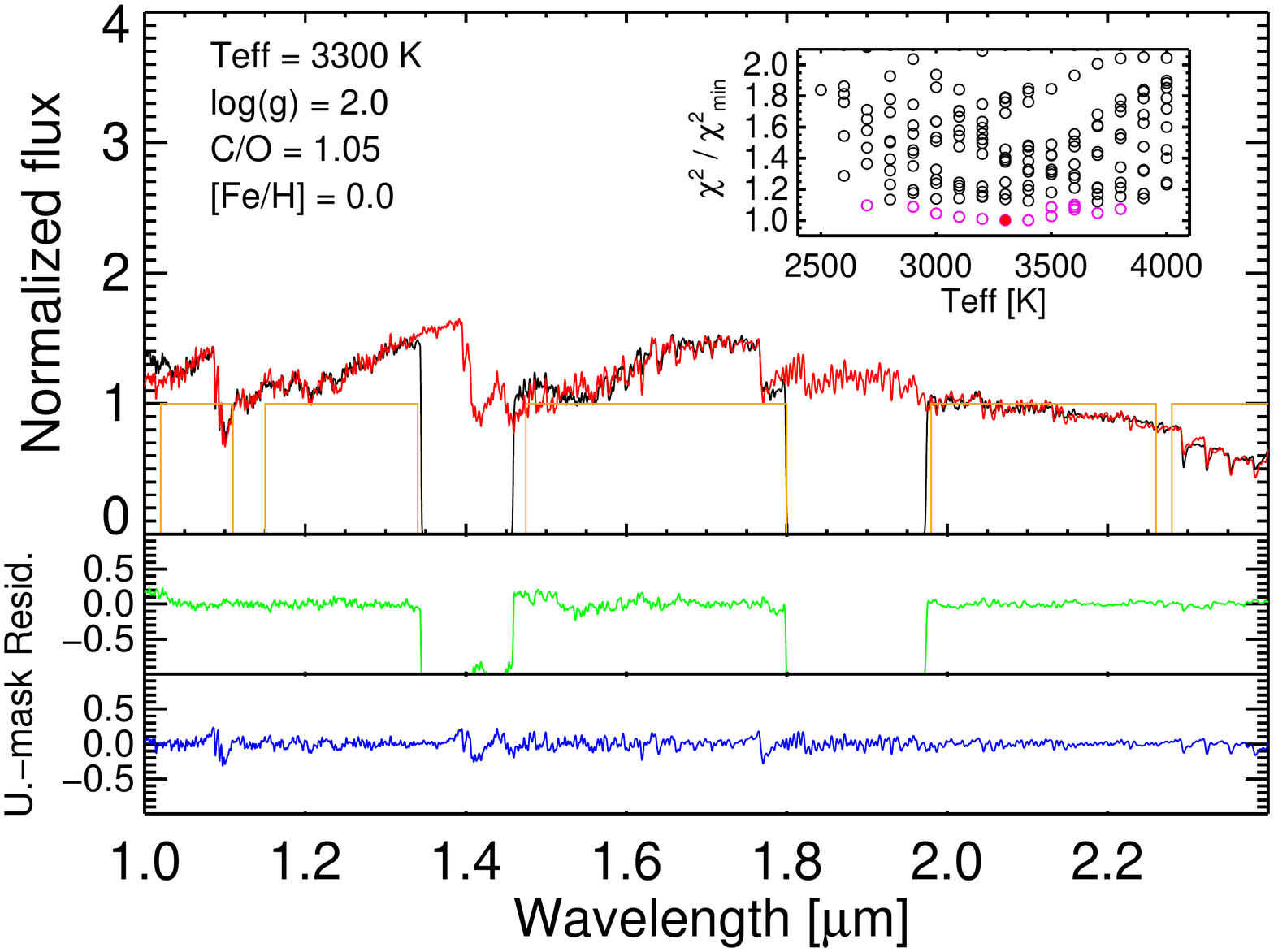}}
\caption{Best-fitting models for SHV 0500412-684054 (Group D). Same legend as for Fig.~\ref{fit_beg_a}.}
\label{fit_beg_d}
\end{figure*}

\begin{figure*}[h] 
\centering
\subfloat[VIS]{\includegraphics[trim=30 10 30 65, clip,width=0.45\hsize]{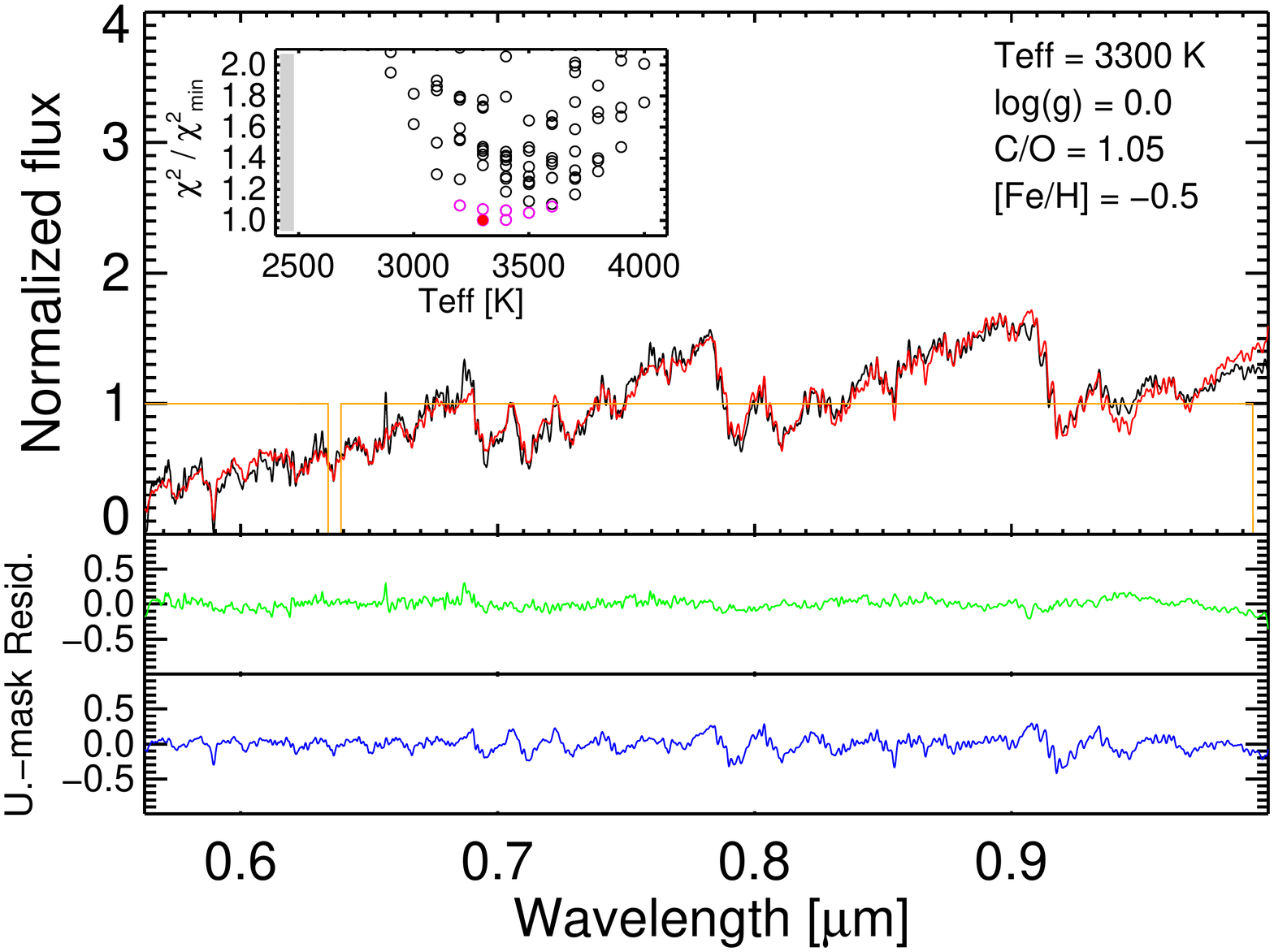}} 
\subfloat[NIR]{\includegraphics[trim=30 10 30 65, clip,width=0.45\hsize]{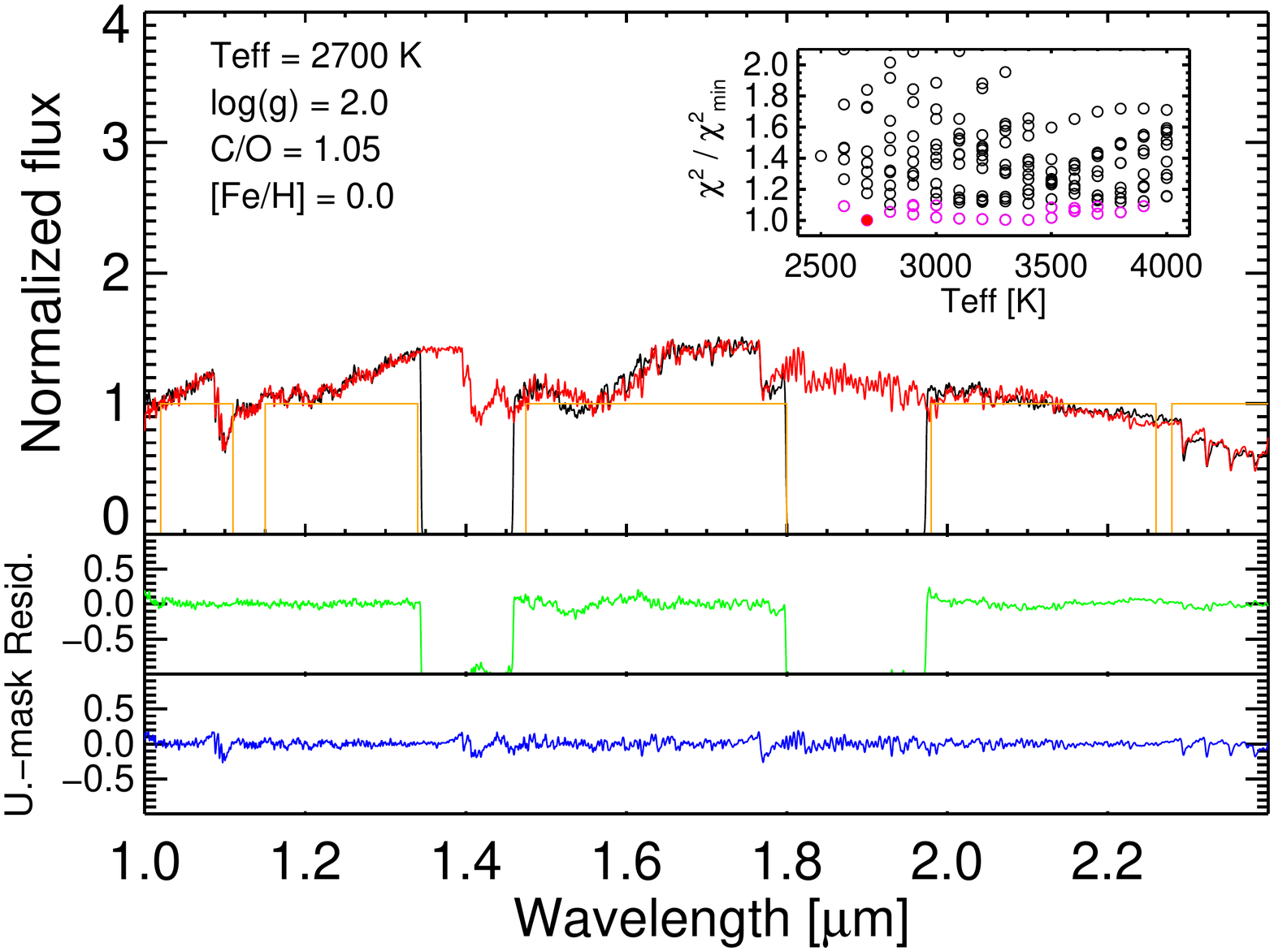}}
\caption{Best-fitting models for SHV 0502469-692418 (Group D). Same legend as for Fig.~\ref{fit_beg_a}.}
\label{ex_grp_d_hcn}
\end{figure*}

\begin{figure*}[h] 
\centering
\subfloat[VIS]{\includegraphics[trim=30 10 30 65, clip,width=0.45\hsize]{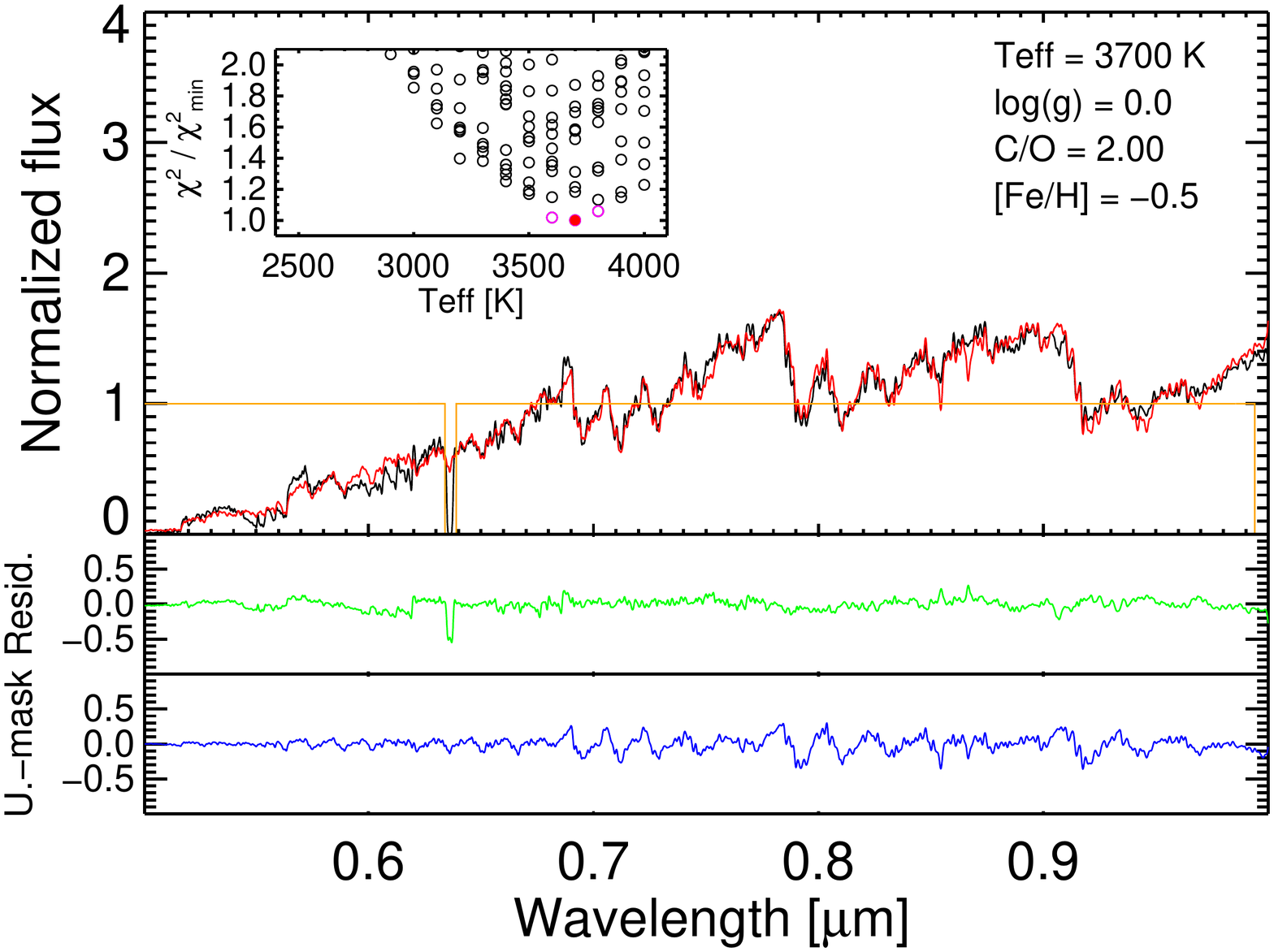}} 
\subfloat[NIR]{\includegraphics[trim=30 10 30 65, clip,width=0.45\hsize]{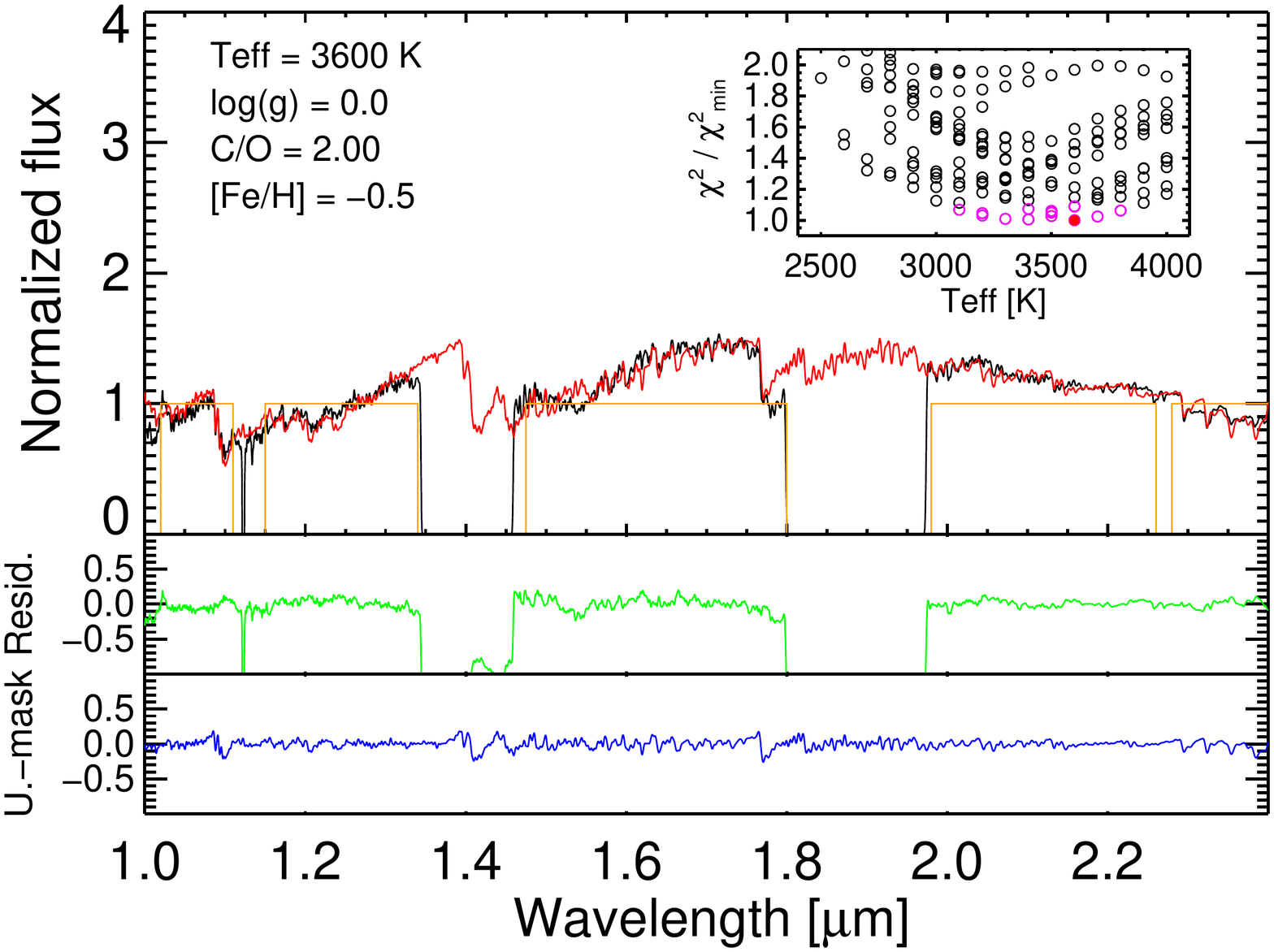}}
\caption{Best-fitting models for SHV 0520505-705019 (Group D). Same legend as for Fig.~\ref{fit_beg_a}.}
\end{figure*}

\begin{figure*}[h] 
\centering
\subfloat[VIS]{\includegraphics[trim=30 10 30 65, clip,width=0.45\hsize]{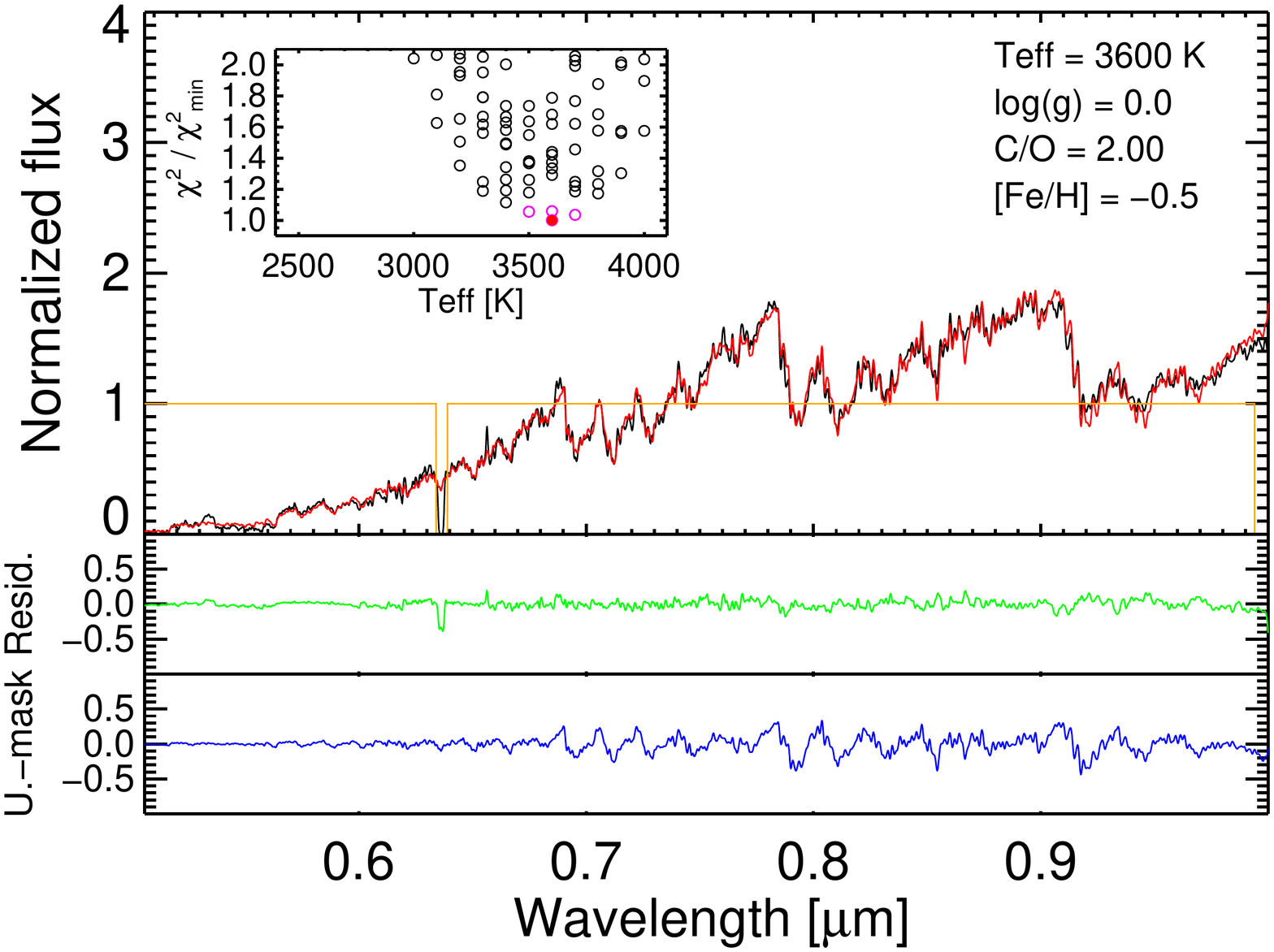}} 
\subfloat[NIR]{\includegraphics[trim=30 10 30 65, clip,width=0.45\hsize]{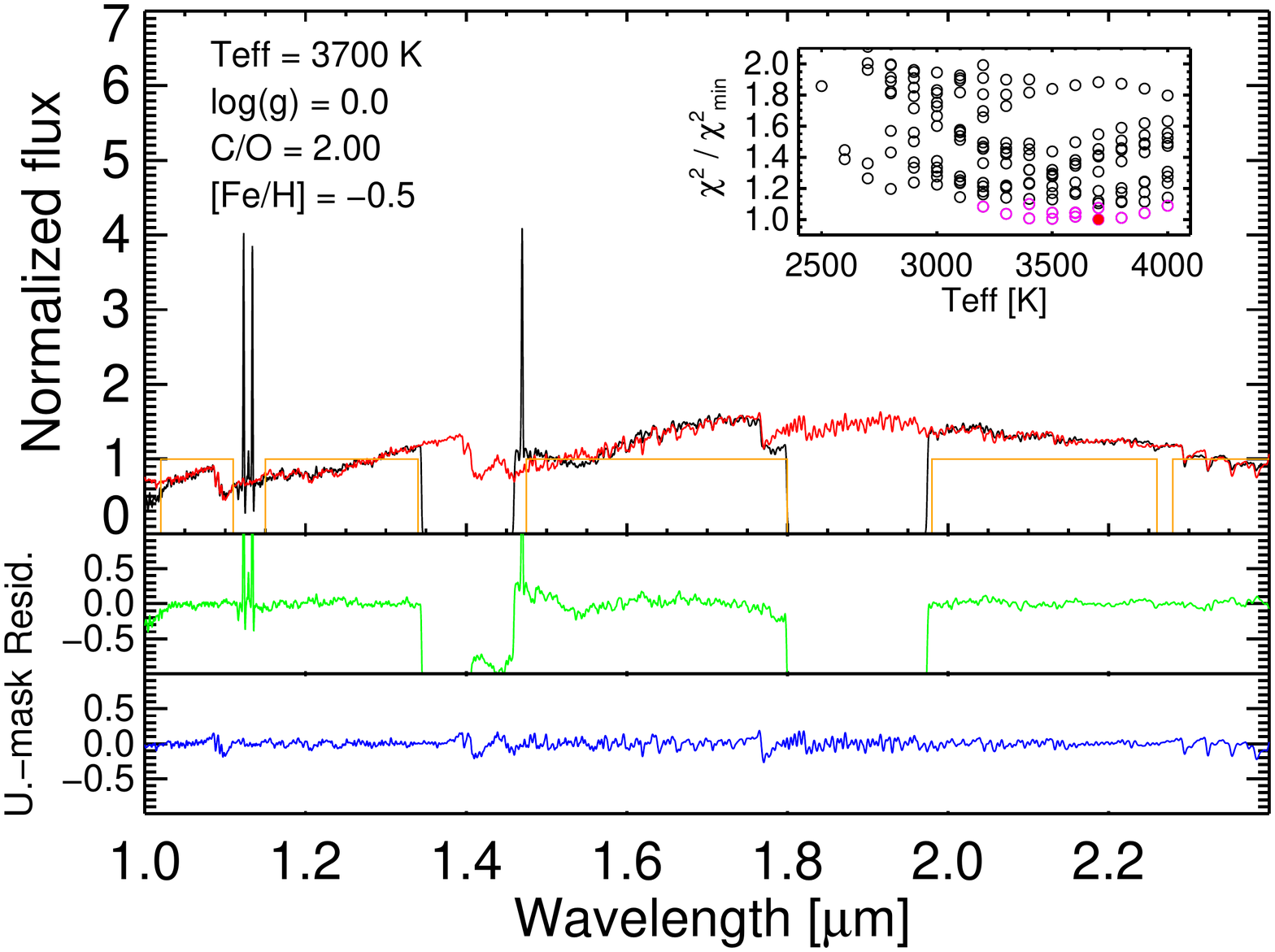}}
\caption{Best-fitting models for SHV 0518222-750327 (Group D). Same legend as for Fig.~\ref{fit_beg_a}.}
\end{figure*}

\begin{figure*}[h] 
\centering
\subfloat[VIS]{\includegraphics[trim=30 10 30 65, clip,width=0.45\hsize]{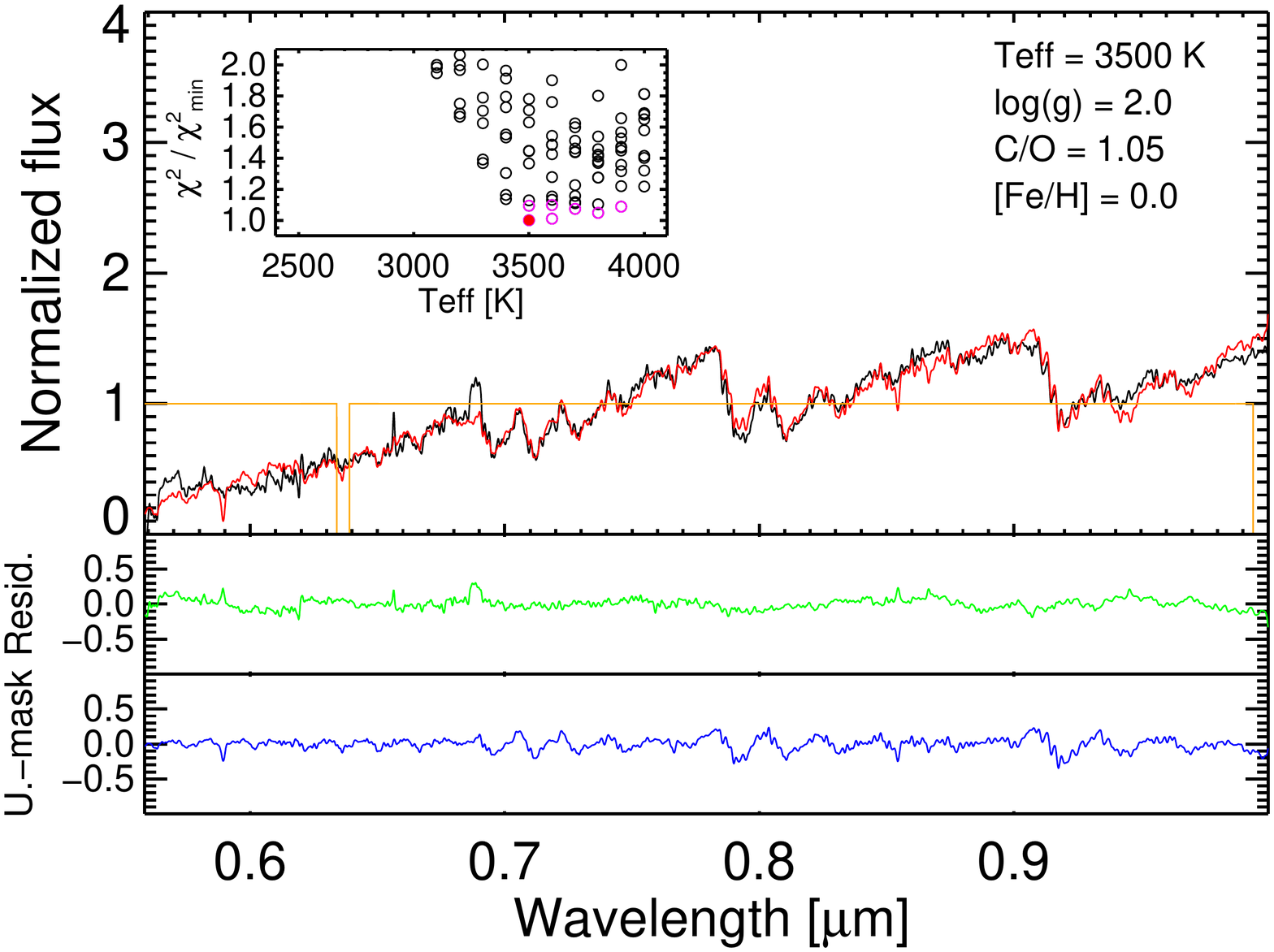}} 
\subfloat[NIR]{\includegraphics[trim=30 10 30 65, clip,width=0.45\hsize]{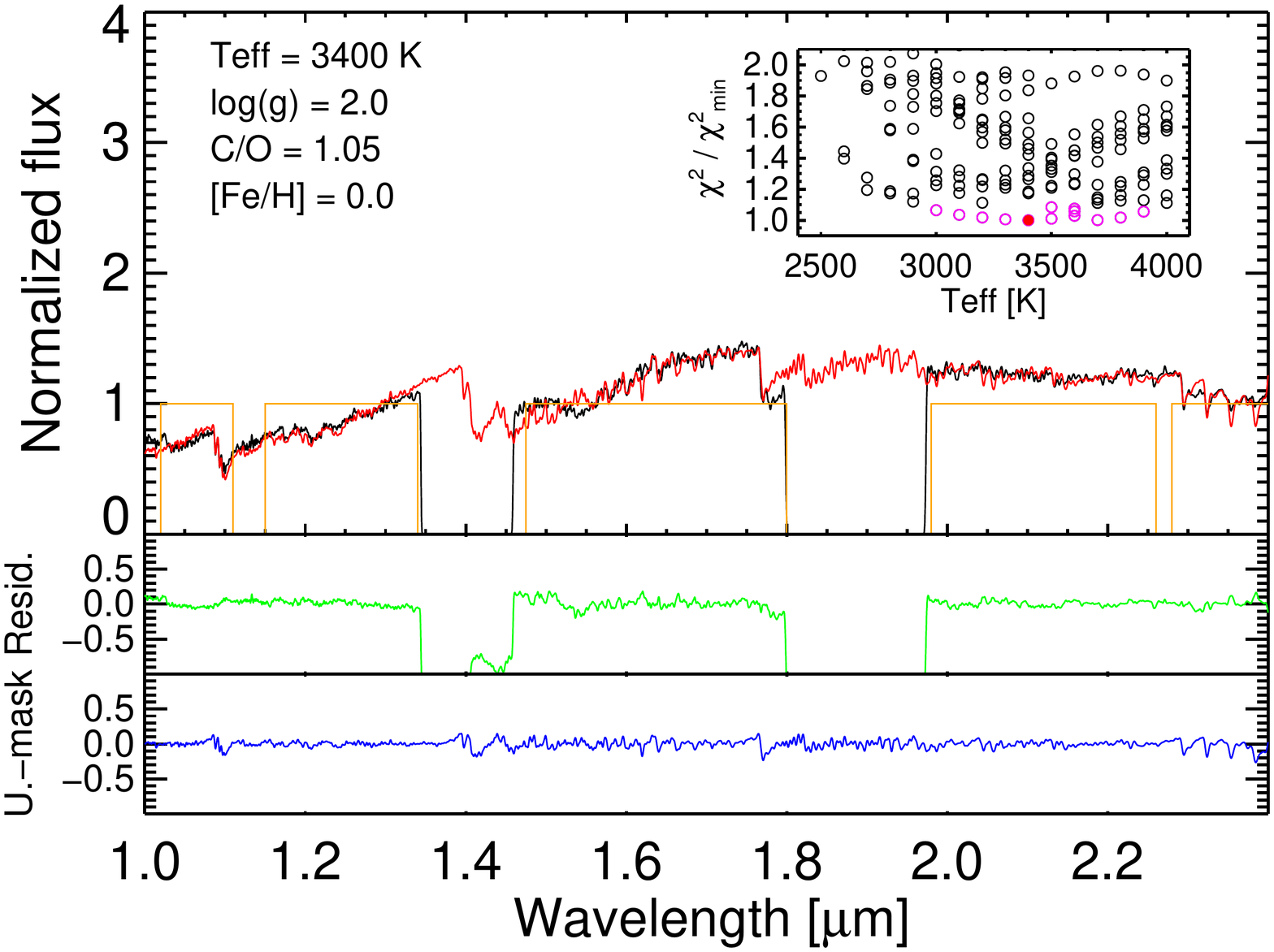}}
\caption{Best-fitting models for SHV 0527072-701238 (Group D). Same legend as for Fig.~\ref{fit_beg_a}.}
\end{figure*}

\begin{figure*}[h] 
\centering
\subfloat[VIS]{\includegraphics[trim=30 10 30 65, clip,width=0.45\hsize]{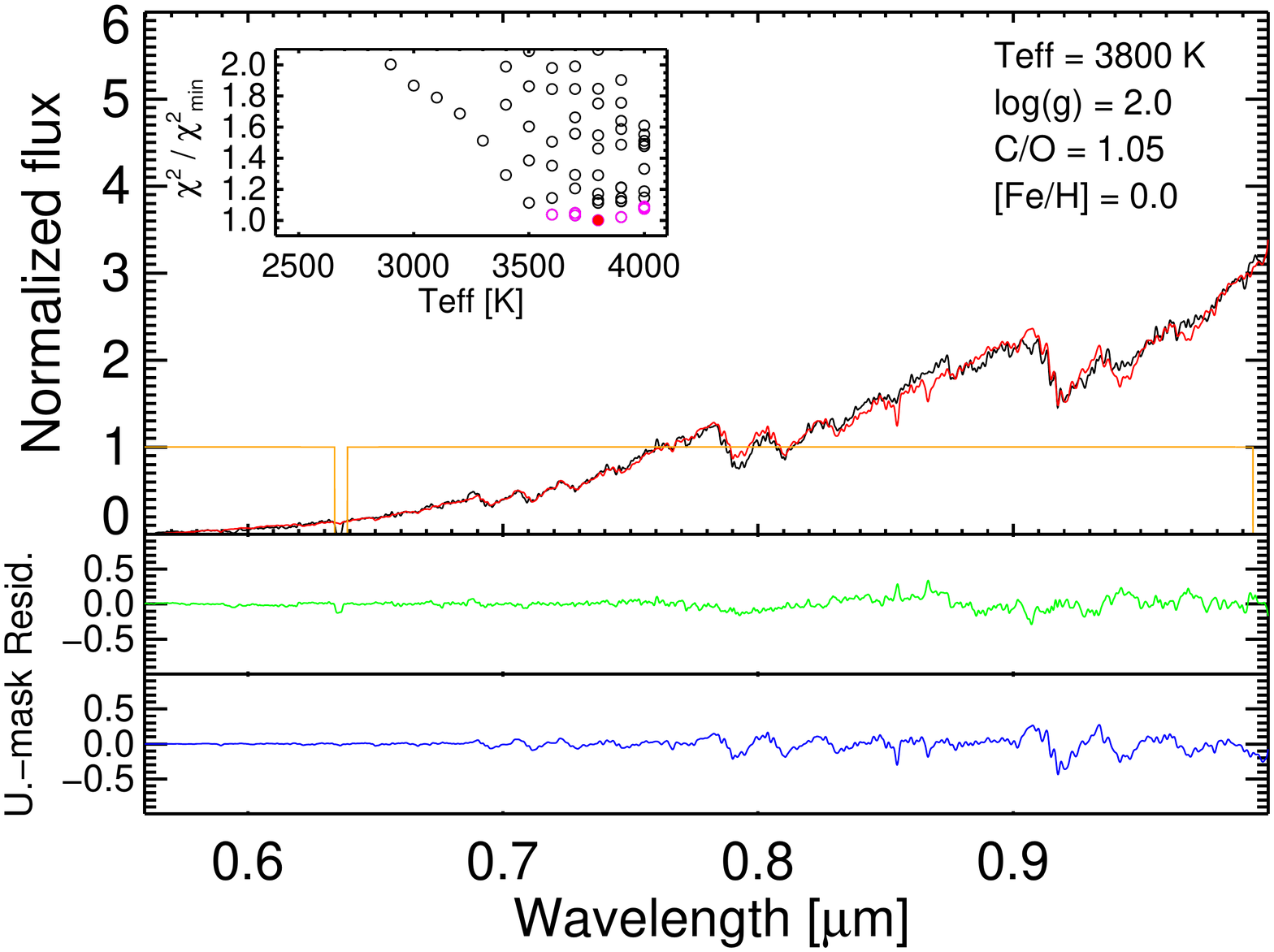}} 
\subfloat[NIR]{\includegraphics[trim=30 10 30 65, clip,width=0.45\hsize]{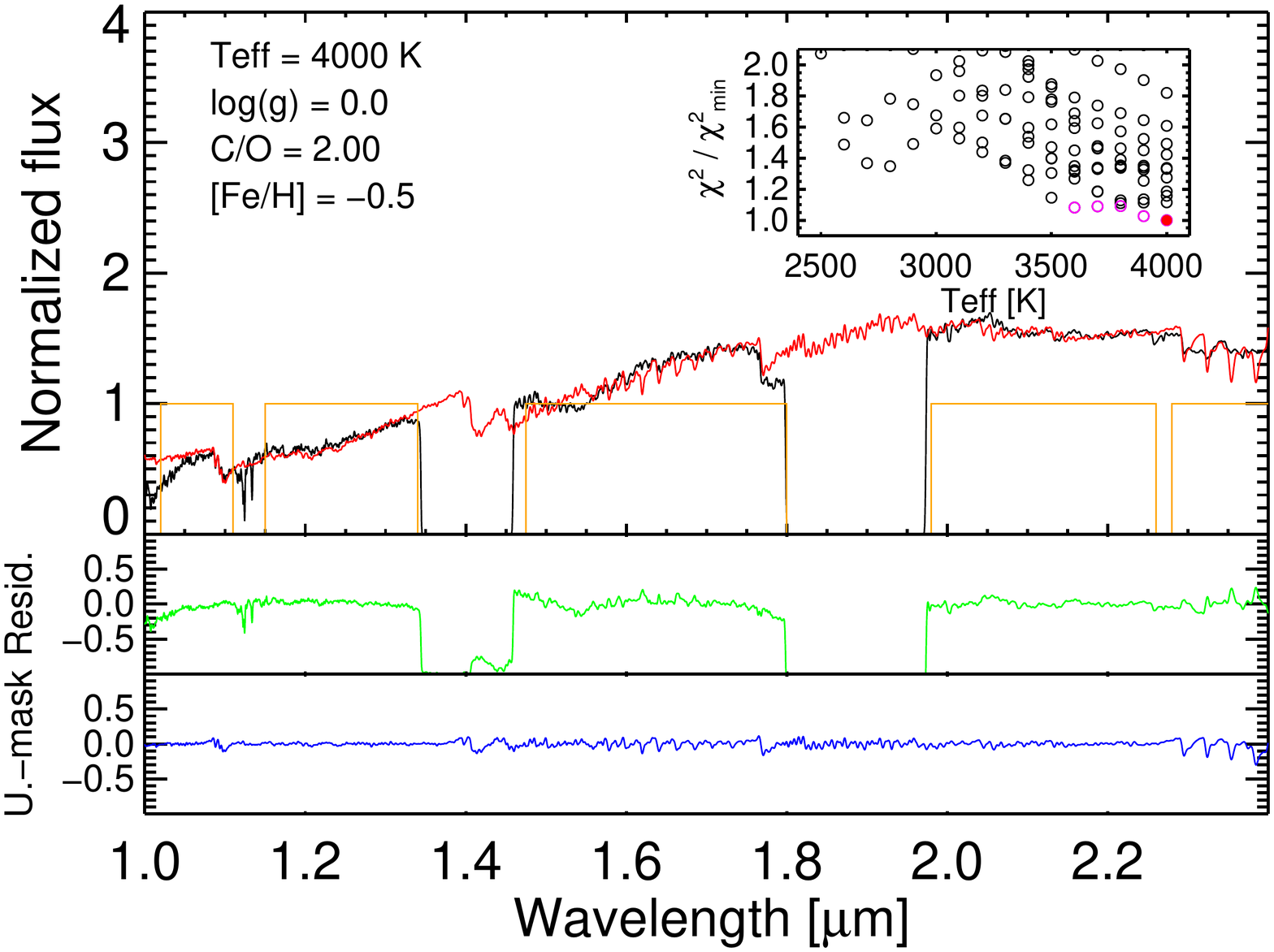}}
\caption{Best-fitting models for SHV 0525478-690944 (Group D). Same legend as for Fig.~\ref{fit_beg_a}.}
\label{ex_grp_d_pb1}
\end{figure*}

\begin{figure*}[h] 
\centering
\subfloat[VIS]{\includegraphics[trim=30 10 30 65, clip,width=0.45\hsize]{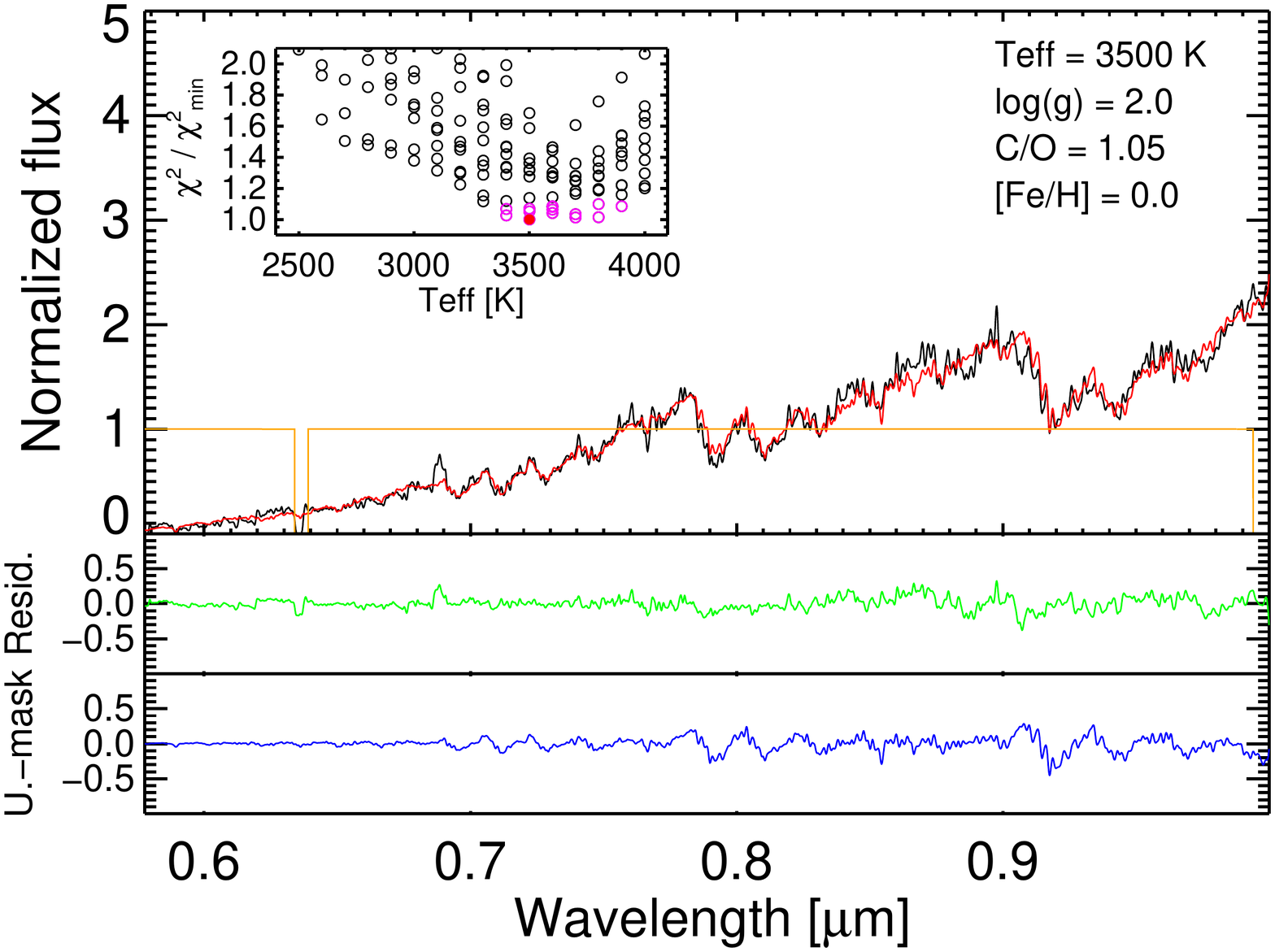}} 
\subfloat[NIR]{\includegraphics[trim=30 10 30 65, clip,width=0.45\hsize]{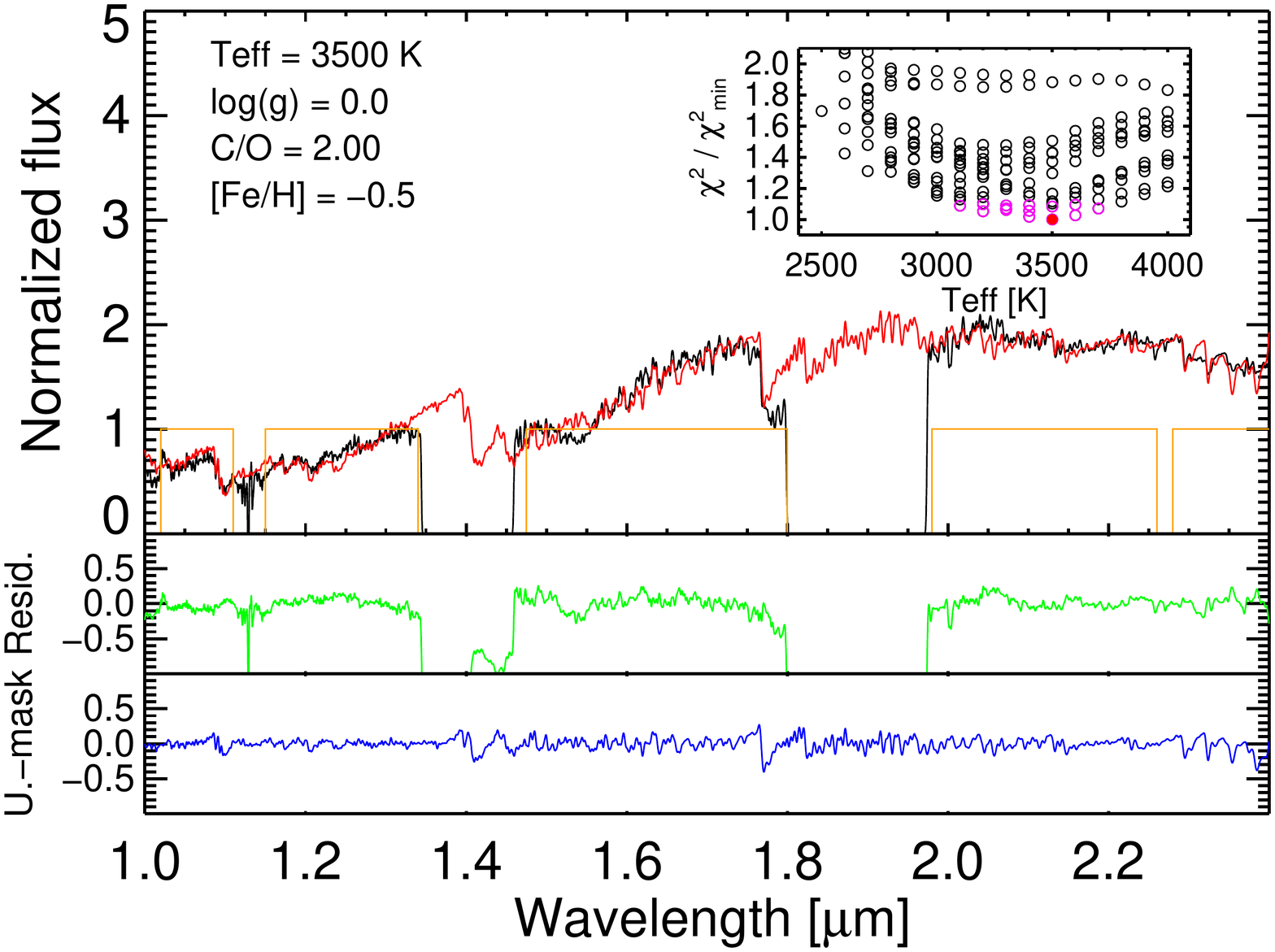}}
\caption{Best-fitting models for SHV 0536139-701604 (Group D). Same legend as for Fig.~\ref{fit_beg_a}.}
\end{figure*}


\begin{figure*}[h] 
\centering
\subfloat[VIS]{\includegraphics[trim=30 10 30 65, clip,width=0.45\hsize]{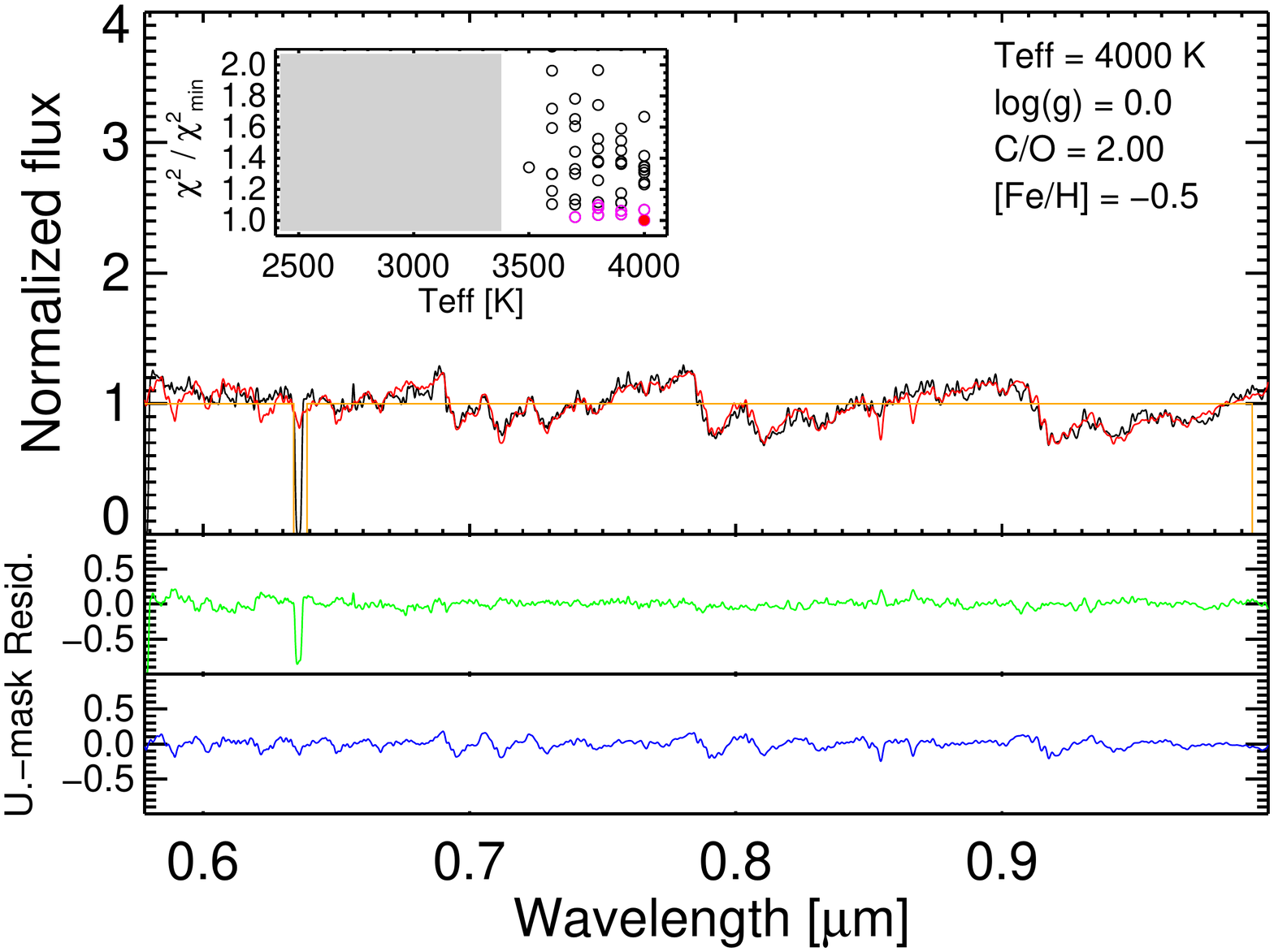}} 
\subfloat[NIR]{\includegraphics[trim=30 10 30 65, clip,width=0.45\hsize]{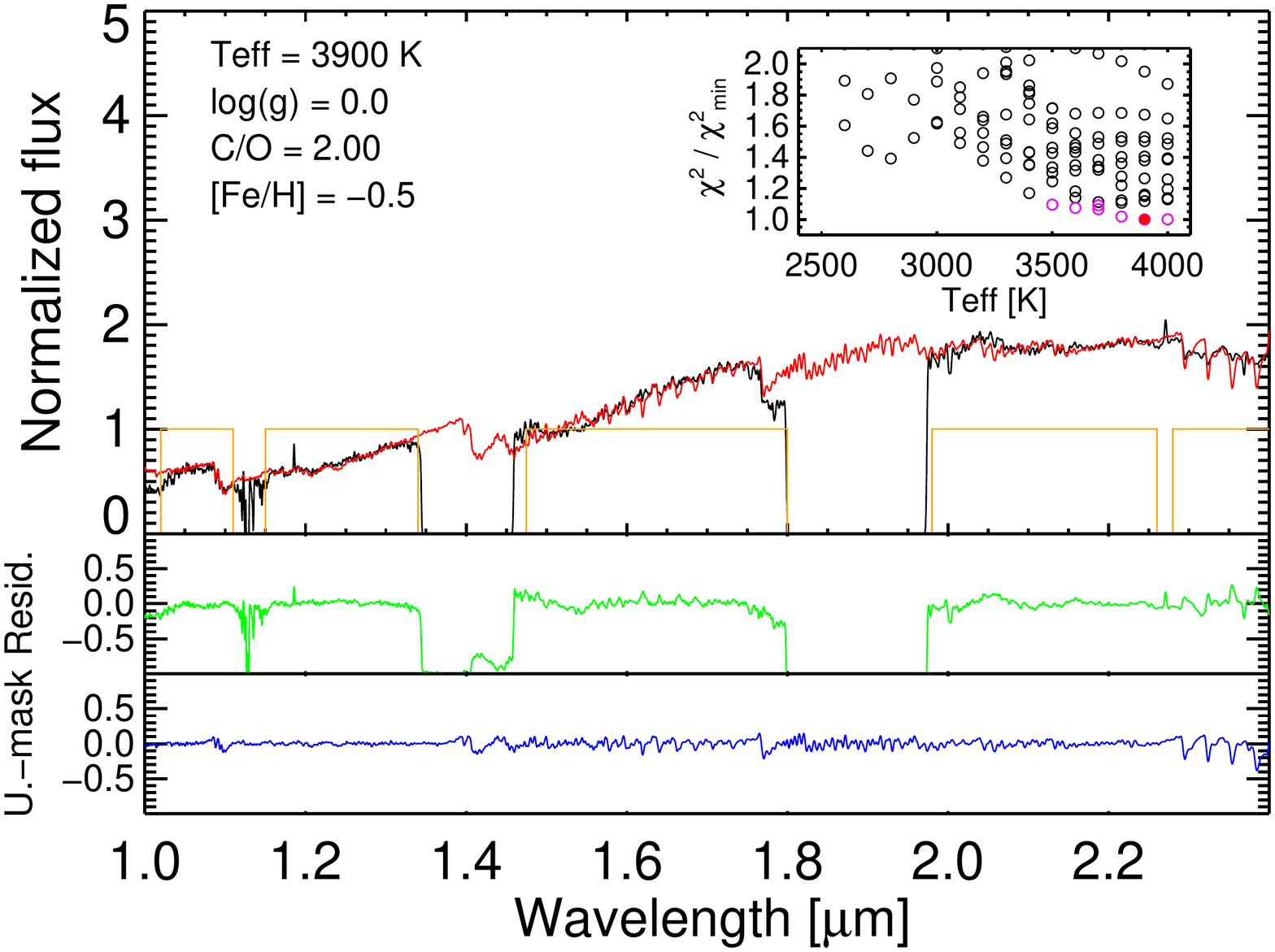}}
\caption{Best-fitting models for SHV 0528537-695119 (Group D). Same legend as for Fig.~\ref{fit_beg_a}.}
\label{fit_end_d}
\end{figure*}


\section{Color-color plots}

\begin{figure*}[h]
	\begin{center}
		\includegraphics[trim=40 40 0 440]{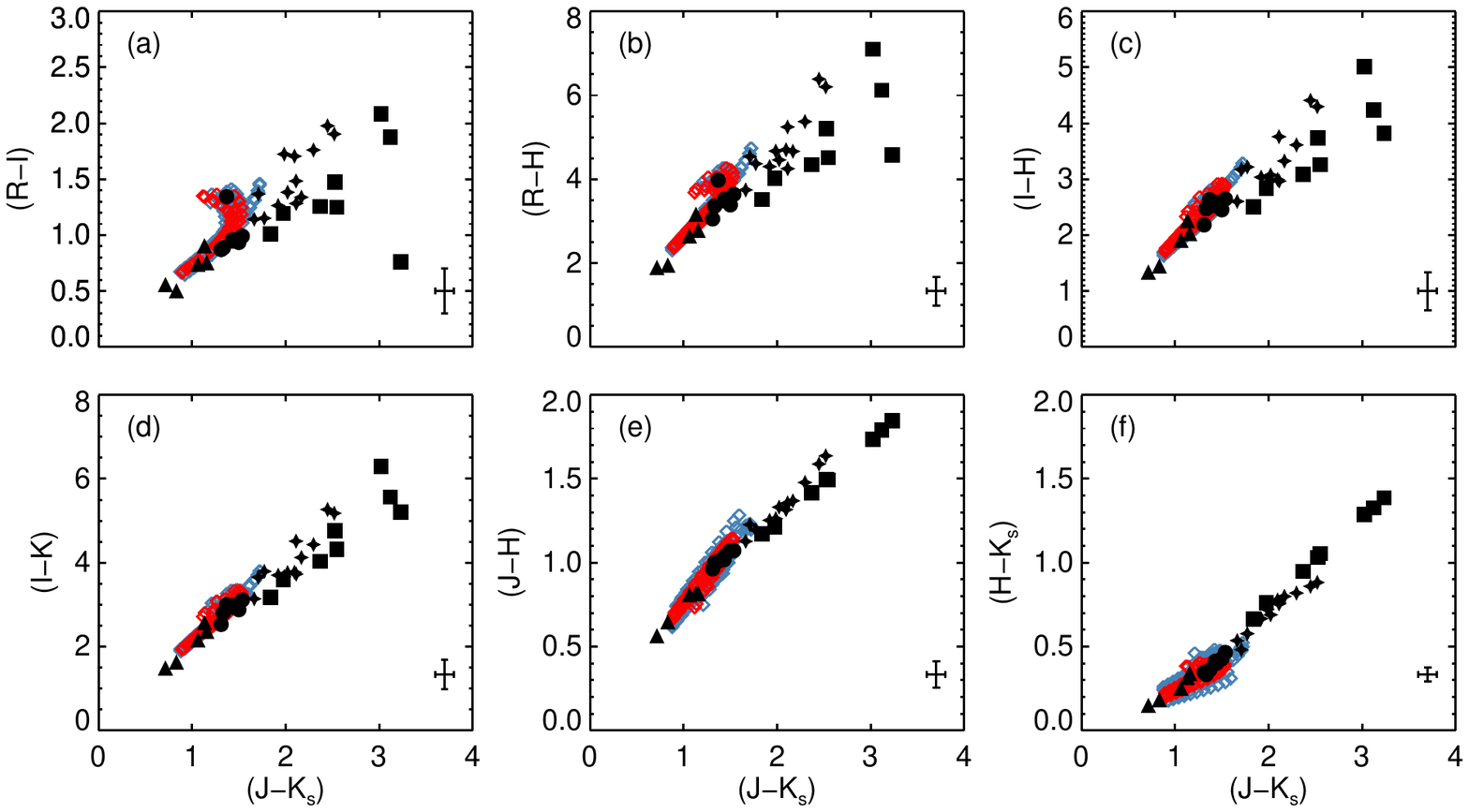}
                \caption{Some color-color plots containing our
                  sample of observed carbon stars (black symbols) and the models (colored diamonds). The triangles are for stars from Group A, the circles for Group B, the stars for Group C, and the squares for Group D. The open diamonds represent the models at solar (in blue) and subsolar metalliticity (red).
The bars show the $\pm1\sigma$ root-mean-square deviation
 of our photometry with respect to the literature (large-amplitude variables 						  excluded). This is an upper limit of the uncertainties 
 in the flux calibration and any possible residual variability.}
	     \label{plot_color_color}
    	\end{center}
\end{figure*}

Figure~\ref{plot_color_color} shows the color-color plot for our observations and the grid of models.

\clearpage

\end{appendix}


\end{document}